%% file: entwurf_newnew.tex
\newcommand{\pmu}{\partial_{\mu}}

\newcommand{\nmu}{\nabla_{\mu}}
\newcommand{\nmo}{\nabla^{\mu}}
\newcommand{\nnu}{\nabla_{\nu}}

\newcommand{\MAmu}{{\mathcal A}_{\mu}}

\newcommand{\MAnu}{{\mathcal A}_{\nu}}
\newcommand{\MAlu}{{\mathcal A}_{\lambda}}

\newcommand{\exMAmu}{{}^{(ex)}\!{\mathcal A}_{\mu}}

\newcommand{\exAmu}{{}^{(ex)}\!A_{\mu}}
\newcommand{\exAnu}{{}^{(ex)}\!A_{\nu}}

\newcommand{\Aaomu}{A^{\alpha}_{\;\;\mu}}

\newcommand{\Agomu}{A^{\gamma}_{\;\;\mu}}
\newcommand{\Aamu}{A^{a}_{\;\;\mu}}
\newcommand{\Aemu}{A^{1}_{\;\;\mu}}
\newcommand{\Aenu}{A^{1}_{\;\;\nu}}
\newcommand{\Azmu}{A^{2}_{\;\;\mu}}
\newcommand{\Aznu}{A^{2}_{\;\;\nu}}
\newcommand{\Bmu}{B_{\mu}}
\newcommand{\Bsmu}{B^{*}_{\;\;\mu}}
\newcommand{\Bmo}{B^{\mu}}
\newcommand{\Bsmo}{B^{*\mu}}
\newcommand{\Bnu}{B_{\nu}}

\newcommand{\Bsnu}{B_{\nu}^{*}}

\newcommand{\Cabg}{C^{\alpha}_{\;\;\beta\gamma}}

\newcommand{\Cbga}{C^{\beta}_{\;\;\gamma\alpha}}

\newcommand{\MDmu}{{\mathcal D}_{\mu}}
\newcommand{\MDnu}{{\mathcal D}_{\nu}}
\newcommand{\MDmo}{{\mathcal D}^{\mu}}

\newcommand{\MDlu}{{\mathcal D}_{\lambda}}

\newcommand{\Dmu}{D_{\mu}}
\newcommand{\Dmo}{D^{\mu}}

\newcommand{\MFmunu}{{\mathcal F}_{\mu\nu}}

\newcommand{\exFmunu}{{}^{(ex)}\!F_{\mu \nu}}

\newcommand{\Famunu}{F^{a}_{\;\;\mu\nu}}
\newcommand{\Faomunu}{F^{\alpha}_{\;\;\mu\nu}}

\newcommand{\Femunu}{F^{1}_{\;\;\mu\nu}}
\newcommand{\Fzmunu}{F^{2}_{\;\;\mu\nu}}

\newcommand{\exfnu}{{}^{(ex)}\!f_{\nu}}

\newcommand{\Sfnu}{{}^{(S)}\!f_{\nu}}

\newcommand{\Gmunu}{G_{\mu\nu}}
\newcommand{\Gmonu}{G^{\mu}_{\;\;\nu}}
\newcommand{\gmunu}{g_{\mu\nu}}

\newcommand{\glunu}{g_{\lambda\nu}}
\newcommand{\Gsmunu}{G^{*}_{\;\;\mu\nu}}
\newcommand{\Gsmonu}{G^{*\mu}_{\;\;\;\;\nu}}

\newcommand{\MHmu}{{\mathcal H}_{\mu}}

\newcommand{\MHmo}{{\mathcal H}^{\mu}}
\newcommand{\MHnu}{{\mathcal H}_{\nu}}

\newcommand{\hmu}{h_{\mu}}

\newcommand{\hsmu}{h^{*}_{\;\mu}}

\newcommand{\MI}{{\mathcal I}}
\newcommand{\MJmu}{{\mathcal J}_{\mu}}

\newcommand{\MJnu}{{\mathcal J}_{\nu}}

\newcommand{\jmu}{j_{\mu}}

\newcommand{\jalomu}{j^{\alpha}_{\;\;\mu}}
\newcommand{\jaomu}{j^{a}_{\;\;\mu}}
\newcommand{\jaumu}{j_{a\mu}}
\newcommand{\jeomu}{j^{1}_{\;\;\mu}}

\newcommand{\jzomu}{j^{2}_{\;\;\mu}}

\newcommand{\jdomu}{j^{3}_{\;\;\mu}}
\newcommand{\jvomu}{j^{4}_{\;\;\mu}}
\newcommand{\jeonu}{j^{1}_{\;\;\nu}}
\newcommand{\jzonu}{j^{2}_{\;\;\nu}}
\newcommand{\jdonu}{j^{3}_{\;\;\nu}}
\newcommand{\jvonu}{j^{4}_{\;\;\nu}}

\newcommand{\jalumu}{j_{\alpha\mu}}
\newcommand{\jalumo}{j_{\alpha}^{\;\;\mu}}
\newcommand{\jeumu}{j_{1\mu}}
\newcommand{\jzumu}{j_{2\mu}}
\newcommand{\jdumu}{j_{3\mu}}
\newcommand{\jvumu}{j_{4\mu}}
\newcommand{\jbumu}{j_{\beta\mu}}
\newcommand{\jalonu}{j^{\alpha}_{\;\;\nu}}
\newcommand{\jbomu}{j^{\beta}_{\;\;\mu}}

\newcommand{\exjmu}{{}^{(ex)}\!j_{\mu}}
\newcommand{\exjnu}{{}^{(ex)}\!j_{\nu}}

\newcommand{\Kaubu}{K_{\alpha\beta}}

\newcommand{\Kaobo}{K^{\alpha\beta}}

\newcommand{\keumu}{k_{1\mu}}
\newcommand{\kzumu}{k_{2\mu}}
\newcommand{\Tmunu}{T_{\mu\nu}}

\newcommand{\MFu}{{\mathfrak u}}
\newcommand{\Mvaumu}{\upsilon_{\alpha\mu}}
\newcommand{\Mvbumu}{\upsilon_{\beta\mu}}

\newcommand{\Gmu}{\Gamma_{\mu}}
\newcommand{\Gmo}{\Gamma^{\mu}}
\newcommand{\gmu}{\gamma_{\mu}}
\newcommand{\gmo}{\gamma^{\mu}}
\newcommand{\Gnu}{\Gamma_{\nu}}
\newcommand{\Glu}{\Gamma_{\lambda}}
\newcommand{\Smunu}{\Sigma_{\mu\nu}}

\newcommand{\Smono}{\Sigma^{\mu\nu}}
\newcommand{\talu}{\tau_{\alpha}}
\newcommand{\tbu}{\tau_{\beta}}
\newcommand{\tgu}{\tau_{\gamma}}

\newcommand{\wmu}{\omega_{\mu}}

\newcommand{\wabm}{\omega^{\alpha}_{\;\;\beta\mu}}
\newcommand{\wbam}{\omega^{\beta}_{\;\;\alpha\mu}}
\newcommand{\wbal}{\omega^{\beta}_{\;\;\alpha\lambda}}

\newcommand{\I}{{\mathcal{I}}}

\newcommand{\M}{{\mathcal{M}}}

\newcommand{\X}{X}

\newcommand{\vvecr}{(\vec{r})}

\newcommand{\Sa}{{}^1\mathrm{S}_0}
\newcommand{\Sb}{{}^3\mathrm{S}_1}

\newcommand{\exAo}{{}^{(ex)}\!A_0}
\newcommand{\eAo}{{}^{(1)}\!A_0}
\newcommand{\zAo}{{}^{(2)}\!A_0}

\newcommand{\aAo}{{}^{(a)}\! A_0}

\newcommand{\jalunu}{j_{\alpha\nu}}

\newcommand{\alo}{{}^{(a)}\!l_0}

\newcommand{\jbeunu}{j_{\beta\nu}}

\newcommand{\kaumu}{k_{a\mu}}
\newcommand{\eko}{{}^{(1)}\! k_0}
\newcommand{\zko}{{}^{(2)}\! k_0}
\newcommand{\ako}{{}^{(a)}\! k_0}

\newcommand{\pmo}{\partial^{\mu}}
\newcommand{\pno}{\partial^{\nu}}

\newcommand{\aphipm}{{}^{(a)}\!\phi_{\pm}(\vec{r})}

\newcommand{\ephip}{{}^{(1)}\!\phi_+(\vec{r})}
\newcommand{\ephim}{{}^{(1)}\!\phi_-(\vec{r})}
\newcommand{\zphip}{{}^{(2)}\!\phi_+(\vec{r})}
\newcommand{\zphim}{{}^{(2)}\!\phi_-(\vec{r})}

\newcommand{\zetajml}{\zeta^{j,m}_{\;\; l}}
\newcommand{\zetappn}{\zeta^{\frac{1}{2},\frac{1}{2}}_{\;\;\, 0}}
\newcommand{\zetappe}{\zeta^{\frac{1}{2},\frac{1}{2}}_{\;\;\, 1}}
\newcommand{\zetapmn}{\zeta^{\frac{1}{2},-\frac{1}{2}}_{\;\;\, 0}}
\newcommand{\zetapme}{\zeta^{\frac{1}{2},-\frac{1}{2}}_{\;\;\, 1}}

\newcommand{\Ta}{T_{{\bf (}a\!\!\! a\!\!\! a{\bf )}}}

\newcommand{\nlu}{\nabla_{\!\lambda}}

\newcommand{\MBmu}{\mathcal{B}_{\mu}}

\newcommand{\Bkomu}{B^k_{\;\;\mu}}
\newcommand{\Bskomu}{B^{\ast k}_{\;\;\;\;\mu}}

\newcommand{\kamu}{k_{a\mu}}
\newcommand{\aRsp}{{}^{(a)}\!R^{\ast}_+}
\newcommand{\aSsp}{{}^{(a)}\!S^{\ast}_+}
\newcommand{\aSp}{{}^{(a)}\!S_+}
\newcommand{\aRsm}{{}^{(a)}\!R^{\ast}_-}
\newcommand{\aSsm}{{}^{(a)}\!S^{\ast}_-}
\newcommand{\aSm}{{}^{(a)}\!S_-}
\newcommand{\eRsp}{{}^{(1)}\!R^{\ast}_+}
\newcommand{\eSsp}{{}^{(1)}\!S^{\ast}_+}
\newcommand{\zSp}{{}^{(2)}\!S_+}
\newcommand{\eRsm}{{}^{(1)}\!R^{\ast}_-}
\newcommand{\eSsm}{{}^{(1)}\!S^{\ast}_-}
\newcommand{\zSm}{{}^{(2)}\!S_-}
\newcommand{\eSp}{{}^{(1)}\!S_+}
\newcommand{\eSm}{{}^{(1)}\!S_-}
\newcommand{\eSpm}{{}^{(1)}\!S_{\pm}}
\newcommand{\zSpm}{{}^{(2)}\!S_{\pm}}

\newcommand{\zAtu}{{}^{(2)}\!A_{\vartheta}}

\newcommand{\zAru}{{}^{(2)}\!A_r}

\newcommand{\zApu}{{}^{(2)}\!A_{\varphi}}
\newcommand{\Btu}{B_{\vartheta}}
\newcommand{\Bru}{B_r}
\newcommand{\Bpu}{B_{\varphi}}

\newcommand{\aRpm}{{}^{(a)}\!R_{\pm}}
\newcommand{\aSpm}{{}^{(a)}\!S_{\pm}}
\newcommand{\aRp}{{}^{(a)}\!R_+}
\newcommand{\aRm}{{}^{(a)}\!R_-}
\newcommand{\eRpm}{{}^{(1)}\!R_{\pm}}
\newcommand{\eRp}{{}^{(1)}\!R_+}
\newcommand{\eRm}{{}^{(1)}\!R_-}
\newcommand{\zRpm}{{}^{(2)}\!R_{\pm}}
\newcommand{\zRp}{{}^{(2)}\!R_+}
\newcommand{\zRm}{{}^{(2)}\!R_-}
\newcommand{\MatTTmunu}{{}^{(M)}\!\mathcal{T}_{\mu\nu}}
\newcommand{\Za}{Z_{{\bf (}a\!\!\! a\!\!\! a{\bf )}}}
\newcommand{\Ze}{Z_{\bf (1)}}
\newcommand{\Zz}{Z_{\bf (2)}}

\newcommand{\Falomunu}{F^{\alpha}_{\;\;\mu\nu}}
\newcommand{\MatTmunu}{{}^{(M)}\!T_{\mu\nu}}
\newcommand{\MMatTmunu}{{}^{(M)}\!\mathcal{T}_{\mu\nu}}
\newcommand{\MFmono}{{\mathcal F}^{\mu\nu}}
\newcommand{\MFmolo}{{\mathcal F}^{\mu\lambda}}
\newcommand{\jdomo}{j^{3\mu}}
\newcommand{\jvomo}{j^{4\mu}}
\newcommand{\Mvamu}{{\upsilon_{a\mu}}}
\newcommand{\Cgab}{C^{\gamma}_{\;\;\alpha\beta}}

\newcommand{\ket}[1]{|#1\rangle}
\newcommand{\Lfnu}{{}^{(L)}\!f_{\nu}}
\newcommand{\jbeonu}{j^\beta_{\;\;\nu}}
\newcommand{\Smulu}{\Sigma_{\mu\lambda}}
\newcommand{\keunu}{k_{1\nu}}
\newcommand{\kzunu}{k_{2\nu}}
\newcommand{\akn}{{}^{(a)}\!k_0}

\newcommand{\fdhf}{\delta_{\mathrm{HF}}E=\frac{1}{2}\delta_{\mathrm{St}}E}
\newcommand{\fdst}{\delta_{\mathrm{St}}E}
\newcommand{\fSa}{{}^1\mathrm{S}_0}
\newcommand{\fSb}{{}^3\mathrm{S}_1}
\newcommand{\fSc}{\mathrm{S}_z=+1}
\newcommand{\fSd}{-1=\mathrm{S}_z}
\newcommand{\fSe}{{}^1\mathrm{S}_0}
\newcommand{\fSf}{{}^3\mathrm{S}_1}

\documentclass[amssymb,amsmath,preprint,eqsecnum,aps,epsfig,graphics,nofootinbib,showpacs]{revtex4}
\usepackage{epsfig,amssymb,amsmath,stmaryrd,dsfont,color}
\usepackage[a4paper]{geometry}

\begin{document}

\title{\LARGE{Helium Multiplet Structure \\ in \\ Relativistic Schr\"odinger Theory}\\[6ex]}
\author{R. Gr\"abeldinger, T. Beck, M. Mattes and M. Sorg}
\affiliation{ II.\ Institut f\"ur Theoretische Physik der
Universit\"at Stuttgart\\Pfaffenwaldring 57\\ D-70550 Stuttgart\\ Germany\\ 
{\rm e-mail:} {\tt sorg@theo2.physik.uni-stuttgart.de}}

\begin{abstract}
%\begin{center}\Large Abstract\end{center}
\indent The emergence of a multiplet structure of the helium-like ions is studied within Relativistic Schr\"odinger Theory (RST), a fluid-dynamic approach to the relativistic quantum theory of the many-particle systems. The fluid-dynamic character of RST demands to specify the electronic current densities $\jmu$ for any $N$-particle configuration which is exemplified here by considering the helium singlet (${}^1S_0$) and triplet (${}^3S_1$) states in great detail. Since the use of densities in RST is based upon the concept of wave functions, the new theory appears as a certain kind of (relativistic) unification of the conventional wave function formalism and the density functional theory, which both are the most prominent theoretical tools in atomic and molecular physics. As a demonstration of the practical usefulness of RST, the energy difference $\Delta E_{1\backslash 2}$ of the helium singlet states $2s^2 \ {}^1S_0$ and $1s^2 \ {}^1S_0$ is calculated for a large range of nuclear charge numbers $z_{ex}$ ($2\leq z_{ex}\leq 100$), whereas the corresponding experimental values are available only up to $z_{ex}=42$ (molybdenum). The deviations of these RST results from the observational data is less than $0,3 \%$. \vspace{3cm}
\end{abstract}

\maketitle

\section{Introduction and Survey of Results}

Surely, from the classical conception of the world, the strangest feature of the quantum phenomena refers to the so-called \textit{wave-particle duality} \cite{b7}, \cite{b42}. Indeed, such a kind of "duality" represents a certain dilemma for theoretical physics which always tries to erect a mathematically unique and logically coherent view upon the material world; whereas the notorious wave-particle duality forces us to describe the elementary matter sometimes in terms of fluid dynamics (wave picture) and sometimes in terms of point particle dynamics (particle picture) with its associated probabilistic interpretation. The latter picture is favourable when dealing with statistical ensembles of particles \cite{b30}, and the wave picture is to be preferred for the treatment of quantum liquids \cite{b39}.\\
\indent In view of such an ambiguous theoretical situation, any progress towards a unification of both pictures should be welcome; no matter whether it refers more to the particle picture or rather to the wave conception of matter. Probably, greater efforts have been undertaken in the last decades to further develop and elaborate the particle picture in connection with quantum statistics, but with the recent experimental realization of the Bose-Einstein condensates \cite{b40} the interest in using fluid-dynamic concepts in quantum theory has been revived.\\
\indent Concerning the competition of both conceptual lines of thinking with respect to the production of testable predictions for the outcomes of real experiments, it should be especially interesting to consider those fields of "intersection" where the predictive power of both theoretical approaches may be directly opposed to each other and to the observational data. Naturally such a field could be \textit{atomic physics} where experimental spectroscopy can establish very precisely the energy level systems of the many-electron systems. Selecting this field of physics as the arena of competition provides us with an additional advantage, namely the possibility to oppose the potentiality of both approaches to each other, especially when one has to take into account also the relativistic effects which for higher nuclear charge number ($z_{ex}\gtrsim 30$, say) can no longer be neglected. Furthermore, within the conventional (i.e. particle) approach, one can take into account also the QED corrections in order to shift the predictions closer to the experimental data. This then must necessarily force the proponents of the fluid-dynamic picture to think about the incorporation of self-energy effects into their theory. Thus the arena for the competition seems to be well-prepared; but where are the competitors?\\
\indent On the conventional (i.e. point-particle) side, it is somewhat dissapointing to see that there does not exist a generally applicable and well-working relativistic quantum mechanics for many-particle systems. A possible candidate could have been the Bethe-Salpeter type of equations which however are afflicted with many deficiencies (see the critical evaluation of the Bethe-Salpeter equations in the literature, e.g. ref.s \cite{a23,b31}. Here the main problem seems to refer to the generalization of the non-relativistic {\it probabilistic} concepts to the {\it relativistic} domain, which would imply to encode into the mathematical apparatus the tensor product of one-particle Hilbert spaces as the N-particle Hilbert space. As a consequence, there exist merely some auxiliary constructions in order to take account for the relativistic effects in atoms and molecules: Dirac-Fock approximation \cite{b15}, multi-configuration Dirac-Fock method (MCDF) \cite{b15,a3}, all-order technique in relativistic many-body perturbation theory (MBPT) \cite{a5}, or relativistic $\frac{1}{Z}$-expansion method \cite{a6}. For an overview of approximation methods (being based upon the tensor product construction), see ref. \cite{b54}.\\
\indent However, the incorporation of the relativistic laws is no problem for a \textit{fluid-dynamic} theory, as is demonstrated by the recently established \underline{R}elativistic \underline{S}chr\"odinger \underline{T}heory (RST) \cite{a16,a9,a43}, which is based upon the Whitney sum of the one-particle bundles in place of the tensor product construction. Some preliminary tests of RST in the field of atomic physics have already demonstrated that the numerical predictions of this new theory are rather better than worse in comparison to the standard quantum theory, even if the QED self-energy effects are taken into account (see ref.s \cite{a43,u5}).\\
\indent But there seems to arise a certain type of problem for the Whitney sum construction, which is not present for the conventional tensor product formalism and whose solution is the aim of study of the present paper: this refers to the multiplet structure of the many-electron configurations in atoms. In order to have a simple demonstration of how to proceed in more complicated situations, one may consider here the singlet (${}^{1}S_0$) and triplet (${}^{3}S_1$) configurations of the helium-like ions with arbitrary nuclear charge numbers $z_{ex}$ ($2\leq z_{ex}\leq 100$). Neglecting first the spin-spin interactions and the relativistic effects, one can start with some Schr\"odinger energy level of conventional notation ($ns, n's$) where the principal quantum numbers are denoted by $n,n'$ and the orbital angular momentum is zero ($l=l'=0$), so that both (non-relativistic) electrons are in an $s$-state (see fig.3 below). Switching now on the spin-spin interactions (and also the relativistic effects) lets the four different spin states of the Schr\"odinger configurations ($ns, n's$) split up into one singlet state (of conventional notation $n_1s, n_2s\ {}^{1}S_0$) and three degenerate triplet states ($n_1s, n_2s\ {}^{3}\!S_1$). This multiplet splitting of the energy levels emerges in a very natural way within the (non-relativistic) tensor product formalism. \\
\indent Thus, the question arises now for the new theory: In what way does this multiplet structure emerge in RST? Observe here that the spin phenomenon in RST is traced back to some rotational flow of the wave function, principally not being different from the mechanism for the generation of the {\it orbital} angular momentum. Therefore the specification of the RST two-electron multiplet structure must consist in fixing the geometric pattern of the currents due to the singlet (${}^{1}S_0$) and triplet (${}^{3}S_1$) configurations. Our proposition for the solution of this problem is the following: 

\pagebreak

\begin{figure}
\begin{center}
\input{ellips.pstex_t}
\end{center}
\end{figure}

\noindent The two triplet configurations with conventional notation $|S=1; S_z=\pm 1>$ are described in RST by letting both electronic currents encircle the $z$-axis, either in the positive or negative sense. However, the "exotic" spin state $|S=1; S_z=0>$ is due to a rotational flow in the plane containing the $z$-axis so that no non-trivial $z$-component of the total electronic spin $\vec{S}$ can arise ($S_z=0$); but nevertheless there does exist a rotational flow ($S=1$)! On the other hand, the singlet states $\Sa$ are assumed to be due to the two currents encircling the $z$-axis in opposite directions such that both the total spin quantum number $S$ and its $z$-component $S_z$ are zero: $|S=0; S_z=0>$, see the illustration above.\\
\indent The intention of the present paper is to elaborate this proposition from the mathematical point of view and test its physical consequences with respect to the spectroscopic data. The energy eigenvalue equations for the singlet and triplet states are set up within the RST formalism; but the triplet case is too complicated to obtain a numerical solution which could be used in order to demonstrate the degeneracy of the three states $|S=1; S_z=0, \pm 1>$. However, for the isotropic subset $ns^2\ \Sa$ of the singlet states $n_1s,n_2s\ \Sa$ one encounters a one-dimensional eigenvalue problem which can be solved numerically so that a comparison to the experimental data \cite{m1} is possible. More concretely, we numerically solve the RST eigenvalue problem for the singlet states $2s^2\ \Sa$ and $1s^2\ \Sa$ in order to calculate the corresponding energy difference ($\Delta E_{1\backslash 2}$) by taking the value of the RST energy functional upon the numerical solutions. Both situations, with included and neglected self-interactions, are considered, and it is found that the RST predictions deviate from the corresponding experimental values \cite{m1} by less than $0.3 \% $, see table $I$ and fig.4 below. This result is somewhat amazing because the non-relativistic limit of RST coincides with the Hartree-Fock approach which misses the exchange effects for the singlet states (see fig.3 below and the discussion in ref.\cite{u6}). Thus, the conclusion is that the exchange effect does not act in a completely symmetric way upon the singlet and triplet states (as suggested by the standard theory), but is weakened for the singlet states. The mechanism of the RST self-interactions is discussed in detail; this elucidates the result that the inclusion of the self-interactions can hardly improve the "semiclassical" value for the energy difference $\Delta E_{1/2}$.\\
\indent The results are elaborated by the following arrangement:\\
\indent In {\bf{Sect.II}} a brief sketch of the fundamental RST features is presented in order to realize clearly the differences from the standard quantum theory. Here it should become obvious that RST is a rather general field theoretical framework whose indispensible features refer only to the (local) conservation of energy-momentum and charge, but otherwise leave great freedom to reify the general structure. Special realizations of RST refer to the Klein-Gordon theory of scalar particles and to the Dirac theory of spin-$1/2$ electrons; the latter realization is taken as the basis for the subsequent discussions.\\
\indent {\bf{Sect.III}} establishes the general RST eigenvalue problem for the helium-like ions, see equations (\ref{323})-(\ref{326d}) below. This eigenvalue problem must be complemented by the specification of an energy functional ($E_T$, say), whose values upon the solutions of the eigenvalue problem yield the desired atomic energy levels. The general structure of this RST energy functional is typical for the Whitney sum construction to be used for the fluid-dynamic approach in place of the tensor product construction: the total energy $E_T$ equals the sum of mass eigenvalues ($\sum M_ac^2$) minus those interaction contributions ($\Delta E_T$) which are already included in each one of the mass eigenvalues $M_ac^2$ and are therefore counted twice if one simply forms the sum of the mass eigenvalues, see equations (\ref{335})-(\ref{336}) below.\\
\indent In {\bf{Sect.IV}}, the general two-particle theory is cut down to the triplet configurations $\Sb$ (\textit{ortho-helium}), whose RST structure is then studied in great detail. Naturally both states $|S=1; S_z=\pm 1>$ with non-zero spin-component $S_z$ have a very similar geometric structure, so that it is sufficient to write explicitely down only one of both configurations (i.e. $|S=1; S_z=+1>$). In contrast to these relatively simple configurations which admit a convenient non-relativistic approximation, the state $|S=1; S_z=0>$ with vanishing spin component $S_z$ has a distinctly different geometric structure (see the figure above) so that it seems very difficult to obtain an approximate solution to the corresponding eigenvalue problem (see equations (\ref{48})-(\ref{413b}) below). Therefore the question of degeneracy of all three states $|S=1; S_z=0, \pm 1>$ must be left unclarified for the moment (the degeneracy of both states $|S=1; S_z=\pm 1>$ is self-evident).\\
\indent {\bf{Sect.V}} presents a detailed study of the singlet states $\Sa$ (\textit{para-helium}). These RST counterparts of the conventional states $|S=0; S_z=0>$ have the peculiarity of vanishing exchange density, so that no exchange interactions (of the "electric" type) do exist between the two electrons. Since the "electric" interactions are mostly much stronger than their "magnetic" counterparts, the para-helium levels are not shifted (in lowest-order approximation) by the RST exchange interactions, which is in contrast to the predictions of the standard quantum theory being based upon the tensor product formalism, see fig.3 below. Therefore it may appear as if RST, and the Hartree-Fock approach as its non-relativistic limit, would produce inaccurate predictions of the para-helium energy levels.\\
\indent This supposition is then tested explicitely in {\bf{Sect.VI}} by numerically solving the RST eigenvalue problem of Sect.V for the \textit{isotropic} states $ns^2\ \Sa$ of para-helium for principal quantum numbers $n=1$ and $n=2$. This admits us to compute the corresponding energy difference $\Delta E_{1\backslash 2}$ for the para-helium states   
\begin{equation}\label{11}
\Delta E_{1\backslash 2} = \ E \Big\vert_{2s^2 \, \sf {}^1S_0} - E \Big\vert_{1s^2 \, \sf {}^1S_0}
\end{equation}
\noindent and to compare our RST result to the experimental values
\cite{m1}. Unfortunately the analogous results have been omitted by the other theoretical
approaches being conveniently available in the literature, i.e. the all-order technique in
MBPT \cite{a5} and the $\frac{1}{Z}$-expansion method \cite{a6}. The result of this
comparison is very favourable for the present fluid-dynamic approach: its predictions for
the energy difference $\Delta E_{1\backslash 2}$ deviate from the experimental values
\cite{m1} by less than $0.3 \%$ and do approach them even better for increasing nuclear
charge number $z_{ex}$ (fig.4). This signals that the missing of the "electric" exchange
energy for the singlet states in the RST and Hartree-Fock approaches may not merely be an
artefact of the formalism but actually points to the existence of a real effect in the
atomic structure! The RST predictions for the energy difference $\Delta E_{1\backslash 2}$
(\ref{11}) are extended beyond the largest, experimentally available nuclear charge number
(i.e. $z_{ex}=42$, molybdenum) where it must be left to future experiments to verify or
falsifiate the RST results in this ultra-relativistic regime (Table II).
  
\section{RST Dynamics}

Similar to most of the modern gauge theories of elementary matter, the fundamental equations of motion represent also in RST a coupled system of \textit{matter} and \textit{gauge} fields. In the present context, the gauge field equations are considered to be even more fundamental than the matter field equations since the latter type becomes subjected to certain restrictions in order to guarantee the conservation laws (e.g. for charge and energy-momentum); and these restrictions directly originate from the gauge field dynamics. Here it will turn out that the well-known wave equations of conventional quantum theory (i.e. of Dirac and Klein-Gordon type, respectively) do represent only a restricted set of possible matter equations which however are sufficient to ensure the desired conservation laws. Though this restrictive set has an acceptable non-relativistic limit (i.e. the Hartree-Fock equations \cite{a16,a15}), it may be nevertheless insufficient to cover all relativistic many-particle situations. Subsequently it is demonstrated that RST is able to provide a more general class of relativistic wave equations; however, for our present treatment of ortho- and para-helium we are satisfied with the \textit{Dirac realization} of RST.\\
\indent The present procedure of generalizing the classical wave equations of Dirac and Klein-Gordon is based upon the use of a new field variable, i.e. the \textit{Hamiltonian} $\MHmu$, which itself is a dynamical object of the theory and obeys its own field equations (\textit{Hamiltonian dynamics}). This object $\MHmu$ may be conceived as some kind of hidden variable which guides the matter through the "landscape" of gauge fields just in such a way that the desired conservation laws for matter are ensured. For the original restricted set of field equations, the Hamiltonian $\MHmu$ can be again eliminated, so that the conventional wave equations of Dirac and Klein-Gordon do reappear. However, for the more general field configurations (and especially for the relativistic \textit{mixtures}) the Hamiltonian $\MHmu$ cannot be eliminated and remains an essential part of the dynamical system.\\
\indent According to the logical dominance of the gauge system, we first write down the corresponding dynamical equations in order to deduce thereof the restrictions for the motion of matter; and afterwards we introduce the Hamiltonian in such a way that these restrictions (i.e. conservation laws) are actually obeyed.

\subsection{Gauge Field Equations}

The interactions among the material constituents of an N-fermion system are described by the bundle connection (i.e.  "gauge potential") $\MAmu$, which adopts its values in the gauge algebra $\MFu(N)$ being itself spanned by the $N^2$ generators $\talu$ ($\alpha=1,\dots,N^2$)\cite{a16,a43}
\begin{equation}
\label{21}
\MAmu = -i\ \exAmu \cdot \mathds{1} + \Aaomu\talu \doteqdot \exMAmu + {}^{(s)}\MAmu \ .
\end{equation}
\noindent Here the external potential $\exAmu$ refers to some external source, not being part of the dynamical system (e.g. fixed nucleus). For the electromagnetic interactions, the original gauge group $U(1)\times U(N)$ is spontaniously broken down to its maximal Abelian subgroup $U(1)\times \dotsc \times U(1)$ which then describes the proper electromagnetic interactions, whereas the frozen gauge degrees of freedom refer to the {\it exchange} interactions. Correspondingly, one splits up the internal connection ${}^{(S)}\MAmu$ into its electromagnetic part (${}^{(em)}\mathcal{A}_{\mu}$) and exchange part ($\MBmu$),
\begin{subequations}
\begin{align}
             {}^{(S)}\MAmu & = {}^{(em)}\mathcal{A}_{\mu} + \MBmu\label{22a}\\
{}^{(em)}\mathcal{A}_{\mu} & = \Aamu \tau_a\label{22b}\\
                     \MBmu & = \Bkomu \chi_k - \Bskomu \bar{\chi}_k \ , \label{22c}
\end{align}
\end{subequations}
\noindent where the $N$ \textit{electromagnetic generators} $\tau_a$ ($a=1,\dots,N$) are adopted to be anti-Hermitian ($\bar{\tau}_a=-\tau_a$) and commuting ($[\tau_a,\tau_b]=0$), in contrast to the \textit{exchange generators} $\chi_k$ ($k=1,\dots,\frac{N(N-1)}{2}$). For instance, for a two-fermion system ($N=2$) one has the decomposition 
\begin{equation}\label{23}
\MAmu = -i\ \exAmu \cdot \mathds{1} + \Aemu \tau_1 + \Azmu \tau_2 + \Bmu\chi - \Bsmu\bar{\chi},
\end{equation}
\noindent with the commutation relations
\begin{subequations}
\begin{alignat}{2}
   [\tau_1,\tau_2] & =0 & \qquad [\chi,\bar{\chi}] & =-i\ (\tau_1-\tau_2)\label{24a}\\
     [\tau_1,\chi] & =i\chi & \qquad [\tau_1,\bar{\chi}] & =-i\bar{\chi}\label{24b}\\
     [\tau_2,\chi] & =-i\chi & \qquad [\tau_2,\bar{\chi}] & =i\bar{\chi} \ . \label{24c}
\end{alignat}
\end{subequations}
\indent Now, in order to specify the field equation for the bundle connection $\MAmu$, one first considers its curvature $\MFmunu$ ("field strength"),
\begin{equation}\label{25}
\MFmunu \doteqdot \nmu\MAnu - \nnu\MAmu + [\MAmu,\MAnu],
\end{equation}
\noindent which, e.g., for a two-fermion system decomposes as follows
\begin{subequations}
\begin{align}
\MFmunu & =  -i\ \exFmunu \cdot \mathds{1} + \Faomunu \tau_{\alpha}\label{26a}\\
        & =  -i\ \exFmunu \cdot \mathds{1} + \Famunu \tau_a + \Gmunu \chi -
                 \Gsmunu\bar{\chi} \ ,\label{26b}
\end{align}
\end{subequations}
\noindent with the curvature components being given in terms of the connection components as follows \cite{a16}:
\begin{subequations}
\begin{align}
         \exFmunu & = \nmu \ \exAnu - \nnu \ \exAmu\label{27a}\\
          \Femunu & = \nmu \Aenu - \nnu \Aemu + i\ (\Bmu \Bsnu - \Bnu \Bsmu)\label{27b}\\
          \Fzmunu & = \nmu \Aznu - \nnu \Azmu - i\ (\Bmu \Bsnu - \Bnu \Bsmu)\label{27c}\\
           \Gmunu & = \nmu \Bnu - \nnu \Bmu + i\ (\Aemu - \Azmu) \Bnu - i\ (\Aenu - \Aznu) \Bmu\label{27d}\\
          \Gsmunu & = \nmu \Bsnu - \nnu \Bsmu - i\ (\Aemu - \Azmu) \Bsnu + i\ (\Aenu - \Aznu) \Bsmu \ .\label{27e}
\end{align}
\end{subequations}
\noindent Once the curvature $\MFmunu$ of the connection $\MAmu$ has been introduced, the field equation for $\MAmu$ is selected to be the usual (non-Abelian) Maxwell equation:
\begin{equation}       
\MDmo \MFmunu  = -4\pi i\alpha_s\MJnu\label{28}
\end{equation}
\begin{equation*}
(\MDlu \MFmunu  \doteqdot \nlu \MFmunu + [\MAlu, \MFmunu]) \ .
\end{equation*}
\noindent Decomposing here the (Hermitian) current operator $\MJmu$ with respect to the chosen Lie algebra basis $\left\{\tau_{\alpha}\right\}$ as
\begin{equation}\label{29}
\MJmu = \exjmu \cdot \mathds{1} + i\jalomu \tau_{\alpha}\ ,
\end{equation}
\noindent i.e. more concretely for our two-fermion system
\begin{equation}\label{210}
\MJmu = \exjmu \cdot \mathds{1} + i\jaomu \tau_a + i(\hsmu \chi - \hmu \bar{\chi})\ ,
\end{equation}
\noindent one finds the component version of the operator-valued RST-Maxwell equations (\ref{28}) to be of the following form:
%linksbndig anordnen?
\begin{subequations}\label{211}
\begin{align}
                                                     \nmo \ \exFmunu & =  4\pi\,\alpha_s\,\exjnu\label{211a}\\
                       \nmo \Femunu + i\ (\Bmo \Gsmunu - \Bsmo \Gmunu) & =  4\pi\,\alpha_s\,\jeonu\label{211b}\\
                       \nmo \Fzmunu - i\ (\Bmo \Gsmunu - \Bsmo \Gmunu) & =  4\pi\,\alpha_s\,\jzonu\label{211c}\\
       \nmo \Gmunu + i\ (\Aemu-\Azmu)\Gmonu - i\ (\Femunu-\Fzmunu)\Bmo & =  4\pi\,\alpha_s\,\jdonu\label{211d}\\
    \nmo \Gsmunu - i\ (\Aemu-\Azmu)\Gsmonu + i\ (\Femunu-\Fzmunu)\Bsmo & =  -4\pi\,\alpha_s\,\jvonu\label{211e}\ .
\end{align}
\end{subequations}
\indent At this stage of the development of the theory, one necessarily becomes confronted with the first restriction upon the motion of matter which itself is thought to produce in some way those {\it Maxwell currents} $\jalomu$ (\ref{29}) which enter the Maxwell equations (\ref{28}). The point here is namely that any bundle curvature $\MFmunu$ (\ref{25}) must obey the identity
\begin{equation}\label{212}
\left[\MDmu,\MDnu\right]\MFmono \equiv 0
\end{equation}
which by adopting that link (\ref{28}) of the Maxwell equations to the current $\MJmu$ immediately implies the "continuity equation" in operator form
\begin{equation}\label{213}
\MDmo \MJmu \equiv 0 \ .
\end{equation}
Actually this is a severe restriction upon the motion of matter which becomes more obvious when written in component form, e.g. for the present two-fermion system:
\begin{subequations}\label{214}
\begin{align}
\nmo \ \jeomu &= i \ (\Bmo \ \jvomu + \Bsmo \ \jdomu)\label{214a}\\
\nmo \ \jzomu &= -i \ (\Bmo \ \jvomu + \Bsmo \ \jdomu)\label{214b}\\
\nmo \ \jdomu &= -i \ (\Aemu -\Azmu) \jdomo + i \ \Bmo \ (\jeomu - \jzomu)\label{214c}\\
\nmo \ \jvomu &= i \ (\Aemu -\Azmu) \jvomo + i \ \Bsmo \ (\jeomu - \jzomu)\label{214d}.
\end{align}
\end{subequations}
This source system is to be interpreted in the following way: for given Maxwell currents $\jalomu$ with sources $(\nmo \, \jalomu)$ as specified by the present equations (\ref{214a})-(\ref{214d}), the connection components $\{\Aamu;\Bmu\}$ must be thought to be solutions of the RST-Maxwell system (\ref{211b})-(\ref{211e}) with just these currents $\jalomu$ emerging on the right-hand sides! (see fig.1).
\begin{figure*}
\setlength{\unitlength}{1cm}
\label{fig1}
\begin{picture}(20,8)
{\thicklines
  \put(0,5){\line(1,0){3}}
  \put(3,5){\line(0,1){2}}
  \put(3,7){\line(-1,0){3}}
  \put(0,7){\line(0,-1){2}}

  \put(5,5){\line(1,0){5}}
  \put(10,5){\line(0,1){2}}
  \put(10,7){\line(-1,0){5}}
  \put(5,7){\line(0,-1){2}}

  \put(12,5){\line(1,0){3}}
  \put(15,5){\line(0,1){2}}
  \put(15,7){\line(-1,0){3}}
  \put(12,7){\line(0,-1){2}}

  \put(4.5,0.5){\line(1,0){6}}
  \put(10.5,0.5){\line(0,1){2}}
  \put(10.5,2.5){\line(-1,0){6}}
  \put(4.5,2.5){\line(0,-1){2}} }

  \put(0.7,6.2){\large sources}
  \put(0.9,5.5){$\nmo \; \jalomu $}
  \put(5.8,6.2){\large matter currents}
  \put(7.2,5.5){$\jalomu $}
  \put(12.5,6.2){\large potentials}
  \put(13.2,5.5){$\Aaomu$}
  \put(5.8,1.7){\large source equations}
  \put(6.7,1){(\ref{214a})-(\ref{214d})}
  \put(10.6,6.2){(\ref{211})}

{\thicklines
  \put(1.5,5){\vector(1,-1){3}}
  \put(7.5,5){\vector(0,-1){2.5}}
  \put(13.5,5){\vector(-1,-1){3}}
  \put(5,6){\vector(-1,0){2}}
  \put(10,6){\vector(1,0){2}}
}
\end{picture}
\flushleft{{\large Fig. 1:\quad{\it Self-Consistency of Currents
and Potentials (\ref{214})}}}
\end{figure*}

A similar self-consistency problem for the coupled matter and gauge field
dynamics arises for the energy-momentum density of matter
(${}^{({\rm M})}\Tmunu$, say): According to the general logical
structure of the gauge field theories, one expects that the
well-known Lorentz force density $\Lfnu$ should act as the
generator of the matter energy-momentum, i.e.
\begin{equation}\label{215}
\nmo \, {}^{({\rm M})}\Tmunu = \Lfnu\ .
\end{equation}
For the present situation of a composite system (N particles), the Lorentz force will consist of an external part $\exfnu$ and an internal part $\Sfnu$
\begin{equation}\label{216}
\Lfnu=\exfnu+\Sfnu \ .
\end{equation}
\begin{figure}
\label{fig2}
\setlength{\unitlength}{1cm}
\begin{picture}(20,8)
{\thicklines
  \put(4.5,5){\line(1,0){6}}
  \put(10.5,5){\line(0,1){2.5}}
  \put(10.5,7.5){\line(-1,0){6}}
  \put(4.5,7.5){\line(0,-1){2.5}}}
  \put(10.5,6.5){\line(-1,0){6}}
  \put(6.5,6.5){\line(0,-1){1.5}}
{\thicklines
  \put(0,5){\line(1,0){3}}
  \put(3,5){\line(0,1){1.5}}
  \put(3,6.5){\line(-1,0){3}}
  \put(0,6.5){\line(0,-1){1.5}}

  \put(0,1.5){\line(1,0){3}}
  \put(3,1.5){\line(0,1){2}}
  \put(3,3.5){\line(-1,0){3}}
  \put(0,3.5){\line(0,-1){2}}

  \put(12,4.2){\line(1,0){3.5}}
  \put(15.5,4.2){\line(0,1){2.3}}
  \put(15.5,6.5){\line(-1,0){3.5}}
  \put(12,6.5){\line(0,-1){2.3}}

  \put(4.5,0.5){\line(1,0){6}}
  \put(10.5,0.5){\line(0,1){2}}
  \put(10.5,2.5){\line(-1,0){6}}
  \put(4.5,2.5){\line(0,-1){2}}}

  \put(0.4,5.9){field strengths}
  \put(1.2,5.3){$\Falomunu$}
  \put(0.5,2.7){force density}
  \put(0.6,2){$\Sfnu$ (\ref{219})}
  \put(6.7,6.8){\large Matter}
  \put(4.9,6){currents}
  \put(5.3,5.4){$\jalomu $}
  \put(7,6){energy-momentum}
  \put(7.9,5.4){$\MatTmunu$}
  \put(12.25,5.8){energy-momentum}
  \put(13.3,5.3){source}
  \put(13,4.6){$\nmo {}^{(M)}\Tmunu$}
  \put(5.6,1.7){\large energy-momentum}
  \put(6.3,1){\large balance (\ref{215})}
  \put(3.3,6){(\ref{211})}
{\thicklines
  \put(4.5,5.75){\vector(-1,0){1.5}}
  \put(1.5,5){\vector(0,-1){1.5}}
  \put(3,2){\vector(1,0){1.5}}
  \put(10.5,5.75){\vector(1,0){1.5}}
  \put(13.75,4.2){\line(0,-1){2.2}}
  \put(13.75,2){\vector(-1,0){3.25}}}
\end{picture}
\flushleft{{\large Fig. 2:\quad{\it Energy-Momentum Balance
(\ref{215})}}}
\end{figure}

Here the external part $\exfnu$ is assumed to be built up by the external field strength $\exFmunu$ and the \textit{total electromagnetic current} $\jmu$ as the sum of the electromagnetic gauge currents $\jaomu (a=1,...,N)$
\begin{equation}\label{217}
\jmu \doteqdot -\sum_{a=1}^N \jaomu\ ,
\end{equation}
i.e. one puts
\begin{equation}\label{218}
\exfnu = \hbar c \, \exFmunu \ j^{\mu} \ .
\end{equation}
A similar construction suggests itself for the internal part $\Sfnu$
\begin{equation}\label{219}
\Sfnu = \hbar c \ \Falomunu \ \jalumo
\end{equation}
which however obliges us to introduce a covariantly constant fibre metric $\Kaubu$ for the Lie algebra bundle in order to lower and lift the gauge algebra indices \cite{a43}
\begin{subequations}\label{220}
\begin{align}
\jalomu & =\Kaobo \ \jbumu \label{220a} \\
\jalumu & =\Kaubu \ \jbomu \quad \mbox{etc}. \label{220b}
\end{align}
\end{subequations}
But here it should be self-evident, that the requirement of
energy-momentum balance (\ref{215}) poses again the same logical
problem as it already did emerge from the source equations for the
currents (\ref{214a})-(\ref{214d}): namely to let the matter move
in such a specific way that the source $\Sfnu$ of its
energy-momentum content $\MatTmunu$ becomes the product of its
currents $\jalumu$ and the field strengths $\Falomunu$ being
generated by just those currents $\jalumu$ via the RST-Maxwell
equations (\ref{211b})-(\ref{211e})!(fig.2)\\
\indent A very pleasant feature of RST is now that both self-consistency
problems (\ref{214}) and (\ref{215}) for the motion of matter can
be solved in one step, namely by setting up the right equation of
motion together with a suitable definition of currents $\jalomu$
and energy-momentum $\MatTmunu$. The interesting point here is
that the conventional field equations of matter (i.e. Dirac and
Klein-Gordon) do reappear only as a restricted subset of a larger
set of possibilities.

\subsection{Matter Dynamics}

The RST description of matter is based upon the intensity matrix $\MI$, which is a Hermitian ($4N\times 4N$)-matrix ($\bar{\MI}=\MI$) for a system of $N$ fermions. If this system is to be described by a pure state $\Psi$, the intensity matrix $\MI$ degenerates to the tensor product of $\Psi$:
\begin{equation}\label{221}
\MI\Rightarrow\Psi\otimes\bar{\Psi} \ .
\end{equation}
\noindent The field equation for $\MI$ is the \underline{R}elativistic von \underline{N}eumann \underline{E}quation (RNE)
\begin{align}
\MDmu\ \MI & = \frac{i}{\hbar c}[\MI\ \bar{\MHmu} - \MHmu\ \MI] \label{222}\\
(\MDmu\ \MI & \doteqdot \pmu\ \MI + [\MAmu , \MI]) \ ,\nonumber
\end{align}
\noindent or for the situation of a pure state $\Psi$, respectively, the \underline{R}elativistic \underline{S}chr\"odinger \underline{E}quation (RSE)
\begin{align}
i\hbar c \, \MDmu \Psi & = \MHmu \Psi\label{223}\\
(\MDmu \Psi & \doteqdot \pmu \Psi + \MAmu \Psi).\nonumber
\end{align}
\noindent The intensity matrix $\MI (x)$ and the wave function $\Psi (x)$ are conceived as sections of appropriate fiber bundles with the "potential" $\MAmu$ (\ref{21}) working as the bundle connection. Therefore the first task is to ensure the generally valid bundle identities
\begin{subequations}
\begin{align}
 [\MDmu \MDnu - \MDnu \MDmu]\ \MI & = [\MFmunu, \MI]\label{224a}\\
[\MDmu \MDnu - \MDnu \MDmu]\Psi & = \MFmunu \Psi \label{224b}
\end{align}
\end{subequations}
\noindent by an appropriate choice of the field equations for the Hamiltonian $\MHmu$ (\textit{Hamiltonian dynamics}). Consequently, the first field equation for the Hamiltonian is the \textit{integrability condition}
\begin{align}
\MDmu \MHnu - \MDnu \MHmu & + \frac{i}{\hbar c}[\MHmu,\MHnu] = i\hbar c \MFmunu\label{225}\\
(\MDmu \MHnu & \doteqdot \nmu \MHnu + [\MAmu, \MHnu])\nonumber
\end{align}
\noindent which in combination with the matter dynamics (\ref{221})-(\ref{222}) just ensures the validity of those bundle identities (\ref{224a})-(\ref{224b}).
   
\subsection{Charge Conservation}

The second field equation for the Hamiltonian $\MHmu$, which is a $\mathfrak{gl}(4N,\mathds{C})$-valued one-form, must now refer to the guarantee of the conservation laws discussed in the preceding subsection. For this purpose, one first introduces the total velocity operator $\Gmu$ as a Hermitian one-form ($\bar{\Gamma}_{\mu}=\Gmu)$ which takes its values also in the Lie algebra $\mathfrak{gl}(4N,\mathds{C})$. By means of this object, one alternatively defines the total electromagnetic current $\jmu$ (\ref{217}) through
\begin{equation}
\jmu \doteqdot tr(\MI\cdot \Gmu)\label{226}
\end{equation}
\noindent and then one requires this current to obey the conservation law
\begin{equation}
\nmo \jmu = 0\label{227}.
\end{equation}
\noindent Indeed, adding up both source equations (\ref{214a}) and (\ref{214b}) for the two-particle situation ($N=2$) actually yields a result which just coincides with the present conservation law (\ref{227}). Thus, in this special respect, our choice of the matter dynamics would be consistent with the gauge field dynamics, provided we suceed in deducing the conservation law (\ref{227}) {\it also} from the proposed matter dynamics (\ref{222}). However, this can easily be done by simply transcribing the derivative (\ref{227}) of the current $\jmu$ (\ref{226}) to the intensity matrix $\MI$ and using the RNE (\ref{222}) (or the RSE (\ref{223}) for the case of a pure state) which ultimately yields
\begin{equation}
0 = \nmo \jmu = tr\left\{\MI \cdot \left(\MDmo \Gmu + \frac{i}{\hbar c}\left[\bar{\mathcal{H}}^\mu \Gmu - \Gmo \MHmu \right]\right)\right\} \ .\label{228}
\end{equation}
\noindent Thus, in order to ensure the validity of the desired conservation law (\ref{227}) for any matter arrangement $\MI$, we require the total velocity operator $\Gmu$ to obey the following operator equation 
\begin{equation}
\MDmo \Gmu + \frac{i}{\hbar c}[\bar{\mathcal{H}^{\mu}} \Gmu -\Gmu \MHmo] = 0\label{229} \ .
\end{equation}
\noindent This equation is to be understood in the sense that for given Hamiltonian $\MHmu (x)$ (as solution of the Hamiltonian dynamics) the total velocity operator $\Gmu$ must be a solution of its source equation (\ref{229}).\\
\indent Next, one turns to the operator-valued continuity equation (\ref{213}) which reads in component form
\begin{equation}\label{230}
D^{\mu} \jalumu = 0 \ ,
\end{equation}
with its contravariant two-particle realization being given by equations (\ref{214a})-(\ref{214d}). Here the covariant derivative (D) of the gauge objects is defined through
\begin{subequations}\label{231}
\begin{align}
 D_{\mu} \, \jalunu &\doteqdot \nmu \, \jalunu - \wbam \, \jbeunu \label{231a} \\
 D_{\mu} \, \jalonu &\doteqdot \nmu \, \jalonu + \wabm \, \jbeonu \ , \label{231b}
\end{align}
\end{subequations}
where the connection one-form $\wmu=\left\lbrace \wabm \right\rbrace$ takes its values in the adjoint representation of the gauge algebra, i.e.
\begin{equation}\label{232}
\wbam = \Cbga \Agomu \ ,
\end{equation}
and the structure constants $\Cabg$ are defined through the usual commutation relations
\begin{equation}\label{233}
\left[\talu, \tbu \right] = \Cgab \ \tgu \ .
\end{equation}
Now in order to ensure the desired validity of the source equation (\ref{230}) for the gauge currents $\jalumu$, one introduces the {\it gauge velocity operators} $\Mvaumu$ which then build up the currents $\jalumu$ through
\begin{equation}\label{234}
\jalumu = tr \left( \I \cdot \Mvaumu \right),
\end{equation}
similarly to the construction of the total current $\jmu$ (\ref{226}). Accordingly, the transcription of the derivative ($D$) of $\jalumu$ in equation (\ref{230}) to the gauge velocity operators $\Mvaumu$ (\ref{234}) yields a result quite analogous to equation (\ref{228}), i.e.
\begin{equation}\label{235}
\Dmo \jalumu = tr \left\lbrace \MI \left(\MDmo \Mvaumu+ \frac{i}{\hbar c} \left(\bar{\MHmo} \cdot \Mvaumu - \Mvaumu \cdot \MHmo\right)\right)\right\rbrace.
\end{equation}
Therefore, in order to satisfy the requirement (\ref{230}), one adopts the velocity operators $\Mvaumu$ to be solutions of the following source equation:
\begin{equation}\label{236}
\MDmo \Mvaumu + \frac{i}{\hbar c} \left(\bar{\MHmo} \cdot \Mvaumu - \Mvaumu \cdot \MHmo\right) = 0,
\end{equation}
where the covariant derivative of the gauge operators is defined through
\begin{equation}\label{237}
\MDlu \Mvaumu = \nlu \Mvaumu + \left[\MAlu, \Mvaumu \right] - \wbal \Mvbumu \ .
\end{equation}
\indent However, in order to not introduce too many new objects requiring their own field equations, one would like to express those gauge velocity operators $\Mvaumu$ in terms of the already existing objects $\talu$ and $\Gmu$, i.e. we try to identify $\Mvaumu$ as the anticommutator of the gauge algebra generators $\talu$ and the total velocity $\Gmu$:
\begin{equation}\label{238}
\Mvaumu=\frac{i}{2} \left\lbrace \talu, \Gmu \right\rbrace \ .
\end{equation}
Since the electromagnetic generators $\tau_a$ ($a=1,\dots,N$) are adopted to be anti-Hermitian ($\bar{\tau}_a=-\tau_a$) and the velocity operator is taken to be Hermitian ($\bar{\Gamma}_{\mu}=\Gmu$), the electromagnetic velocity operators $\upsilon_{a \mu}$ ($a=1,\dots,N$) are also Hermitian ($\Mvamu = \bar{\upsilon}_{a\mu}$) and consequently the electromagnetic currents $\jaumu$ ($a=1,\dots,N$) (\ref{234}) are real, just as the electromagnetic potentials $\Aamu$ and their field strengths $\Famunu$. In contrast to this, the remaining N(N-1) exchange generators ($\chi_k,\bar{\chi}_k$; $k=1,\dots,\frac{N(N-1)}{2}$) are arranged in $\frac{N(N-1)}{2}$ pairs of Hermitian conjugate operators, so that the exchange potentials $B^k_{\;\;\mu}$ and their field strengths $G^k_{\;\;\mu\nu}$ are complex, as well as the corresponding exchange currents $h^k_{\;\;\mu}$; see for instance the two-particle realization (\ref{211b})-(\ref{211e}) of such an arrangement.\\
\indent Introducing the chosen ansatz (\ref{238}) for the velocity operators $\Mvaumu$ into their field equation (\ref{236}) must then necessarily yield some condition upon the total velocity operator $\Gmu$, namely one finds the following combination of commutators $\left[\, ..\, ,\, ..\, \right]$ and anti-commutators $\left\lbrace\, ..\, ,\, ..\, \right\rbrace$ :
\begin{eqnarray}\label{239}
\frac{i}{2} \left\lbrace\talu, \MDmo\Gmu + \frac{i}{\hbar c}\ \left( \bar{\MHmo}\Gmu-\Gmu\MHmo \right)\right\rbrace
+ \frac{1}{2\hbar c}\left[\talu, \bar{\MHmo}\Gmu + \Gmu\MHmo\right] - \\ \nonumber
-\frac{1}{2\hbar c} \left\lbrace \bar{\MHmo} \left[\talu, \Gmu \right]+\left[ \talu,\Gmu \right] \MHmo \right\rbrace = 0 \ .
\end{eqnarray}
Thus, summarizing the situation with the charge conservation laws, it becomes clear that both conservation laws (\ref{227}) and (\ref{230}) can simultaneously be satisfied by requiring the total velocity operator $\Gmu$ to obey the former equation (\ref{229}) and let it commute with the gauge algebra generators $\talu$
\begin{equation}\label{240}
\left[\talu,\Gmu\right]=0 \ .
\end{equation}
However, observe here that these two conditions are actually sufficient to guarantee the total charge conservation (\ref{227}) but leave us with one remaining term for the gauge continuity equation (\ref{230}). More concretely, introducing the (Hermitian) {\it mass operator} $\M$ ($=\bar\M$) through
\begin{equation}\label{241}
 \bar{\MHmo}\Gmu + \Gmu\MHmo \doteqdot 2\, \M c^2 \ ,
\end{equation}
then the source equation for the gauge velocity operators $\Mvaumu$ becomes in place of the desired result (\ref{236}):
\begin{equation}\label{242}
 \MDmo \Mvaumu + \frac{i}{\hbar c}\left[ \bar{\MHmo}\Mvaumu - \Mvaumu\MHmo \right] = \frac{c}{\hbar} \, \left[\talu,\M \right] \ ,
\end{equation}
and consequently the source equation (\ref{230}) for the gauge currents $\jalumu$ acquires an inhomogeneous term:
\begin{equation}\label{243}
 \Dmo \jalumu = tr \left\lbrace \MI \cdot \left[\talu, \frac{\M c}{\hbar} \right] \right\rbrace \ .
\end{equation}
Clearly, one can here easily bring down to zero the right-hand side in order to attain the desired homogeneous equation, namely by simply letting the mass operator $\M$ being proportional to unity
\begin{equation}\label{244}
 \M \Rightarrow M\cdot \bf{1} \ .
\end{equation}

\subsection{Energy-Momentum Conservation}

Before one can make a final decision for the choice of the total
velocity operator $\Gmu$ as a solution of its source equation
(\ref{229}), one must first specify the second field equation
(\textit{conservation equation}) for the Hamiltonian $\MHmu$. Here
one expects that this equation for $\MHmu$ will emerge in connection
with an energy-momentum conservation law for matter. More
concretely, one tries to introduce the energy-momentum density of
matter ($\MatTmunu$, say) in such a way that its source is just
the well-known Lorentz force density $\Lfnu$ (\ref{216}) due to
the external and internal gauge fields. This idea may be realized
by first defining an energy-momentum operator for fermionic matter
($\MatTTmunu$, say) through
\begin{equation}
\MatTTmunu = \frac{1}{4}\left\lbrace\Gmu \MHnu +
\bar{\mathcal{H}}_{\nu}\Gmu + \Gnu \MHmu + \bar{\mathcal{H}}_{\mu}
\Gnu \right\rbrace\label{245} \ ,
\end{equation}
\noindent and then the energy-momentum density for matter
$\MatTmunu$ is obtained quite similarly as for the current
densities (\ref{226}) or (\ref{234}) by putting
\begin{equation}
\MatTmunu = tr\left(\MI \cdot \MMatTmunu \right)\label{246} \ .
\end{equation}
\noindent Whether (or not) this is a successful construction is
decided by looking for the divergence of this tensor:
\begin{equation}\label{247}
\nmo \, \MatTmunu = tr\left\lbrace\MI \left(\MDmo\, \MMatTmunu +
\frac{i}{\hbar c}\left[\bar{\mathcal{H}}^{\mu}\,\MMatTmunu -
\MMatTmunu\, \MHmo\right]\right)\right\rbrace\ .
\end{equation}
\noindent As mentioned above, cf.(\ref{215}), this divergence of
the matter tensor $\MatTmunu$ should equal the Lorentz force
density $\Lfnu$ which surely will be true when the "RST
divergence" of the operator $\MMatTmunu$ yields just the
corresponding Lorentz force operator ${}^{(L)}\mathfrak{f}_{\nu}$, i.e.
\begin{equation}
\MDmo\ \MMatTmunu + \frac{i}{\hbar c}\left[\bar{\mathcal{H}}^{\mu}\ \MMatTmunu - \MMatTmunu \MHmo \right] = {}^{(L)}\mathfrak{f}_{\nu}\label{248} \ ,
\end{equation}
\noindent so that the desired force density ${}^{(L)}f_{\nu}$ (\ref{216}) emerges in the usual way as
\begin{equation}
\Lfnu = tr\left(\MI\cdot {}^{(L)}\mathfrak{f}_{\nu}\right) \ . \label{249}
\end{equation}
\indent However, the desired RST divergence of the operator $\MatTTmunu$ can be computed in a straight-forward manner under use of the integrability condition (\ref{225}) and is then ultimately found to look as follows:
\begin{align}\label{250}
&\MDmo \, \MatTTmunu + \frac{i}{\hbar c}\,\left[\bar{\mathcal{H}}^{\mu} \ \MatTTmunu - \MatTTmunu \MHmo\right] \\ \nonumber
& = \frac{1}{4}\,\bigg\lbrace i\hbar c\, \Big( \Gmo \MFmunu + \MFmunu \Gmo \Big) + \Gnu \Big( \MDmo \MHmu - \frac{i}{\hbar c}\,\MHmo \MHmu \Big) + \Big(\MDmo \bar{\mathcal{H}}_{\mu} + \frac{i}{\hbar c} \, \bar{\mathcal{H}}^{\mu} \bar{\mathcal{H}}_{\mu} \Big) \Gnu \\ \nonumber
& + \Big(\MDmo \Gmu + \frac{i}{\hbar c}\,[\bar{\mathcal{H}}^{\mu} \Gmu - \Gmu \MHmo]\Big)\cdot \MHnu + \bar{\mathcal{H}}_{\nu}\cdot \Big(\MDmo \Gmu + \frac{i}{\hbar c}\,[\bar{\mathcal{H}}^{\mu} \Gmu - \Gmu \MHmo]\Big) \\ \nonumber
& + \left[\MDmu \Gnu - \MDnu \Gmu \right]\cdot \MHmo + \bar{\mathcal{H}}^{\mu}\cdot \left[\MDmu \Gnu - \MDnu \Gmu \right] + \MDnu \left[\Gmo \MHmu - \bar{\mathcal{H}}^{\mu} \Gmu \right]\bigg\rbrace \ . \nonumber
\end{align}
\noindent Here, according to the original idea (\ref{248}), the right-hand side must yield the Lorentz-force operator ${}^{(L)}\mathfrak{f}_{\nu}$ which is to be conceived as the product of velocity operators and curvature components, see the first term. Obviously, in order to achieve this goal, one can resort to the previous equation (\ref{229}) which puts the RST divergence of the total velocity operator $\Gmu$ to zero; but this is of course not sufficient, and further assumptions concerning $\Gmu$ and $\MHmu$ must be adopted. Any such choice, which then validates the charge conservation laws (\ref{227}), (\ref{230}) and the energy-momentum conservation (or balance, resp.) law (\ref{215}), is considered as a particular \textit{realization of RST}. We now turn to the most important one of these realizations.

\subsection{Dirac Realization of RST}

\indent Evidently the requirement of conservation laws can be expressed in terms of certain conditions upon the velocity operator $\Gmu$ and the Hamiltonian $\MHmu$. First consider the case of $\Gmu$ which must have vanishing RST divergence (\ref{229}) and is required to commute with the gauge algebra generators $\talu$, cf.(\ref{240}), in order that the charge conservation laws be satisfied. An additional condition arises from the requirement of energy-momentum balance (\ref{248}) and (\ref{250}) where we assume now also the vanishing of the curl of $\Gmu$
\begin{equation}
\MDmu \Gnu - \MDnu \Gmu = 0\label{251}\ .
\end{equation}
\noindent However, we even want to go one step further and put the covariant derivative of $\Gmu$ to zero
\begin{equation}
\MDmu \Gnu = 0\label{252}\ .
\end{equation}
\noindent Fortunately this implies a great simplification of the mathematical formalism, because, on account of the commutation relations (\ref{240}), one can always choose a gauge where (at least locally) the velocity operator $\Gmu$ is absolutely constant:
\begin{equation}
\pmu \Gnu = 0\label{253}\ .
\end{equation}
\noindent Thus, the velocity operator becomes a fixed object, independent of any field kinematics and dynamics. For the \textit{Dirac realization} of RST, one chooses for $\Gmu$ the $N$-fold direct sum of the well-known Dirac matrices $\gmu$ \cite{a16}, i.e.
\begin{equation}
\Gmu = \gmu \oplus \gmu \oplus \dots \oplus \gmu \label{254}\ .
\end{equation}
\indent Naturally, such a choice of the velocity operator $\Gmu$ must imply further consequences. Returning for the moment to the vanishing RST divergence (\ref{229}) of $\Gmu$, one deduces from that equation the following relationship between $\Gmu$ and $\MHmu$:
\begin{equation}\label{255}
\bar\MHmo\cdot\Gmu -\Gmo\cdot\MHmu = 0 \ .
\end{equation}
Effectively this is a condition of Hermiticity and thus one can simplify the former mass relation (\ref{241}) to
\begin{equation}\label{256}
\bar\MHmo\Gmu = \Gmo\MHmu = \M c^2 \ .
\end{equation}
Furthermore, for the Dirac realization of RST for a system of \textit{identical} particles, cf. (\ref{244}), the proper gauge continuity equation (\ref{230}) is implied by the more general relation (\ref{243}). For non-identical particles, one puts the currents and potentials of the exchange type to zero, so that the source system (\ref{230}) becomes reduced to N true conservation laws ($a=1,\dots,N$)
\begin{equation}\label{257}
\nmo \jaomu \equiv 0 \ .
\end{equation}
Indeed, this is obtained from the general relation (\ref{243}) by linear combination of the mass operator $\M$ from the {\it electromagnetic} generators $\tau_a$ alone:
\begin{equation}\label{258}
\M = i \, M^a \tau_a\ ,
\end{equation}
with real-valued particle masses $M_a$. In the present paper, we however treat exclusively identical particles and will therefore always apply the identical-mass relation (\ref{244}) in combination with the Dirac realization.\\
\indent Finally, the source equation of the Hamiltonian $\MHmu$ must be determined in such a way that the right-hand side of the divergence relation (\ref{250}) for the energy-momentum operator $\MatTTmunu$ can really adopt the desired Lorentz form as indicated by the former claim (\ref{248}). However, this goal can be attained by adopting the following source equation for $\MHmu$ {\it (conservation equation)}
\begin{align}\label{259}
\MDmo \MHmu - \frac{i}{\hbar c}\MHmo \cdot \MHmu  &= - i \hbar c \bigg\lbrace \left( \frac{\M c}{\hbar} \right)^2 + \Smono\MFmunu \bigg\rbrace \\
\Big( \Smunu &\doteqdot \frac{1}{4} \left[\Gmu, \Gnu \right] \Big) \ .\nonumber
\end{align}
Indeed, by this choice of the source of $\MHmu$ the RST divergence of the energy-momentum operator $\MatTTmunu$ (\ref{250}) becomes
\begin{align}\label{260}
& \MDmo \, \MatTTmunu + \frac{i}{\hbar c} \left(\bar\MHmo \cdot \MatTTmunu - \MatTTmunu \cdot \MHmo \right) \\
& = \frac{i \hbar c}{4} \left(\MFmolo \left[\Smulu,\Gnu \right] + \left\lbrace \MFmunu, \Gmo \right\rbrace\right) = i \hbar c \, \MFmunu \Gmo \nonumber \ ,
\end{align}
where the following commutation relation of total velocity $\Gmu$ and the Spin(1,3) generators $\Smunu$ is to be observed:
\begin{equation}\label{261}
\left[\Smulu, \Gnu \right]= \Gmu \glunu - \Glu \gmunu\ .
\end{equation}
Thus with this result the Lorentz force operator ${}^{(L)}\mathfrak{f}_{\nu}$ (\ref{248}) is found to be just of the expected form, namely as the product of field strength $\MFmunu$ and velocity $\Gmu$
\begin{equation}\label{262}
{}^{(L)}\mathfrak{f}_{\nu} = i\hbar c \, \MFmunu\cdot \Gmo\ .
\end{equation}
This pleasant result for the force operator ${}^{(L)}\mathfrak{f}_{\nu}$ transcribes then immediately to the corresponding force density $\Lfnu$ (\ref{249})
\begin{eqnarray}\label{263}
\Lfnu &=& \hbar c \left\lbrace \exFmunu \, j^{\mu} + \Faomunu \, \jalumo \right\rbrace \nonumber \\
&\doteqdot& \exfnu + \Sfnu\ ,
\end{eqnarray}
namely by simply inserting the curvature decomposition (\ref{26a}) and observing the trace definitions for the total current $\jmu$ (\ref{226}) and the gauge currents $\jalumu$ (\ref{234}). Thus the former assertions (\ref{218})-(\ref{219}) are actually validated.\\
\indent For the special situation where matter is in a pure state, cf. (\ref{221}), the RSE (\ref{223}) may be converted to the well-known Dirac equation for the N-particle wave function $\Psi$
\begin{equation}\label{264}
i \hbar \ \Gmo \MDmu \Psi = \M c\cdot \Psi\ .
\end{equation}
For this purpose, one merely has to differentiate once more the RSE (\ref{223}) under use of the original form of the conservation equation (\ref{256}). The alternative version (\ref{259}) of the conservation equation can easily be deduced from its original form (\ref{256}) and is needed when one wishes to transcribe the Dirac equation (\ref{264}) to its Klein-Gordon counterpart as a second-order wave equation:
\begin{equation}\label{265}
 \MDmo \MDmu \ \Psi + \left(\frac{\M c}{\hbar}\right)^2 \, \Psi = -\ \Smono \MFmunu \Psi \ .
\end{equation}
For the subsequent applications of the theory to the two-particle systems, it is useful to write down the Dirac equation (\ref{264}) also in component form:
\begin{subequations}\label{266}
\begin{align}
i \hbar \ \gmo \Dmu \psi_1 &= M c \, \psi_1 \label{266a} \\
i \hbar \ \gmo \Dmu \psi_2 &= M c \, \psi_2 \ , \label{266b}
\end{align}
\end{subequations}
where the gauge-covariant derivatives of the single-particle wave functions $\psi_a$ ($a=1,2$) are given by 
\begin{subequations}\label{267}
\begin{align}
\Dmu \psi_1 & = \pmu \psi_1 - i \left[ \exAmu + \Azmu \right] \psi_1 - i \, \Bmu \psi_2 \label{267a}\\
\Dmu \psi_2 & = \pmu \psi_2 - i \left[ \exAmu + \Aemu \right] \psi_2 - i \, \Bsmu \psi_1 \label{267b}\ .
\end{align}
\end{subequations}
This coupled Dirac system will readily be specialized down to its stationary form in order to set up the energy eigenvalue problem for the two-electron atoms.

\section{Energy Eigenvalue Problem}

In order to convince oneself that RST can predict qualitatively the correct multiplet structure of the atomic spectra, one will first consider a very simple situation, i.e. the helium problem (two electrons around the fixed nucleus). Here one may further simplify matters by treating only those stationary two-particle states which have vanishing angular momentum ($L=0$), i.e. the singlet and triplet states $\sf ^1S_0$ and $\sf ^3S_1$.\\
\indent The standard (non-relativistic) quantum mechanics (in lowest order of approximation) says that the four possible combinations of the two spins
\begin{equation}\nonumber
\ket{\!\uparrow\,}\otimes\ket{\!\uparrow\,},\ \ \ket{\!\downarrow\,}\otimes\ket{\!\downarrow\,},\ \ \ket{\!\uparrow\,}\otimes\ket{\!\downarrow\,},\ \ \ket{\!\downarrow\,}\otimes\ket{\!\uparrow\,}
\end{equation}
will be arranged in such a way that there arises a singlet state $\ket{\psi_I}$
\begin{equation}\label{31}
\ket{\psi_I} =\frac{1}{\sqrt{2}}\left\lbrace\ket{\!\uparrow\,}\otimes\ket{\!\downarrow\,}-\ket{\!\downarrow\,}\otimes\ket{\!\uparrow\,}\right\rbrace
\end{equation}
which has vanishing total spin quantum numbers ($S, S_z$) = ($0,0$), being defined through
\begin{subequations}\label{32}
 \begin{align}
 \hat{\vec{S}}^2 \, \ket{\psi_I} &= S(S+1) \, \hbar^2 \, \ket{\psi_I} \label{32a}\\
 \hat{S}_z \, \ket{\psi_I} &= S_z \, \hbar \, \ket{\psi_I}\label{32b}\\
(-S &\leq S_z \leq S) \ , \nonumber
\end{align}
\end{subequations}
and furthermore there arise three triplet states $\ket{\psi_{I\!I\!I}^{(S_z)}}$ with $S=1$ and $S_z=\pm1,0$ :
\begin{subequations}\label{33}
\begin{align}
&S_z=+1:& &\ket{\psi_{I\!I\!I}^{(1)}} = \ket{\!\uparrow\,}\otimes\ket{\!\uparrow\,}\label{33a}\\
&S_z=0:& &\ket{\psi_{I\!I\!I}^{(0)}} = \frac{1}{\sqrt{2}}\left\lbrace\ket{\!\uparrow\,}\otimes\ket{\!\downarrow\,}+\ket{\!\downarrow\,}\otimes\ket{\!\uparrow\,}\right\rbrace \label{33b}\\
&S_z=-1:& &\ket{\psi_{I\!I\!I}^{(-1)}} = \ket{\!\downarrow\,}\otimes\ket{\!\downarrow\,} \ .\label{33c}
\end{align}
\end{subequations}
Since for these two different kinds of states $\ket{\psi_{I}}$ and $\ket{\psi_{I\!I\!I}}$ the single-particle spins $\vec{s}_{(a)}$ ($a=1,2$) do build up the total spin $\vec{S}$ in two different ways (namely "parallel" for the triplet states $S=1$, and "anti-parallel" for the singlet state $S=0$), it does not come as a surprise that the electronic interaction energies of the two spin arrangements are different: the energy eigenvalue of the singlet states $\ket{\psi_{I}}$ (para-helium) is {\it increased} (in lowest perturbation order) by the "exchange energy" $\Delta E_c$:
\begin{equation}\label{34}
\Delta E_c^{(n,n')}=e^2 \int\!\!\!\!\int d^3\vec{r} \, d^3\vec{r}\,' \ \frac{\psi^*_n(\vec{r}) \psi_{n'}(\vec{r}) \ \psi^*_{n'}(\vec{r}\,') \psi_{n}(\vec{r}\,')}{\mid\vec{r}-\vec{r}\,'\mid} \ ,
\end{equation}
whereas the energy of the triplet states $\ket{\psi_{I\!I\!I}}$ (ortho-helium) is {\it lowered} by the same amount $\Delta E_C^{(n,n')}$. Here the raising and lowering of energy refers to the "unperturbed" levels where the spin effect is completely neglected and the corresponding unperturbed wave functions $\psi_n(\vec{r})$ refer to the (spinless) Schr\"odinger eigenvalue problem. Consequently the ortho/para level splitting according to the standard theory $\delta_{St} E$ becomes twice the exchange energy
\begin{equation}\label{35}
\delta_{St}E \doteqdot E_{I\!I\!I} - E_{I} = -\, 2 \ \Delta E_C^{(n,n')} \ ,
\end{equation}
which is in best agreement with the experimental situation (see any standard textbook \cite{b7} and also the discussion in ref. \cite{u6}).\\
\indent Concerning now the situation in RST, which has the well-known Hartree-Fock approach as its non-relativistic limit, it is well-known that the latter approximative method does not always produce useful predictions, though being of great help in many respects (see the discussion of this point in ref. \cite{b25}). For the present case of ortho/para splitting $\delta E$ the ordinary Hartree-Fock method (and therefore also RST) predicts only half of the correct value $\delta_{St} E$ (\ref{35})
\begin{equation}\label{36}
\delta_{H\!F} E = \frac{1}{2} \, \delta_{St} E = - \, \Delta E_C^{(n,n')} \ .
\end{equation}
The reason is here that the ordinary HF method yields an exchange energy $\Delta E_c^{(n,n')}$ (\ref{34}) exclusively for the triplet states $\ket{\psi_{I\!I\!I}}$ (ortho-helium), but not for the singlet states $\ket{\psi_{I}}$, see fig.3. Thus the HF method appears to be helpful for computing the ortho-level system but not for the para-level system of helium. Clearly, since RST is the relativistic generalization of HF, this methodological ambiguity must transcribe to RST which thus could be expected to be a useful technique only when dealing with the ortho-states $\sf ^3S_1$. However the situation becomes more favourable also for the para-states $\sf ^1S_0$ when one considers high-energy situations (i.e. high nuclear charges $Z>>1$) because in this highly relativistic regime it is more important to take into account the relativistic effects than the correlation effects beyond Hartree-Fock (see below and ref. \cite{u5} for a demonstration of this).\\
\indent The ortho/para energy shift comes about mainly through separation/crowding of the electronic charge clouds by (anti-) symmetrizing the spatial parts of the wave functions relative to the symmetry properties of their spin parts, according to the Pauli antisymmetrization postulate for the total wave function.\\
\indent In any case it is necessary and desirable to give an
explicit demonstration of how the singlet and triplet structure of
the helium-like atoms does arise within the framework of RST,
so that the problem of the missing exchange energy
for the singlet states can be elucidated, see the discussion below equation (\ref{418}).

\pagebreak
\begin{figure}
\begin{center}
\label{fig3}
\input{splitting.pstex_t}
\end{center}
{\large{\textbf{Fig.\ 3} \em{Ortho/Para Level Splitting}}}\\
The standard level splitting $\delta_{St}E$ (\ref{35}) of a Schr\"odinger state $ns, n's$ is twice the Hartree-Fock value $\delta_{H\!F} E$ (\ref{36}) because the latter approach treats electrons with different spins $\ket{\!\uparrow\,}$ $\ket{\!\downarrow\,}$ (para-helium) as non-identical particles missing the exchange effect.
\vspace{2cm}
\end{figure}

\subsection{Stationary Two-Particle States}

For a stationary state, one expects that certain physical
observables (e.g. energy) are time-independent, whereas other
objects of the theory are admitted to undergo regular
oscillations. Thus the two components $\psi_a(\vec{r},t)$
($a=1,2$) of the wave function $\Psi$ are written as
\begin{subequations}\label{37}
 \begin{align}
 \psi_1(\vec{r},t) = \exp\left[-\,i\,\frac{M_1 c^2}{\hbar}\,t\right] \cdot
 \psi_1(\vec{r}) \label{37a} \\
 \psi_2(\vec{r},t) = \exp\left[-\,i\,\frac{M_2 c^2}{\hbar}\,t\right] \cdot
 \psi_2(\vec{r}) \ , \label{37b}
 \end{align}
\end{subequations}
where the time-independent Dirac spinors $\psi_a(\vec{r})$ may be
further decomposed into two-component Pauli spinors
$^{(a)}\!\phi_\pm(\vec{r})$ according to ($a=1,2$)
\begin{equation}\label{38}
\psi_a(\vec{r}) = \left(\begin{array}{l}^{(a)}\!\phi_+(\vec{r})\\
^{(a)}\!\phi_- \vvecr\end{array}\right) \ .
\end{equation}
Furthermore, each of the four Pauli spinors $^{(a)}\!\phi_\pm
(\vec{r})$ may be decomposed with respect to an orthonormal basis,
namely either the basis $\zeta^{\pm}_0$ for the upper components
\begin{equation}\label{39}
^{(a)}\!\phi_+(\vec{r}) = {}^{(a)}\!R_+(\vec{r})\cdot
\zeta^{(+)}_0 + {}^{(a)}\!S_+(\vec{r}) \cdot \zeta_0^{(-)} \ ,
\end{equation}
or, resp., the basis $\zeta_1^{(\pm)}$ for the lower components
\begin{equation}\label{310}
^{(a)}\!\phi_-(\vec{r}) = -i \, {}^{(a)}\!R_- \cdot
\zeta^{(+)}_1 - i\, {}^{(a)}\!S_- \cdot \zeta_1^{(-)} \ .
\end{equation}
Here, the two basis systems are defined in terms of the usual
standard basis $\zetajml$ \cite{b27} as
\begin{subequations}\label{311}
\begin{align}
\zeta_0^{(\pm)} = \zetapmn \pm i\,
e^{-i \varphi} \cdot \zetappn \label{311a} \\
\zeta_1^{(\pm)} = \zetapme \pm i\,
e^{-i \varphi} \cdot \zetappe \ . \label{311b}
\end{align}
\end{subequations}
Recall here, that the single-particle standard basis
$\zeta^{\frac{1}{2},\pm\frac{1}{2}}_{\;\;\, 0},
\zeta^{\frac{1}{2},\pm\frac{1}{2}}_{\;\;\, 1}$ has the well-known
eigenvalue properties
\begin{subequations}\label{312}
\begin{align}
\vec{J}^{\,2} \, \zetajml &= j\,(j+1)\, \hbar^2 \,
\zetajml \label{312a}\\ 
J_z \; \zetajml &= m \, \hbar\, \zetajml \label{312b} \\
\vec{L}^{\,2} \, \zetajml &= l\,(l+1) \, \hbar^2 \,
\zetajml \label{312c}\\ 
\vec{S}^{\,2} \, \zetajml &= \frac{1}{2}\,(\frac{1}{2}+1) \,
\hbar^2 \,\zetajml \ . \label{312d}
\end{align}
\end{subequations}
\indent According to such a stationary ansatz for the wave
functions, one expects that also the corresponding gauge fields
are either time-independent or regularly oscillating:
\begin{subequations}\label{313}
\begin{align}
&A^a_{\ \mu}(\vec{r},t) \ \Rightarrow \ \left\lbrace{}^{(a)}\!A_0(\vec{r}),
-\vec{A}_a(\vec{r}) \right\rbrace \label{313a} \\ 
&B_\mu(\vec{r},t) \ \Rightarrow \ \exp \left[\,-i\, \frac{M_1-M_2}{\hbar} \,
c^2 \, t \, \right]
\cdot \left\lbrace B_0(\vec{r}), -\vec{B}(\vec{r}) \right\rbrace \ . \label{313b}
\end{align}
\end{subequations}
Moreover, the RST-Maxwell equations (\ref{211b})-(\ref{211e})
clearly demonstrate that the currents are intimately linked to the
gauge potentials and therefore they may be expected to be both of
a similar form. Thus one assumes that the electromagnetic currents
$\jaomu$ are also time independent:
\begin{equation}\label{314}
\jaomu = - \kaumu \equiv \bar{\psi}_a \, \gmu\, \psi_a \
\Rightarrow \ \left\lbrace {}^{(a)}\!k_0(\vec{r})\, , \,
-\vec{k}_a(\vec{r}) \right\rbrace \,
\end{equation}
whereas the exchange current $h_\mu \ (\,\equiv j_{3\mu} \equiv
-j^4_{\;\;\mu} )$ oscillates countercurrently with respect to the exchange potential
$\Bmu$ (\ref{313b}):
\begin{equation}\label{315}
\hmu = \bar{\psi}_1 \, \gmu \, \psi_2 \ \Rightarrow \exp\left[i\,
\frac{M_1-M_2}{\hbar} \, c^2 \, t \right] \cdot \left\lbrace
h_0(\vec{r})\, ,\, -\vec{h}(\vec{r}) \right\rbrace \ .
\end{equation}
Concerning the external potential $\exAmu$ (\ref{21}), we shall treat the
helium problem where $\exAmu$ is due to the Coulomb potential of
the nucleus
\begin{subequations}\label{316}
\begin{align}
\exAmu \ \Rightarrow \ &\left\lbrace {}^{(ex)}\!A_0(\vec{r})\, ,\, 0
\right\rbrace \label{316a}\\ 
{}^{(ex)}\!A_0(\vec{r}) &= z_{ex} \frac{\alpha_s}{r} \label{316b}\\ 
(\alpha_s &\doteqdot \frac{e^2}{\hbar c}) \ . \nonumber 
\end{align}
\end{subequations}
According to the restriction to such a spherically symmetric potential $\exAmu$, one will resort to the use of spherical polar
coordinates ($r, \vartheta,\varphi$) and will decompose all
three-vectors with respect to such a moving basis system
$\left\lbrace\vec{e}_r, \vec{e}_\varphi, \vec{e}_\vartheta
\right\rbrace$, i.e. one puts
\begin{subequations}\label{317}
\begin{align}
\vec{k}_a(\vec{r}) &= {}^{(a)}\!k_r \, \vec{e}_r +
{}^{(a)}\!k_\vartheta \, \vec{e}_\vartheta + {}^{(a)}\!k_\varphi
\, \vec{e}_\varphi \label{317a}\\ 
\vec{h}(\vec{r}) &= h_r \, \vec{e}_r + h_\vartheta \,
\vec{e}_\vartheta + h_\varphi \, \vec{e}_\varphi \label{317b}\\ 
\vec{A}_a(\vec{r}) &= {}^{(a)}\!A_r \, \vec{e}_r +
{}^{(a)}\!A_\vartheta \, \vec{e}_\vartheta + {}^{(a)}\!A_\varphi
\, \vec{e}_\varphi \label{317c}\\ 
\vec{B}(\vec{r}) &= B_r \, \vec{e}_r + B_\vartheta \,
\vec{e}_\vartheta + B_\varphi \, \vec{e}_\varphi \ . \label{317d}
\end{align}
\end{subequations}

\subsection{Currents}
\indent The current densities $\jalomu = \lbrace\kamu ; \hmu\rbrace$ are especially interesting as far as the \textit{geometric} difference between an ortho-field and a para-field configuration is concerned. Indeed one has to expect that the \textit{physical} difference between both field configurations is ultimately a consequence of the different {\it geometric} patterns of the associated ortho- and para-streamlines. The point here is that the spin in RST is traced back to the rotational flow generated by the wave functions and therefore the spin-spin interaction emerges as the magnetic field energy of those different rotational flows.\\
\indent Inserting the general ansatz (\ref{39})-(\ref{310}) into the RST currents $\jalumu$ (\ref{234})
\begin{subequations}\label{318}
\begin{align}
\jeumu & = - \jzomu \doteqdot \kzumu = \bar{\psi}_2 \gmu \psi_2\label{318a} \\ 
\jzumu & = - \jeomu \doteqdot \keumu = \bar{\psi}_1 \gmu \psi_1 \label{318b} \\
\jdumu & = - \jvomu \doteqdot \hmu = \bar{\psi}_1 \gmu \psi_2 \label{318c} \\
\jvumu & = - \jdomu \doteqdot -\hsmu = -\bar{\psi}_2 \gmu \psi_1 \label{318d}
\end{align}
\end{subequations}
\noindent yields for the electrostatic charge densities ${}^{(a)}\!k_0(\vec{r})$ (\ref{314}) $(a=1,2)$
\begin{equation}\label{319}
{}^{(a)}\!k_0 = \frac{1}{2\pi}\lbrace \aRsp \cdot \aRp + \aSsp \cdot \aSp + \aRsm \cdot \aRm + \aSsm \cdot \aSm \rbrace \ ,
\end{equation}
\noindent and similarly for the corresponding electric three-currents $\vec{k}_a(\vec{r})$ (\ref{317a}) $(a=1,2)$
\begin{subequations}\label{320}
\begin{align}
{}^{(a)}\!k_r & = \frac{i}{2\pi}\lbrace \aRsp \cdot \aRm - \aRp \cdot \aRsm + \aSsp \cdot \aSm - \aSp \cdot \aSsm \rbrace \label{320a}\\
{}^{(a)}\!k_{\vartheta} & = \frac{1}{2\pi}\lbrace \aSsp \cdot \aSm + \aSp \cdot \aSsm - \aRsp \cdot \aRm - \aRp \cdot \aRsm \rbrace \label{320b}\\
{}^{(a)}\!k_{\varphi} & = \frac{i}{2\pi}\lbrace e^{i\vartheta} [\aRsp \cdot \aSm + \aRsm \cdot \aSp] - e^{-i\vartheta} [\aRp \cdot \aSsm + \aRm \cdot \aSsp] \rbrace \label{320c}.
\end{align}
\end{subequations}
\noindent Furthermore the exchange density $h_0(\vec{r})$ becomes
\begin{equation}
h_0 = \frac{1}{2\pi}\lbrace \eRsp \cdot \zRp + \eSsp \cdot \zSp + \eRsm \cdot \zRm + \eSsm \cdot \zSm \rbrace \label{321}
\end{equation}
\noindent and the exchange current $\vec{h}(\vec{r})$ (\ref{317b}) is found as
\begin{subequations}\label{322}
\begin{align}
h_r & = \frac{i}{2\pi}\lbrace \eRsp \cdot \zRm + \eSsp \cdot \zSm - \eRsm \cdot \zRp - \eSsm \cdot \zSp \rbrace \label{322a} \\
h_{\vartheta} & = \frac{1}{2\pi}\lbrace \eSsp \cdot \zSm + \eSsm \cdot \zSp - \eRsp \cdot \zRm - \eRsm \cdot \zRp \rbrace \label{322b} \\
h_{\varphi} & = \frac{i}{2\pi}\lbrace e^{i\vartheta} [\eRsp \cdot \zSm + \eRsm \cdot \zSp] - e^{-i\vartheta}[\eSsp \cdot \zRm + \eSsm \cdot \zRp] \rbrace \ . \label{322c}
\end{align}
\end{subequations}
\noindent These densities and currents are now relevant for the energy eigenvalue problem insofar as they determine the potentials $\Aaomu$ via the solutions of the RST Maxwell equations (\ref{211b}) - (\ref{211e}); these potentials then enter the covariant derivatives $\Dmu \psi_a$ of the wave functions (\ref{267a}) - (\ref{267b}) which in turn enter the eigenvalue equations (\ref{266a}) - (\ref{266b}). This will readily become more evident when writing down now the stationary form of the coupled Dirac system (\ref{266a}) - (\ref{266b}).

\subsection{Mass Eigenvalue Equations}

Naturally, the mass eigenvalues $M_a$ (\ref{37a}) - (\ref{37b}) will form an essential constituent of the atomic energy levels ($E_T$, say), but they are not directly identical with those atomic energies. Rather, one will have to specify the atomic energies $E_T$ in terms of an appropriate energy functional, which, besides by the mass eigenvalues ($M_a c^2$), is built up also by the energy content of the gauge fields. However, the energy contribution of the matter fields is essentially determined by the sum of mass-energies ($M_1c^2+M_2c^2$); and therefore the mass eigenvalue problem must first be solved before the value of the energy functional $E_T$ upon the corresponding solutions of the coupled matter and gauge-field system can be calculated. \\
\indent Once the stationary form of the wave functions has been specified through equations (\ref{37}) - (\ref{312}), one simply inserts this stationary ansatz into the coupled Dirac system (\ref{266a}) - (\ref{266b}) and then finds the corresponding mass eigenvalue equations in terms of the wave amplitudes ${}^{(a)}\!R_{\pm}$ and ${}^{(a)}\!S_{\pm}$. These equations may perhaps look a little bit lengthy when written down in spherical polar coordinates, but it is worthwile to display them explicitly because a lot of information about the peculiarities of the electromagnetic and exchange interactions can be drawn from them. The equation for the wave amplitude $\eRp$ reads
\begin{align} \label{323}
& \frac{\partial \eRp}{\partial r} - \frac{i}{r} \left[\frac{\partial \ \eRp}{\partial \vartheta} + \frac{e^{i\vartheta}}{2\sin\vartheta} \cdot \eRp \right] - \frac{e^{i\vartheta}}{r\sin \vartheta} \left[ \frac{\partial \ \eSp}{\partial \varphi} - \frac{i}{2} \cdot \eSp \right] \\ \nonumber 
& + [\exAo + \zAo]\cdot \eRm + B_0\cdot \zRm + [\zAtu + i \cdot \zAru]\cdot \eRp \\ 
& - ie^{i\vartheta} \cdot \zApu \cdot \eSp + [\Btu + i\Bru]\cdot \zRp - ie^{i\vartheta}\Bpu \cdot \zSp  = -\frac{M_1+M}{\hbar} c \cdot \eRm \ . \nonumber
\end{align}
\noindent Obviously, the "positive-energy" component $\eRp$ of the first particle's wave function couples via the \textit{electrostatic} potential $\zAo$ and electric \textit{exchange} potential $B_0$ to the "negative-energy" components $\eRm$ and $\zRm$, respectively; whereas the coupling to the positive-energy components $\eRp$ and $\zRp$ occurs via the \textit{magnetic} potentials $\zAtu$ and $\zAru$. In contrast to this, the coupling of $\eRp$ to the "secondary" positive-energy components $\aSp$ occurs always via the azimuthal components $\zApu$ and $\Bpu$, respectively. Of course, the coupling of the negative-energy component $\eRm$ follows an analogous pattern:
\begin{align}
& \frac{\partial \ \eRm}{\partial r} + \frac{2}{r}\cdot \eRm + \frac{i}{r}\left[\frac{\partial \ \eRm}{\partial \vartheta} + \frac{e^{i\vartheta}}{2\sin\vartheta} \cdot \eRm \right] + \frac{e^{i\vartheta}}{r\sin \vartheta} \left[ \frac{\partial \ \eSm}{\partial \varphi} - \frac{i}{2} \cdot \eSm \right] \nonumber \\ 
& - [\exAo + \zAo]\cdot \eRp - B_0\cdot \zRp - [\zAtu - i\cdot \zAru]\cdot \eRm \label{324}\\ 
& + ie^{i\vartheta} \cdot \zApu \cdot \eSm - [\Btu - i\Bru]\cdot \zRm + ie^{i\vartheta}\Bpu \cdot \zSm  = \frac{M_1-M}{\hbar} c \cdot \eRp \ .\nonumber
\end{align}  
\noindent Thus it becomes evident that the interactions of the \textit{electric} type couple the wave amplitudes of different energy types, whereas the \textit{magnetic} interactions connect the amplitudes of the same energy type. \\
\indent Another striking feature of these two-particle interactions concerns the coupling of the "primary" amplitudes ${}^{(a)}\!R_{\pm}$ to their "secondary" counterparts ${}^{(a)}\!S_{\pm}$, and vice versa. In order to see this more clearly, one writes down also the eigenvalue equations for the "secondary" components ${}^{(1)}\!S_{\pm}$:
\begin{subequations}\label{325}
\begin{align}\label{325a}
& \frac{\partial \ \eSp}{\partial r} + \frac{i}{r} \left[\frac{\partial \ \eSp}{\partial \vartheta} + \frac{e^{-i\vartheta}}{2\sin\vartheta} \cdot   \eSp \right] + \frac{e^{-i\vartheta}}{r\sin \vartheta} \left[ \frac{\partial \ \eRp}{\partial \varphi} - \frac{i}{2} \cdot \eRp \right] \\ \nonumber 
& + [\exAo + \zAo]\cdot \eSm + B_0\cdot \zSm - [\zAtu - i \cdot \zAru]\cdot \eSp \\ \nonumber 
& + ie^{-i\vartheta} \cdot \zApu \cdot \eRp - [\Btu - i\Bru]\cdot \zSp + ie^{-i\vartheta}\Bpu \cdot \zRp = -\frac{M_1+M}{\hbar} c \cdot \eSm \\ \nonumber 
\end{align}
\begin{align}
& \frac{\partial \ \eSm}{\partial r} + \frac{2}{r} \cdot \eSm - \frac{i}{r}\left[\frac{\partial \ \eSm}{\partial \vartheta} + \frac{e^{-i\vartheta}}{2\sin\vartheta} \cdot \eSm \right] - \frac{e^{-i\vartheta}}{r\sin \vartheta} \left[ \frac{\partial \ \eRm}{\partial \varphi} - \frac{i}{2} \cdot \eRm \right] \nonumber \\
& - [\exAo + \zAo]\cdot \eSp - B_0\cdot \zSp + [\zAtu + i \cdot \zAru]\cdot \eSm \label{325b}\\ 
& - ie^{-i\vartheta} \cdot \zApu \cdot \eRm + [\Btu + i\Bru]\cdot \zSm - ie^{-i\vartheta}\Bpu \cdot \zRm  = \frac{M_1-M}{\hbar} c \cdot \eSp \ . \nonumber
\end{align}
\end{subequations}
\noindent From here it is obvious again that the electric potentials $\lbrace {}^{(a)}\!A_0$, $B_0\rbrace$ mediate between the positive/negative energy components (${}^{(a)}\!S_+\leftrightarrow {}^{(a)}\!S_-$) whereas the magnetic potentials $\lbrace \vec{A}_a$, $\vec{B}\rbrace$ mediate between the components of the same energy type (${}^{(a)}\!S_+\leftrightarrow {}^{(a)}\!S_+$, etc.). However, what may also be seen clearly from the complete set of eigenvalue equations for the first particle ($a=1$) (\ref{323}) - (\ref{325b}) is the fact that the coupling of the primary ($R$) and secondary ($S$) components ($R\leftrightarrow S$) runs along the \textit{azimuthal} direction. This will readily become important when we subdivide the whole set of solutions for the eigenvalue equations into the subsets of ortho-type and para-type, respectively. Finally, let us mention that, for the sake of brevity, we will not write down here the eigenvalue equations for the second particle ($a=2$), because this is obtained from the present first-particle's equations (\ref{323}) - (\ref{325b}) simply by means of the following replacements:
\begin{subequations}\label{326}
\begin{align}\label{326a}
M_1 & \leftrightarrow M_2 \\ \label{326b}
\eRpm \leftrightarrow \zRpm, & \qquad \eSpm \leftrightarrow \zSpm \\ \label{326c}
\eAo \leftrightarrow \zAo, & \qquad B_0 \leftrightarrow B^{\ast}_0 \\ \label{326d}
\vec{A}_1 \leftrightarrow \vec{A}_2, & \qquad \vec{B} \leftrightarrow \vec{B}^{\ast} \ .
\end{align}
\end{subequations}
\indent Two remarks must be made concerning the solutions of the mass eigenvalue problem (\ref{323})-(\ref{326d}). Namely first, these solutions are unique only if one imposes some normalization condition upon the single-particle wave functions $\psi_a(\vec{r})$, i.e. one needs some normalization condition upon the wave amplitudes $\aRpm(\vec{r})$ and $\aSpm(\vec{r})$. The desired condition may be deduced in a very natural way from the non-abelian RST-Maxwell equations (\ref{211b})-(\ref{211e}): Recasting the electromagnetic part (\ref{211b})-(\ref{211c}) of this system into the following form \cite{a43}
\begin{subequations}\label{327}
\begin{align}
\pmo \Femunu = -4 \pi \alpha_s \, {}^{(1)}\!l_\nu &\equiv -4 \pi \alpha_s \left\{ \keunu + \frac{i}{4\pi\alpha_s} \left[\Bmo \Gsmunu - \Bsmo \Gmunu \right]\right\}\label{327a}\\
\pmo \Fzmunu = -4 \pi \alpha_s \, {}^{(2)}\!l_\nu &\equiv -4 \pi \alpha_s \left\{ \kzunu - \frac{i}{4\pi\alpha_s} \left[\Bmo \Gsmunu - \Bsmo \Gmunu \right]\right\} \ ,\label{327b}
\end{align}
\end{subequations}
it becomes easy to see from the self-evident identities 
\begin{equation}\label{328}
\pmo \pno \Famunu \equiv 0
\end{equation}
that both real-valued one-forms ${}^{(a)}\!l_\nu$
\begin{subequations}\label{329}
\begin{align}
{}^{(1)}\!l_\nu \doteqdot \keunu + \frac{i}{4\pi\alpha_s} \left[\Bmo\Gsmunu-\Bsmo \Gmunu \right]\label{329a}\\
{}^{(2)}\!l_\nu \doteqdot \kzunu - \frac{i}{4\pi\alpha_s} \left[\Bmo\Gsmunu-\Bsmo \Gmunu \right]\label{329b}
\end{align}
\end{subequations}
obey a true continuity equation, i.e.
\begin{equation}\label{330}
\pmo {}^{(a)}\!l_\mu  \equiv 0 \ .
\end{equation}
Therefore they can serve to define the desired normalization conditions as
\begin{equation}\label{331}
\int_{(S)} {}^{(a)}\!l_\mu d S^\mu =1
\end{equation}
where the hypersurface ($S$) may be chosen arbitrarily. But clearly, for the present stationary field configurations one will choose a time slice ($t=const$) as the hypersurface ($S$) which cuts the general form (\ref{331}) down to
\begin{equation}\label{332}
\int d^3\vec{r} \ {}^{(a)}\!l_0(\vec{r}) = 1
\end{equation}
with the modified charge densities ${}^{(a)}\!l_0(\vec{r})$ being given by
\begin{subequations}\label{333}
\begin{align}
{}^{(1)}\!l_0(\vec{r}) = {}^{(1)}\!k_0(\vec{r}) + \frac{i}{4\pi\alpha_s} \left[\vec{B}^*(\vec{r}) \cdot \vec{X}(\vec{r}) - \vec{B}(\vec{r}) \cdot \vec{X}^*(\vec{r}) \right]\label{333a}\\
{}^{(2)}\!l_0(\vec{r}) = {}^{(2)}\!k_0(\vec{r}) - \frac{i}{4\pi\alpha_s} \left[\vec{B}^*(\vec{r}) \cdot \vec{X}(\vec{r}) - \vec{B}(\vec{r}) \cdot \vec{X}^*(\vec{r}) \right] \label{333b}\\
\left( \vec{X}(\vec{r}) \equiv \lbrace \X^j(\vec{r}) \rbrace \doteqdot \lbrace G_{0j}(\vec{r})\rbrace \right) \ . \nonumber
\end{align}
\end{subequations}
\indent The second remark refers to the emergence of the gauge potentials $\{ \aAo(\vec{r}), B_0(\vec{r}); \vec{A}_a(\vec{r}), \vec{B}(\vec{r}) \}$ in the mass eigenvalue equations. In order that this system of equations be closed, one has to add the field equations for those potentials, whose general form, however, has been presented already in ref. \cite{a41}. Thus for the present purpose it may be sufficient to merely quote their linearized form 
\begin{subequations}\label{334}
\begin{align}
\Delta \, \aAo(\vec{r}) & = 4 \pi \alpha_s \, \ako(\vec{r}) \label{334a} \\
\Delta \, B_0(\vec{r}) & = - 4 \pi \alpha_s \, h^*_0(\vec{r}) \label{334b} \\
\Delta \, \vec{A_a}(\vec{r}) & = 4 \pi \alpha_s \, \vec{k}_a(\vec{r}) \label{334c} \\
\Delta \, \vec{B}(\vec{r}) & = - 4 \pi \alpha_s \, \vec{h}^*(\vec{r}) \ . \label{334d} 
\end{align}
\end{subequations}
This simplified form is sufficient for a rough estimate of the atomic energy levels (or frequency of spectral lines, resp.). It corresponds to the use of the Coulomb and Breit interactions by the conventional methods in the literature \cite{a5,a6,b54}. However for our numerical caculations (see below) we will rely upon the {\it exact} Poisson equations which are non-linear on account of the non-abelian character of the two-particle theory.

\subsection{Energy Functional}

In order to test the quality of the RST predictions in atomic physics, it is not sufficient to solve the mass eigenvalue system because the mass eigenvalues $M_a c^2$ do in general not coincide with the energy $E_T$ carried by the field configuration. The question of energy functional has been considered in great detail
in some preceding papers \cite{a43,a27} so that it may suffice here to
simply quote the main result:
\begin{equation}\label{335}
E_T = \sum_{a=1}^{2} \hat{z}_a \cdot M_a c^2 - \Delta E_T^{(em)}
- \Delta E_T^{(hg)} \ .
\end{equation}
The general structure of this result is very plausible because it says that the total energy $E_T$ of any stationary RST configuration is the "corrected" sum of mass-energies $M_a c^2$. Indeed the necessity for the emergence of the electromagnetic ($\Delta E_T^{(em)}$) and exchange ($\Delta E_T^{(hg)}$) correction terms arises from the fact that {\it each} mass energy $M_a c^2$ ($a=1,2$) contains already the interaction energy with the other particle so that this interaction energy becomes then counted {\it twice} when one simply forms the sum of mass energies! This is the reason why the mutual interaction energy $\Delta E_T$
\begin{equation}\label{336}
\Delta E_T \doteqdot \Delta E_T^{(em)} + \Delta E_T^{(hg)}
\end{equation}
must be resubtracted from the sum of mass energies. Since both contributions $\Delta E_T^{(em)}$ and $\Delta E_T^{(hg)}$ themselves consist of two terms (i.e. " electric" (e; h) and "magnetic" (m; g)),
\begin{subequations}\label{337}
\begin{align}
 \Delta E_T^{(em)} = \Delta E_T^{(e)} + \Delta E_T^{(m)}\label{337a}\\
 \Delta E_T^{(hg)} = \Delta E_T^{(h)} + \Delta E_T^{(g)}\ ,\label{337b}
\end{align}
\end{subequations}
the total correction energy $\Delta E_T$ (\ref{336}) is built up by four terms:
\begin{equation}\label{338}
\Delta E_T = \Delta E_T^{(e)} + \Delta E_T^{(m)} + \Delta
E_T^{(h)} + \Delta E_T^{(g)} \ ,
\end{equation}
which will now be briefly explained.\\
\indent The energy correction $\Delta E_T^{(e)}$ of the {\it electrostatic} type is the difference of (the sum of) mass equivalents $M_a^{(e)}$ and of the electrostatic gauge field energy $E_R^{(e)}$, i.e.
\begin{equation}\label{339}
\Delta E_T^{(e)} = \sum_{a=1}^2 \hat{z}_a \cdot M_a^{(e)} c^2 -
E_R^{(e)} \ ,
\end{equation}
where the electrostatic mass-energy $M_a^{(e)} c^2$ is defined through
\begin{subequations}\label{340}
\begin{align}
 \hat{z}_1 \cdot M_1^{(e)} c^2 &= - \hbar c \int d^3\vec{r}
 \ \, \zAo(\vec{r}) \cdot \eko(\vec{r})\label{340a} \\
 \hat{z}_2 \cdot M_2^{(e)} c^2 &= - \hbar c \int d^3\vec{r}
 \ \, \eAo(\vec{r}) \cdot \zko(\vec{r})\label{340b} \\
 \Big(\hat{z}_a &\doteqdot \int d^3\vec{r} \ \, \ako(\vec{r})\Big) \ .
 \nonumber
\end{align}
\end{subequations}
Furthermore the electrostatic gauge field energy $E_R^{(e)}$ is given by
\begin{align}\label{341}
 E_R^{(e)} = \frac{\hbar c}{4 \pi \alpha_s} \int d^3\vec{r} \ \,
 \vec{E}_1(\vec{r}) \cdot \vec{E}_2(\vec{r})\\
 \left( \vec{E}_a(\vec{r}) \equiv \{ {}^{(a)}\!E^j(\vec{r})\} \doteqdot
 \{ {}^{(a)}\!F_{0j}(\vec{r}) \} \right) \ . \nonumber
\end{align}
Similarly the exchange correction energy of the electric type $\Delta E_T^{(h)}$ looks as follows
\begin{equation}\label{342}
\Delta E_T^{(h)} = \sum_{a=1}^2 \hat{z}_a \cdot M_a^{(h)}\,c^2 +
E_C^{(h)} \ ,
\end{equation}
with the exchange mass-energies $M_a^{(h)}c^2$ being defined by
\begin{subequations}\label{343}
\begin{align}
 \hat{z}_1 \cdot M_1^{(h)} c^2 &= - \hbar c \int d^3\vec{r}
 \ \, B_0(\vec{r}) \cdot h_0(\vec{r})\label{343a} \\
 \hat{z}_2 \cdot M_2^{(h)} c^2 &= - \hbar c \int d^3\vec{r}
 \ \, B^*_0(\vec{r}) \cdot h^*_0(\vec{r}) \ , \label{343b}
\end{align}
\end{subequations}
and analogously the electric type of exchange energy $E_C^{(h)}$ by
\begin{equation}\label{344}
E_C^{(h)} = \frac{\hbar c}{4\pi \alpha_s} \int d^3\vec{r} \ \,
\vec{X}^{\,*}(\vec{r}) \cdot \vec{X}(\vec{r}) \ .
\end{equation}
Obviously both energy corrections of the electric type $\Delta E_T^{(e)}$ and $\Delta E_T^{(h)}$ are of a very similar structure but
they differ essentially from their magnetic counterparts.\\
\indent Indeed, the point with the magnetic interactions is here that the correction terms $\Delta E_T^{(m)}$ and $\Delta E_T^{(g)}$ are built up by the corresponding gauge field energies alone, i.e. we have for the magnetostatic case
\begin{align}\label{345}
- \Delta E_T^{(m)} \equiv E_R^{(m)} \doteqdot \frac{\hbar c}{4
\pi \alpha_s} \int d^3\vec{r} \ \ \vec{H}_1(\vec{r}) \cdot
\vec{H}_2(\vec{r})\\
\left( \vec{H}_a(\vec{r}) \equiv \{{}^{(a)}\!H^j(\vec{r}) \}
\doteqdot \{\frac{1}{2} \, \epsilon^{jk}_{\ \ l} \, {}^{(a)}\!F_k^{\ l}(\vec{r})\}
\right) \ , \nonumber
\end{align}
and similarly for the exchange case
\begin{align}\label{346}
\Delta E_T^{(g)} \equiv E_C^{(g)} \doteqdot \frac{\hbar c}{4 \pi
\alpha_s} \int d^3\vec{r} \ \ \vec{Y}^*(\vec{r}) \cdot
\vec{Y}(\vec{r})\\
\left( \vec{Y}(\vec{r}) \equiv \{Y^j(\vec{r}) \} \doteqdot
\{\frac{1}{2} \, \epsilon^{jk}_{\ \ l}\, {}^{(a)}\!G_k^{\ l}(\vec{r})\} \right) \ . \nonumber
\end{align}
One can easily show that this somewhat different structure of the electric and magnetic contributions is generated by the combination of the principle of minimal coupling with the Lorentz covariance of the theory. One can also demonstrate that it is just this specific relationship between the electric and
magnetic effects which brings the RST predictions close to the
experimental data.\\
\indent In order to see more clearly the specific way in which the cooperation of minimal coupling and Lorentz covariance generate
these differences of the magnetic and electric energy contributions it may be sufficient to demonstrate this for the
linear approximation of the RST-Maxwell equations where also the linear Poisson equations (\ref{334a})-(\ref{334d}) do hold.
Actually, for this simplified situation, the electrostatic gauge field energy $E_R^{(e)}$ can be converted to the corresponding
mass-energy $M_a^{(e)} c^2$ (\ref{340a})-(\ref{340b}) through a simple integration by parts
\begin{equation}\label{347}
E_R^{(e)} \Rightarrow \ \frac{1}{2} \; \sum_{a=1}^2 \hat{z}_a \cdot
M_a^{(e)} c^2 \ ,
\end{equation}
so that the electrostatic energy correction $\Delta E_T^{(e)}$ (\ref{339}) adopts just the value of this field energy
\begin{equation}\label{348}
\Delta E_T^{(e)} \Rightarrow \ \frac{1}{2} \; \sum_{a=1}^2 \hat{z}_a \cdot
M_a^{(e)} c^2 \equiv E_R^{(e)} \ .
\end{equation}
On the other hand, the magnetic gauge field energy $E_R^{(m)}$ (\ref{345}) can also be converted to its mass-energy equivalent $M_a^{(m)} c^2$, again by simply integrating by parts:
\begin{equation}\label{349}
E_R^{(m)} \Rightarrow \ -\frac{1}{2} \; \sum_{a=1}^2 \hat{z}_a \cdot M_a^{(m)} c^2 \ ,
\end{equation}
with
\begin{subequations}\label{350}
\begin{align}
 \hat{z}_1 \cdot M_1^{(m)} c^2 &\doteqdot \hbar c \int d^3\vec{r} \ \ \vec{k}_1(\vec{r}) \cdot \vec{A}_2(\vec{r}) \label{350a} \\
 \hat{z}_2 \cdot M_2^{(m)} c^2 &\doteqdot \hbar c \int d^3\vec{r} \ \ \vec{k}_2(\vec{r}) \cdot \vec{A}_1(\vec{r}) \ . \label{350b}
\end{align}
\end{subequations}
Consequently, the associated magnetostatic energy correction $\Delta E_T^{(m)}$ (\ref{345}) becomes
\begin{equation}\label{351}
\Delta E_T^{(m)} \Rightarrow \ \frac{1}{2} \; \sum_{a=1}^2 \hat{z}_a \cdot
M_a^{(m)} c^2 \equiv -E_R^{(m)} \ .
\end{equation}
Evidently the difference in sign between the electrostatic case (\ref{348}) and the present magnetostatic case (\ref{351}) is just
necessary in order that the total electromagnetic correction $\Delta E_T^{(em)}$ (\ref{337a}) is built up by the Lorentz
invariant product of the four-potentials $\Aamu$ and currents $\kamu$
\begin{align}\label{352}
\Delta E_T^{(em)} &\Rightarrow \ -\frac{1}{2} \hbar c \int d^3\vec{r} \ \{ \eko \cdot \zAo - \vec{k}_1 \cdot \vec{A}_2 + \zko \cdot \eAo - \vec{k}_2 \cdot \vec{A}_1 \} \\
&\equiv -\frac{1}{2} \hbar c \int d^3\vec{r} \ \{ {}^{(1)}\!k_\mu(\vec{r}) \cdot {}^{(2)}\!A^\mu(\vec{r}) + {}^{(2)}\!k_\mu(\vec{r}) \cdot {}^{(1)}\!A^\mu(\vec{r}) \} \ . \nonumber
\end{align}
\indent A similar argument does apply also to the exchange corrections of the electric type ($\Delta E_T^{(h)}$) and of the magnetic type ($\Delta E_T^{(g)}$). Namely, one finds again that the gauge field energy $E_C^{(h)}$ (\ref{344}) equals the electric exchange mass in the linear approximation 
\begin{equation}\label{353}
E_C^{(h)} \Rightarrow \ -\frac{1}{2} \; \sum_{a=1}^2 \hat{z}_a \cdot
M_a^{(h)} c^2 \ ,
\end{equation}
so that the corresponding exchange corrections $\Delta E_T^{(h)}$
(\ref{342}) becomes half the value hereof
\begin{equation}\label{354}
\Delta E_T^{(h)} \Rightarrow \ \frac{1}{2} \; \sum_{a=1}^2 \hat{z}_a
\cdot M_a^{(h)} c^2 \equiv -E_C^{(h)} \ .
\end{equation}
For the magnetic analogue $E_C^{(g)}$ (\ref{346}) one finds in the linear approximation
\begin{equation}\label{355}
E_C^{(g)} \Rightarrow \ \frac{1}{2} \; \sum_{a=1}^2 \hat{z}_a \cdot
M_a^{(g)} c^2 \equiv \Delta E_T^{(g)} \ ,
\end{equation}
with the magnetic exchange masses $M_a^{(g)}$ being defined through
\begin{subequations}\label{356}
\begin{align}
 \hat{z}_1 \cdot M_1^{(g)} c^2 &\doteqdot \hbar c \int d^3\vec{r}
 \ \ \vec{B}(\vec{r}) \cdot \vec{h}(\vec{r}) \label{356a} \\
 \hat{z}_2 \cdot M_2^{(g)} c^2 &\doteqdot \hbar c \int d^3\vec{r}
 \ \ \vec{B}^*(\vec{r}) \cdot \vec{h}^*(\vec{r}) \ . \label{356b}
\end{align}
\end{subequations}
With this result, the exchange correction $\Delta E_T^{(hg)}$ (\ref{337b}) becomes
\begin{align} \label{357}
\Delta E_T^{(hg)} &\Rightarrow \ \frac{1}{2} \; \sum_{a=1}^2 \hat{z}_a \cdot \left( M_a^{(h)} + M_a^{(g)} \right) c^2 \nonumber \\
&= -\frac{1}{2} \hbar c \int d^3\vec{r} \left\{B_0(\vec{r}) \cdot
h_0(\vec{r}) - \vec{B}(\vec{r}) \cdot \vec{h}(\vec{r}) + B^*_0(\vec{r}) \cdot h^*_0(\vec{r}) - \vec{B}^*(\vec{r}) \cdot
\vec{h}^*(\vec{r}) \right\} \nonumber \\
&\equiv -\frac{1}{2} \hbar c \int d^3\vec{r} \ \left\{\Bmo \hmu + \Bsmo \hsmu \right\} \ , 
\end{align}
which displays an analogous Lorentz invariant structure as its electromagnetic counterpart $\Delta E_T^{(em)}$ (\ref{352}).\\
\indent However as pleasant as the emergence of such a Lorentz invariant structure may appear, one should recall that this result is
only an approximation. However for a comparison  of the RST predictions with the experimental data one is interested in the
highest numerical precision available, and therefore we will rely for our calculations below on the {\it original} definitions of
the energy corrections $\Delta E_T^{(e)}$ (\ref{339}), $\Delta E_T^{(h)}$ (\ref{342}), $\Delta E_T^{(m)}$ (\ref{345}), and
$\Delta E_T^{(g)}$ (\ref{346}). In this way the non-linear and non-abelian effects will be fully included in our numerical
results (table I and fig.4).

\subsection{Mass Functional}

Further insight into the nature of the energy functional $E_T$ (\ref{335}) is gained by a closer inspection of the mass eigenvalues $M_a$ which form the main contribution to the energy $E_T$. Especially, one expects that they should contain also the rest mass energy ($2 M c^2$) of both particles.\\
\indent The desired mass functionals are easily obtained by multiplying both sides of the eigenvalue equations for both particles
(i.e. equations (\ref{323})-(\ref{325b})) by the complex conjugate wave amplitudes ${}^{(a)}\!R^*_{\pm}, {}^{(a)}\!S^*_{\pm}$ (resp.) and integrating over. Thereby the desired mass functionals appear in the following form
\begin{equation}\label{358}
\hat{z}_a \cdot M_a c^2 = Z_{(\mathbf{a})}^2 \cdot M c^2 +
T_{(\mathbf{a})} + \hat{z}_a \cdot \left( M_{a,e}^{(es)} + M_a^{(e)} +
M_a^{(m)} + M_a^{(h)} + M_a^{(g)} \right) c^2 \ .
\end{equation}
Here, the first three terms refer to the one-particle contributions and the remaining mass terms contain the two-particle interaction energies. Observe that the latter kind of contributions has been defined already in equations (\ref{340a})-(\ref{340b}), (\ref{343a})-(\ref{343b}), (\ref{350a})-(\ref{350b}) and (\ref{356a})-(\ref{356b}). However, the one-particle contributions require now some additional explanation. But recall that each mass eigenvalue $M_a c^2$ ($a=1,2$) (\ref{358}) contains the full interaction energy with the other particle so that this would be counted twice if one omitted the correction terms $\Delta E_T^{(...)}$ for the sum of mass eigenvalues (\ref{335})! \\
\indent After the two-particle contributions are clarified now in detail, it is instructive to inspect also the one-particle terms. First observe here that the rest-mass energy ($Mc^2$) appears here in combination with the renormalization constants $Z_{({\bf a})}^2$ being defined by
\begin{equation}\label{359}
 Z_{({\bf a})}^2 = \int d^3\vec{r} \ \, \bar{\psi}_a(\vec{r}) \, \psi_a(\vec{r}) \ .
\end{equation}
These constants cannot be put to unity (i.e. $Z_{({\bf a})}^2 \nRightarrow 1$), otherwise the non-relativistic limit would become incorrect (see the discussion of this point in ref.\cite{a16}). Actually, the renormalization constants $Z({\bf a})^2$ are found to depend upon the {\it kinetic
energy} $T({\bf a})$ of the particles via 
\begin{equation}\label{360}
 Z_{({\bf a})}^2 = 1 - \frac{T({\bf a})}{M c^2} \ .
\end{equation}
Next, turn to the single-particle interaction energy $E_{es}^{(e)}$ which specifies the electrostatic interaction energy of the two-particle system as a whole with the external source (e.g. nucleus)
\begin{equation}\label{361}
 E_{es}^{(e)} = -\hbar c \int d^3\vec{r} \ \, \exAo(\vec{r}) \cdot j_0(\vec{r}) \doteqdot \sum_{a=1}^2 \hat{z}_a M_{a,e}^{(es)} \ .
\end{equation}
Here the (electrostatic) external source is characterized by the time-component ($\exAo$) of its four-potential $\exAmu$, cf. (\ref{21}), and the total charge density $j_0(\vec{r})$ is the sum of both single-particle contributions $\ako(\vec{r})$, cf. (\ref{217}),
\begin{equation}\label{362}
 j_0(\vec{r}) = \eko(\vec{r}) + \zko(\vec{r}) \ ,
\end{equation}
with the single-particle currents ${}^{(a)}\!k_0(\vec{r})$ being defined in terms of the wave function $\psi_a(\vec{r})$ through
equations (\ref{318a})-(\ref{318b}).\\
\indent Surely, the most  intricate one-particle contribution to the mass functional $M_a c^2$ (\ref{358}) is the kinetic energy
T({\bf a}). With respect to the spherical polar coordinates ($r, \vartheta, \varphi$), this contribution is the sum of three terms
\begin{equation}\label{363}
T({\bf a}) = {}^{(a)}\!T_r + {}^{(a)}\!T_\vartheta + {}^{(a)}\!T_\varphi
\end{equation}
which refer to the motion in the radial, longitudinal, and azimuthal directions, resp. In terms of the ansatz functions $\aRpm, \aSpm$ (\ref{39})-(\ref{310}), the radial part reads
\begin{align}\label{364}
{}^{(a)}\!T_r = \frac{\hbar c}{2\pi} \int d^3\vec{r} \ \{ \aRsp \cdot
\frac{\partial \, \aRm}{\partial r} + \aRp \cdot
\frac{\partial \, \aRsm}{\partial r} + \frac{2}{r} (\aRsp \cdot \aRm + \aRp \cdot \aRsm) \nonumber \\
+ \aSsp \cdot \frac{\partial \aSm}{\partial r} + \aSp \cdot
\frac{\partial \aSsm}{\partial r} + \frac{2}{r} (\aSsp \cdot \aSm
+ \aSp \cdot \aSsm) \} \ .
\end{align}
The longitudinal part ${}^{(a)}\!T_\vartheta$ turns out as the most complicated one and looks as follows:
\begin{align}\label{365}
{}^{(a)}\!T_\vartheta &= \frac{i \hbar c}{2\pi} \int \frac{d^3\vec{r}}{r} \{ \frac{\aRsp}{e^{i\vartheta / 2}\sqrt{\sin{\vartheta}}} \cdot \frac{\partial}{\partial \vartheta} [ e^{i\vartheta / 2}\sqrt{\sin{\vartheta}}\cdot \aRm] \nonumber\\
&-\frac{\aRp}{e^{-i\vartheta / 2} \sqrt{\sin{\vartheta}}} \cdot \frac{\partial}{\partial \vartheta}[e^{-i\vartheta / 2} \sqrt{\sin{\vartheta}} \cdot \aRsm] - \frac{\aSsp}{e^{-i\vartheta / 2}\sqrt{\sin{\vartheta}}} \cdot \frac{\partial}{\partial \vartheta} [ e^{-i\vartheta / 2}\sqrt{\sin{\vartheta}}\cdot \aSm] \nonumber\\
&+ \frac{\aSp}{e^{i\vartheta / 2} \sqrt{\sin{\vartheta}}} \cdot \frac{\partial}{\partial \vartheta}[e^{i\vartheta / 2} \sqrt{\sin{\vartheta}} \cdot \aSsm] \ .
\end{align}
And finally, the azimuthal part ${}^{(a)}\!T_\varphi$ of the kinetic energy reads in terms of the azimuthal current ${}^{(a)}\!k_\varphi$ (\ref{320c}):
\begin{align}\label{366}
{}^{(a)}\!T_\varphi = -\frac{\hbar c}{2} \int \frac{d^3\vec{r}}{r} \ \frac{{}^{(a)}\!k_\varphi}{\sin{\vartheta}} + \frac{\hbar c}{2\pi} \int \frac{d^3\vec{r}}{r\sin{\vartheta}} \ \Big\{ e^{i\vartheta} [ \aRsp \cdot \frac{\partial \, \aSm}{\partial \varphi} + \aRsm \cdot \frac{\partial \, \aSp}{\partial \varphi}] \nonumber \\ 
+ e^{-i\vartheta} [ \aRp \cdot \frac{\partial \, \aSsm}{\partial \varphi}+ \aRm \cdot \frac{\partial \, \aSsp}{\partial \varphi}]\Big\} \ .
\end{align}
\indent With the eigenvalue equations and the explicit form of the energy functional $E_T$ being at hand now, one becomes able to test the potentiality of RST in the field of the two-electron ions. More concretely, one will first solve the mass-eigenvalue equations (\ref{323})-(\ref{325b}), etc., together with the Poisson equations (\ref{334a})-(\ref{334d}); then one will normalize the solutions according to the prescription (\ref{332}), and finally one will compute the desired atomic energy level $E_T$ by considering the value of the energy functional $E_T$ (\ref{335}) upon the obtained solutions. However, in order to
compare these results rigorously to both the experimental data \cite{m1} and to the results of other theoretical approaches (e.g. 1/Z-expansion method \cite{a6}, all-order technique in many-body perturbation theory \cite{a5}), we will treat the original non-linear and non-abelian eigenvalue problem whose
linearized version (\ref{334a})-(\ref{334d}) we presented here only for the sake of brevity.

\section{Triplet Structure}

If RST is equipped with physical meaning, this theory must be able to demonstrate the emergence of two kinds of helium: namely (i) ortho-helium (parallel spins) with a triplet structure of its energy levels, and (ii) para-helium (anti-parallel spins) with a singlet structure (see also fig.3). This means that the totality of solutions of the RST eigenvalue system (\ref{323})-(\ref{325b}) must be shown to subdivide into two subsets, the triplet solutions ($J=1$) and the singlet solutions ($J=0$), where however the one triplet member $\sf ^3S_1$ splits itself up into the three sublevels with conventional notation $\ket{S=1; S_z=0, \pm1}$ being degenerated when no external magnetic field $\vec{H}_{ex}$ is present. Thus the first task is to identify that subset of solutions to the RST eigenvalue system which corresponds to the conventional triplet states $\ket{S=1;S_z=0,\pm1}$.

\subsection{Triplet fields $\ket{S=1;S_z=\pm1}$}

It should be evident that, when the total magnetic quantum number $S_z$ ($=s_{z(1)}+s_{z(2)}$) is extremal (i.e. $S_z=\pm1$), then both single-particle spins $s_{z(a)}$ should point either to the positive or negative z-direction ($s_{z(1)}=s_{z(2)}= \pm\frac{1}{2}$). Accordingly, restricting ourselves to the positive z-direction, we reparametrize for this case the eight ansatz functions $\aRpm, \aSpm$ in our general ansatz (\ref{39})-(\ref{310}) in terms of only four {\it real-valued} wave amplitudes $R_\pm(r, \vartheta), S_\pm(r,\vartheta)$ in the following way:
\begin{subequations}\label{41}
\begin{align}
\eRp = - \eSp &= -\frac{i}{2} e^{i\varphi} \cdot R_+ \label{41a} \\
\eRm = - \eSm &= -\frac{i}{2} e^{i\varphi} \cdot R_- \label{41b} \\
\zRp = - \zSp &= -\frac{i}{2} e^{i\varphi} \cdot S_+ \label{41c} \\
\zRm = - \zSm &= -\frac{i}{2} e^{i\varphi} \cdot S_- \ . \label{41d}
\end{align}
\end{subequations}
Thus the Pauli-spinors ${}^{(a)}\!\phi_\pm(\vec{r})$ (\ref{38})-(\ref{310}) become
\begin{subequations}\label{42}
\begin{align}
{}^{(1)}\!\phi_+(\vec{r}) \ &\Rightarrow \ R_+ \cdot \zetappn \label{42a}\\
{}^{(1)}\!\phi_-(\vec{r}) \ &\Rightarrow \ - i \, R_- \cdot \zetappe \label{42b}\\
{}^{(2)}\!\phi_+(\vec{r}) \ &\Rightarrow \ S_+ \cdot \zetappn \label{42c}\\
{}^{(2)}\!\phi_-(\vec{r}) \ &\Rightarrow \ - i \, S_+ \cdot \zetappe \ . \label{42d}
\end{align}
\end{subequations}
\indent Once the wave functions $\psi_a(\vec{r})$ have thus been specified, one can turn to the associated ortho-currents. From our equation (\ref{319}) one easily deduces the charge densities ${}^{(a)}\!k_0(\vec{r})$ as
\begin{subequations}\label{43}
\begin{align}
{}^{(1)}\!k_0 &= \frac{1}{4 \pi} \{ R_+^{\;2} + R_-^{\;2} \} \label{43a}\\
{}^{(2)}\!k_0 &= \frac{1}{4 \pi} \{ S_+^{\;2} + S_-^{\;2} \} \ . \label{43b}
\end{align}
\end{subequations}
Similarly, one finds from the component equations (\ref{320a})-(\ref{320c}) that both currents $\vec{k}_a(\vec{r})$ (\ref{317a}) encircle the z-axis in the same direction, i.e.
\begin{subequations}\label{44}
\begin{align}
\vec{k}_1 &= \frac{1}{2 \pi} \, R_+ R_- \, \sin\vartheta \cdot \vec{e}_\varphi \doteqdot k_1(r, \vartheta) \, \sin\vartheta \cdot \vec{e}_\varphi \label{44a}\\
\vec{k}_2 &= \frac{1}{2 \pi} \, S_+ S_- \, \sin\vartheta \cdot \vec{e}_\varphi \doteqdot k_2(r, \vartheta) \, \sin\vartheta \cdot \vec{e}_\varphi \ , \label{44b}
\end{align}
\end{subequations}
see the introductory figure. And finally the exchange density $h_0(\vec{r})$ (\ref{321}) becomes
\begin{equation}\label{45}
 h_0 = \frac{1}{4\pi} \{ R_+ S_+ + R_- S_- \} \ ,
\end{equation}
with the corresponding exchange current $\vec{h}(\vec{r})$ (\ref{322a})-(\ref{322c}) being found as
\begin{align}\label{46}
\vec{h} &= \frac{i}{4\pi} \{ R_+ S_- - R_- S_+ \} \, \vec{e}_r + \frac{1}{4\pi} \{ R_+ S_- + R_- S_+ \} \, \vec{e}_\varphi \\ &\doteqdot i \, \eta(r,\vartheta) \, \vec{e}_r + h(r,\vartheta) \, \sin\vartheta \, \vec{e}_\varphi \ . \nonumber
\end{align}
Moreover, since the gauge potentials are coupled to the currents, see the linear approximation (\ref{334a})-(\ref{334d}) hereof, the functional form of the potentials will be qualitatively the same as for the currents, i.e. we assume that the radial and longitudinal components of the gauge fields vanish (${}^{(a)}A_r = {}^{(a)}A_\vartheta = B_\vartheta \equiv 0$) and are then left with
\begin{subequations}\label{47}
\begin{align}
\vec{A}_a(\vec{r}) &= r \sin \vartheta \, A_a(r,\vartheta) \cdot \vec{e}_\varphi \label{47a}\\
\vec{B}(\vec{r}) &= i \, \beta(r,\vartheta) \cdot \vec{e}_r + r \sin\vartheta \, B(r,\vartheta) \cdot \vec{e}_\varphi \label{47b} \ .
\end{align}
\end{subequations}
Here we have assumed that all triplet objects depend exclusively upon the radial variable $r$ and spherical polar angle $\vartheta$ so that the triplet configuration is SO(2) symmetric around the z-axis.

\subsection{Mass Eigenvalue Equations ($S=1, S_z=+1$)}

With the general functional form of all static triplet fields being specified, one can introduce these now into the general mass-eigenvalue system (\ref{323})-(\ref{326d}) in order to obtain the corresponding eigenvalue system for the triplet states. The point here is that, through the triplet ansatz (\ref{41a})-(\ref{41d}), the set of eight complex-valued wave amplitudes ${}^{(a)}\!R_\pm(r,\vartheta, \varphi), {}^{(a)}\!S_\pm(r,\vartheta, \varphi)$ becomes reduced to only four real-valued wave amplitudes $R_\pm(r,\vartheta), S_\pm(r,\vartheta)$. Consequently, there arises for each particle an ordinary Dirac equation for the radial motion, which is however complemented by a separate equation for the longitudinal degree of freedom (described by the polar angle $\vartheta$). Thus the radial equation for the positive-energy amplitude $R_+$ of the first particle ($a=1$) is found as
\begin{align}\label{48}
\frac{\partial R_+}{\partial r} + \left[\exAo + \zAo \right]\cdot R_- + B_0 \cdot S_- - \beta \cdot S_+& \nonumber \\
- r \sin^2\vartheta \left[A_2\cdot R_+ + B \cdot S_+ \right] &= - \frac{M_1+M}{\hbar} c \cdot R_- \ ,
\end{align}
which is to be complemented by its longitudinal counterpart
\begin{equation}\label{49}
\frac{\partial R_+}{\partial \vartheta} = r^2 \sin\vartheta \cos\vartheta \left[A_2\cdot R_+ + B \cdot S_+ \right] \ .
\end{equation}
Of course the same procedure does apply also to the negative-energy amplitude $R_-$, which yields the radial equation as
\begin{align}\label{410}
\frac{\partial R_-}{\partial r} + \frac{2}{r} R_- - \left[\exAo + \zAo \right]\cdot R_+ - B_0 \cdot S_+ - \beta \cdot S_-& \nonumber \\
+ r \sin^2\vartheta \left[A_2\cdot R_- + B \cdot S_- \right] &= \frac{M_1-M}{\hbar} c \cdot R_+ \ ,
\end{align}
and its longitudinal counterpart as
\begin{equation}\label{411}
\frac{\partial R_-}{\partial \vartheta} = r^2 \sin\vartheta \cos\vartheta \left[A_2\cdot R_- + B \cdot S_- \right] \ .
\end{equation}
The analogous equations for the second particle ($a=2$) can easily be found from here by simply applying the particle permutation operation (\ref{326a})-(\ref{326d}) and are then found to look as follows for the positive-energy amplitude $S_+$:
\begin{subequations}\label{412}
\begin{align}
\frac{\partial S_+}{\partial r} + \left[\exAo + \eAo \right]\cdot S_- + B_0 \cdot R_- + \beta \cdot R_+& \nonumber \\
- r \sin^2\vartheta \left[A_1\cdot S_+ + B \cdot R_+ \right] &= - \frac{M_2+M}{\hbar} c \cdot S_- \label{412a} \\[2ex]
\frac{\partial S_+}{\partial \vartheta} = r^2 \sin\vartheta \cos\vartheta \left[A_1\cdot S_+ + B \cdot R_+ \right] \ , \label{412b}
\end{align}
\end{subequations}
and similarly for the negative-energy amplitude $S_-$
\begin{subequations}\label{413}
\begin{align}
\frac{\partial S_-}{\partial r} + \frac{2}{r} S_- - \left[\exAo + \eAo \right]\cdot S_+ - B_0 \cdot R_+ - \beta \cdot R_-& \nonumber \\
+ r \sin^2\vartheta \left[A_1\cdot S_- + B \cdot R_- \right] &= \frac{M_2-M}{\hbar} c \cdot S_+ \label{413a} \\[2ex]
\frac{\partial S_-}{\partial \vartheta} = r^2 \sin\vartheta \cos\vartheta \left[A_1\cdot S_- + B \cdot R_- \right] \ . \label{413b}
\end{align}
\end{subequations}
\indent As was mentioned already in connection with the general form of the mass eigenvalue system (see end of Sect. 3c), the preceding eigenvalue equations (\ref{48})-(\ref{413b}) must be complemented by the Poisson equations for the gauge potentials ${}^{(a)}\!A_0(r,\vartheta)$, $A_a(r,\vartheta)$, $B(r,\vartheta)$ and $\beta(r,\vartheta)$. For the high-precision calculations in atomic physics, the linear approximations (\ref{334a})-(\ref{334d}) are not sufficient and one has to evoke the exact non-linear form of the Poisson equations, see ref.\cite{a41}.

\subsection{Lowest-Order Approximation}

It should not come as a surprise that it is very hard (if not impossible) to find \textit{analytic} solutions of the preceding mass-eigenvalue problem for the ortho-system. Therefore one must be satisfied with discussing the problem of degeneration of the triplet states $\ket{S=1; S_z=0,\pm1}$ in an approximative manner. This may be done by considering the value of the energy functional $E_T$ (\ref{335}) upon certain {\it approximate} solutions of the present eigenvalue problem, both for the present states $\ket{S=1;S_z=\pm1}$ and for the state $\ket{S=1; S_z=0}$ which is to be specified hereafter. In this way one can then decide, at least in such a lowest-order approximation, whether or not all three states $\ket{S=1; S_z=0,\pm1}$ are actually degenerated in RST.\\
\indent Our choice of lowest-order approximation simply consists in neglecting all the interelectronic interactions, i.e. we put to zero the gauge potentials ${}^{(a)}\!A_0(r,\vartheta)$, $A_a(r,\vartheta)$, $B_0(r,\vartheta)$, $\beta(r,\vartheta)$ and $\vec{B}(\vec{r})$. This results in a great simplification of the preceding eigenvalue system (\ref{48})-(\ref{413b}); namely the angular equations (\ref{49}), (\ref{411}), (\ref{412b}) and (\ref{413b}) merely imply that the two-particle wave amplitudes $R_\pm$, $S_\pm$ become independent of the polar angle $\vartheta$ (i.e. $R_\pm(r,\vartheta), S_\pm(r,\vartheta) \ \rightarrow \  \tilde{R}_\pm(r), \tilde{S}_\pm(r)$, and furthermore the radial equations (\ref{48}), (\ref{410}), (\ref{412a}) and (\ref{413a}) are reduced to the following single-particle form
\begin{subequations}\label{414}
\begin{align}
& \frac{d\tilde{R}_+}{dr} + \exAo \cdot \tilde{R}_- = - \frac{\tilde{M}_1+M}{\hbar}c\cdot \tilde{R}_- \label{414a}\\
& \frac{d\tilde{R}_-}{dr} + \frac{2}{r}\tilde{R}_- - \exAo \cdot \tilde{R}_+ = \frac{\tilde{M}_1-M}{\hbar}c\cdot \tilde{R}_+ \label{414b}\\
& \frac{d\tilde{S}_+}{dr} + \exAo \cdot \tilde{S}_- = - \frac{\tilde{M}_2+M}{\hbar}c\cdot \tilde{S}_- \label{414c}\\
& \frac{d\tilde{S}_-}{dr} + \frac{2}{r}\tilde{S}_- - \exAo \cdot \tilde{S}_+ = \frac{\tilde{M}_2-M}{\hbar}c\cdot \tilde{S}_+ \label{414d} \ .
\end{align}
\end{subequations}
\noindent The solutions for these two decoupled one-particle configurations are well-known in the literature (see, e.g. ref. \cite{b27}) and may be parametrized by the principal quantum numbers $n_a$ ($a=1,2$) of both particles. For instance, the one-particle mass eigenvalues $\tilde{M}_a$ are given by 
\begin{equation}
\tilde{M}_a = \frac{M}{\sqrt{1+(\frac{z_{ex}\alpha_s}{n_a-1+\sqrt{1-(z_{ex}\alpha_s)^2}})^2}} \ \ . \label{415}
\end{equation}
\begin{equation*}
(n_a = 1,2,3,\dots)
\end{equation*}
\indent Thus one can determine the approximate two-particle energy spectrum ($\tilde{E}_T(n_1,n_2)$, say) by taking the value of the \textit{exact} energy functional $E_T$ (\ref{335}) upon the \textit{approximate} solutions $\lbrace \tilde{R}^{n_1}_{\pm},\tilde{S}^{n_2}_{\pm} \rbrace $ of the reduced eigenvalue system (\ref{414a})-(\ref{414d}), yielding the desired energy spectrum in a lowest-order approximation.
\indent Naturally one expects that, for such a rough estimate, it will also be sufficient to resort to the linear approximation of the Poisson equations for the gauge potentials, as specified by equations (\ref{334a})-{\ref{334d}). Using the standard boundary conditions, the formal solutions of these linear Poisson equations are given by
\begin{subequations}\label{416}
\begin{align}
     \aAo(\vec{r}) & = -\alpha_s \cdot \int d^3\vec{r}\;' \ \frac{\ako(\vec{r}\;')}{|\vec{r}-\vec{r}\;'|} \label{416a}\\
\vec{A}_a(\vec{r}) & = -\alpha_s \cdot \int d^3\vec{r}\;' \ \frac{\vec{k}_a(\vec{r}\;')}{|\vec{r}-\vec{r}\;'|} \label{416b}\\
      B_0(\vec{r}) & = \alpha_s \cdot \int d^3\vec{r}\;' \ \frac{h^{\ast}_{\;0}(\vec{r}\;')}{|\vec{r}-\vec{r}\;'|} \label{416c}\\
  \vec{B}(\vec{r}) & = \alpha_s \cdot \int d^3\vec{r}\;' \ \frac{\vec{h}^{\ast}(\vec{r}\;')}{|\vec{r}-\vec{r}\;'|} \label{416d} \ .
\end{align}
\end{subequations}
\noindent Here one concludes from equation (\ref{45}) that the exchange density $h_0$ is real-valued ($h_0=h^{\ast}_0$) which then also holds for the "electric" exchange potential $B_0(\vec{r})$ (\ref{416c}). With this lowest-order arrangement, the electrostatic energy correction $\Delta E^{(e)}_T$ (\ref{348}) becomes
\begin{align}
\Delta E^{(e)}_T & \Rightarrow \Delta \tilde{E}^{(e)}_T = e^2 \cdot \int \int d^3\vec{r} d^3\vec{r}\;' \ \frac{\eko (\vec{r}) \zko (\vec{r}\;')}{|\vec{r}-\vec{r}\;'|} \nonumber\\
& = \frac{e^2}{16\pi^2} \cdot \int \int d^3\vec{r} d^3\vec{r}^{'} \ \frac{(\tilde{R}^2_+(r)+\tilde{R}^2_-(r))\cdot (\tilde{S}^2_+(r')+\tilde{S}^2_-(r'))}{|\vec{r}-\vec{r}\;'|} \ , \label{417}
\end{align}
\noindent and similarly for the electric exchange correction $\Delta E^{(h)}_T$ (\ref{354})
\begin{align}
&\Delta E^{(h)}_T \Rightarrow \Delta \tilde{E}^{(h)}_T = -e^2 \cdot \int \int d^3\vec{r} d^3\vec{r}\;' \ \frac{h^{\ast}_0(\vec{r})h_0(\vec{r}\;')}{|\vec{r}-\vec{r}\;'|} \nonumber\\
& = -\frac{e^2}{16\pi^2} \cdot \int \int d^3\vec{r} d^3\vec{r}\;' \ \frac{(\tilde{R}_+(r)\tilde{S}_+(r)+\tilde{R}_-(r)\tilde{S}_-(r))(\tilde{R}_+(r')\tilde{S}_+(r')+\tilde{R}_-(r')\tilde{S}_-(r'))}{|\vec{r}-\vec{r}\;'|} \ . \label{418}
\end{align}
\noindent Observe here that the non-relativistic limit (where the "negative-energy" components $\tilde{R}_-,\tilde{S}_-$ are neglected against their positive-energy counterparts $\tilde{R}_+,\tilde{S}_+$) of this exchange correction $\Delta E^{(h)}_T$ just coincides with the conventional exchange integral $\Delta E^{(n,n^{'})}_C$ (\ref{34}). The wave amplitude $\tilde{R}_+(r)$ ($\tilde{S}_+(r)$) corresponds to the first (second) particle's wave function $\psi_1(\vec{r})$ ($\psi_2(\vec{r})$), up to the angular normalization factor of $(4\pi)^{-1/2}$ (\ref{43a})-(\ref{43b}). Thus, as far as this lowest-order approximation is concerned, RST is found to be in agreement with the Hartree-Fock approach, see the discussion of the level shift of ortho-helium caused by the exchange interactions in fig.3.\\
\indent Both energy corrections $\Delta E^{(e)}_T$ (\ref{417}) and $\Delta E^{(h)}_T$ (\ref{418}) of the "electric" type are much greater than their "magnetic" counterparts $\Delta E^{(m)}_T$ (\ref{351}) and $\Delta E^{(g)}_T$ (\ref{355}). Therefore it makes sense to define an \textit{"electric" degeneracy} for the triplet states $|S=1;S_z=0,\pm 1>$, namely through the requirement that their energy be identical only upon the kinetic and electric parts ($\tilde{E}_T$, say) of the exact functional $E_T$:
\begin{align}
\tilde{E}_T & = \sum_{a=1}^2\lbrace \Za^2 \cdot Mc^2 + \Ta + \hat{z}_a\cdot M_{a,e}^{(es)}c^2 \rbrace + \frac{1}{2} \sum_{a=1}^2 \hat{z}_a (M_a^{(e)}+M_a^{(h)})c^2 \ . \label{419}
\end{align}
\noindent In contrast to this, the complete (but linearized) energy functional would contain also the "magnetic" terms, i.e.
\begin{align}
E_T & \Rightarrow \sum_{a=1}^2 \left( \Za^2 \cdot Mc^2 + \Ta + \hat{z}_a\cdot M_{a,e}^{(es)}c^2 \right)\nonumber \\
    & + \frac{1}{2} \sum_{a=1}^2 \hat{z}_a \lbrace M_a^{(e)}+M_a^{(m)}+M_a^{(h)}+M_a^{(g)} \rbrace c^2 \ , \label{420}
\end{align}
\noindent where the renormalization constants $\Za$ (\ref{359}) read in terms of the wave amplitudes $R_{\pm},S_{\pm}$ of the ortho-system (\ref{42a})-(\ref{42b})
\begin{subequations}\label{421}
\begin{align}
\Ze ^2 & = \int d^3 \vec{r} \ \frac{R_+^2-R_-^2}{4\pi} \label{421a}\\
\Zz ^2 & = \int d^3 \vec{r} \ \frac{S_+^2-S_-^2}{4\pi} \label{421b} \ .
\end{align}
\end{subequations}
\indent Finally, the kinetic energies $\Ta$ (\ref{363}) of both particles have to be specified in terms of the wave amplitudes $R_{\pm},S_{\pm}$ in order that the total energy $\tilde{E}_T$ (\ref{419}) is completely expressed in terms of wave amplitudes and gauge potentials. Thus introducing the triplet ansatz (\ref{41a})-(\ref{41d}) into the kinetic energies (\ref{364})-(\ref{366}) immediately yields for the first particle ($a=1$)
\begin{subequations}\label{422}
\begin{align}
          {}^{(1)}T_r & = \frac{\hbar c}{4\pi} \int d^3 \vec{r} \ \lbrace R_+ \cdot \frac{\partial R_-}{\partial r} - R_- \cdot \frac{\partial R_+}{\partial r} + \frac{2}{r}R_+R_- \rbrace \label{422a}\\
{}^{(1)}T_{\vartheta} & = -\frac{\hbar c}{4\pi} \int \frac{d^3 \vec{r}}{r} \ R_+R_- \label{422b}\\
  {}^{(1)}T_{\varphi} & = \frac{\hbar c}{4\pi} \int \frac{d^3 \vec{r}}{r} \ R_+R_- \label{422c} \ ,
\end{align}
\end{subequations}  
\noindent and similarly for the second particle ($a=2$)
\begin{subequations}\label{423}
\begin{align}
          {}^{(2)}T_r & = \frac{\hbar c}{4\pi} \int d^3 \vec{r} \ \lbrace S_+ \cdot \frac{\partial S_-}{\partial r} - S_- \cdot \frac{\partial S_+}{\partial r} + \frac{2}{r}S_+S_- \rbrace \label{423a}\\
{}^{(2)}T_{\vartheta} & = -\frac{\hbar c}{4\pi} \int \frac{d^3 \vec{r}}{r} \ S_+S_- \label{423b}\\
  {}^{(2)}T_{\varphi} & = \frac{\hbar c}{4\pi} \int \frac{d^3 \vec{r}}{r} \ S_+S_- \label{423c} \ .
\end{align}
\end{subequations}  
\noindent As expected, the "orbital energy" $T_O$ vanishes for both particles
\begin{equation}
{}^{(a)}T_O \doteqdot {}^{(a)}T_{\vartheta} + {}^{(a)}T_{\varphi} = 0 \label{424}
\end{equation}
\noindent which is a consequence of the vanishing {\it orbital} angular momentum of the presently considered triplet states ${}^3S_1$.\\
\indent Summarizing, the RST configurations of definite spin direction $|S=1,S_z=\pm 1>$ are well-defined now in their lowest-order approximation, namely through their eigenvalue equations (\ref{414a})-(\ref{415}), potentials (\ref{416a})-(\ref{416d}) and energy functional $\tilde{E}_T$ (\ref{419}). Therefore the question of their degeneracy with the states $|S=1,S_z=0>$ of indefinite spin direction can be studied by explicitely working out, as the next task, all the RST fields (especially the corresponding energy functional) due to those curious states which have spin quantum number $S=1$ but vanishing $z$-component $S_z=0$. Since we are satisfied here with a demonstration of the degeneracy only in the lowest-order approximation, we can restrict ourselves to showing that the approximate mass eigenvalue equations for the exotic states $S_z=0$ do agree with the present ones (\ref{414a})-(\ref{415}), because in this case the kinetic ($\Ta$) and interaction energies ($M_a^{(\dots)}c^2$) entering the energy functional $\tilde{E}_T$ (\ref{419}) are identical.

\subsection{Indefinite Spin Direction $\ket{S=1;S_z=0}$}

The proposal for the exotic configurations with indefinite spin direction consists in disposing of the wave amplitudes $\aRpm , \aSpm$, due to the general ansatz (\ref{39})-(\ref{310}), in the following way:
\begin{subequations}\label{425}
\begin{align}
&\eSp = \eSm = \zSp = \zSm = 0 \label{425a}\\
&\eRp = R_+(r,\vartheta) \ e^{i \, \varphi/2} \label{425b}\\
&\eRm = R_-(r,\vartheta) \ e^{i \, \varphi/2} \label{425c}\\
&\zRp = S_+(r,\vartheta) \ e^{i \, \varphi/2} \label{425d}\\
&\zRm = S_-(r,\vartheta) \ e^{i \, \varphi/2} \ . \label{425e}
\end{align}
\end{subequations}
Thus the Pauli spinors $^{(a)}\!\phi_\pm(\vec{r})$ (\ref{39}) become for the exotic states
\begin{subequations}\label{426}
\begin{align}
&^{(1)}\!\phi_+(\vec{r}) = e^{i \, \varphi/2} \, R_+(r,\vartheta) \cdot \zeta_0^{(+)} \label{426a} \\
&^{(1)}\!\phi_-(\vec{r}) = -i \, e^{i \, \varphi/2} \, R_-(r,\vartheta) \cdot \zeta_1^{(+)} \label{426b} \\
&^{(2)}\!\phi_+(\vec{r}) = e^{i \, \varphi/2} \, S_+(r,\vartheta) \cdot \zeta_0^{(+)} \label{426c} \\
&^{(2)}\!\phi_-(\vec{r}) = -i \, e^{i \, \varphi/2} \, S_-(r,\vartheta) \cdot \zeta_1^{(+)} \ . \label{426d}
\end{align}
\end{subequations}
Here one assumes that the four wave amplitudes $R_\pm(r,\vartheta)$, $S_\pm(r,\vartheta)$ are unique, complex-valued functions over three-space so that the corresponding Pauli-spinors $^{(a)}\!\phi_\pm$ become ambiguous with respect to their sign:
\begin{equation}\label{427}
 ^{(a)}\!\phi_\pm(r,\vartheta,\varphi+2\pi)=-{}^{(a)}\!\phi_\pm(r,\vartheta,\varphi) \ ,
\end{equation}
which is frequently the case for spinors, because they transform under $SU(2)$ in place of $SO(3)$. \\
\indent However the corresponding physical densities (as "observables" of the theory) must be real-valued and unique. Actually, inserting the present triplet ansatz (\ref{425a})-(\ref{425e}) into the electromagnetic currents (\ref{319})-(\ref{320c}) yields first for the charge densities of both particles the unique result
\begin{subequations}\label{428}
\begin{align}
\eko &= \frac{R_+^*R_+ + R_-^*R_-}{2\pi} \label{428a} \\
\zko &= \frac{S_+^*S_+ + S_-^*S_-}{2\pi} \label{428b} \ .
\end{align}
\end{subequations}
Similarly, the electromagnetic current becomes for the first particle ($a=1$)
\begin{subequations}\label{429}
\begin{align}
^{(1)}\!k_r &= i \, \frac{R_+^*R_- - R_+ R_-^*}{2\pi} \label{429a} \\
^{(1)}\!k_\vartheta &= - \frac{R_+^*R_- + R_+R_-^*}{2\pi} \label{429b} \ ,
\end{align}
\end{subequations}
and for the second particle ($a=2$)
\begin{subequations}\label{429n}
\begin{align}
^{(2)}\!k_r &= i \, \frac{S_+^*S_- - S_+ S_-^*}{2\pi} \label{429na} \\
^{(2)}\!k_\vartheta &= - \frac{S_+^*S_- + S_+S_-^*}{2\pi} \label{429nb} \ ,
\end{align}
\end{subequations}
whereas both currents $\vec{k}_a(\vec{r})$ have vanishing azimuthal component ($a=1,2$):
\begin{equation}\label{431}
^{(a)}\!k_\varphi \equiv 0 \ .
\end{equation}
Thus the currents $\vec{k}_a(\vec{r})$ (\ref{317a}) of the exotic states do not encircle the z-axis (as the symmetry axis of the field configuration), but they build up a rotational flow in the two-planes containing the z-axis, see the introductory figure. A similar geometric pattern does emerge also for the exchange current $\vec{h}(\vec{r})$ (\ref{317b}) which is found to have vanishing azimuthal component (\ref{322c})
\begin{equation}\label{432}
h_\varphi \equiv 0 \ ,
\end{equation}
whereas the radial and longitudinal components (\ref{322a})-(\ref{322b}) are obtained as
\begin{subequations}\label{433}
\begin{align}
h_r &= \frac{i}{2\pi} \, \left\{R_+^*S_- - R_-^* S_+ \right\} \label{433a} \\
h_\vartheta &= - \frac{1}{2\pi} \left\{R_+^*S_- + R_-^* S_+\right\} \label{433b} \ .
\end{align}
\end{subequations}
Finally, the exchange density $h_0$ (\ref{321}) is found to be of the following form
\begin{equation}\label{434}
h_0 =\frac{R_+^*S_+ + R_-^*S_-}{2\pi} \ .
\end{equation}
Concerning the functional form of the associated gauge fields, one will adapt this again to the geometric pattern of the currents. Thus one puts to zero the azimuthal components of the magnetic vector potentials (${}^{(a)}\!A_\varphi = B_\varphi \equiv 0$) which reduces their general forms (\ref{317c})-(\ref{317d}) to
\begin{subequations}\label{435}
\begin{align}
\vec{A}_a(\vec{r})&= {}^{(a)}\!A_r \vec{e}_r + {}^{(a)}\!A_\vartheta \vec{e}_\vartheta \label{435a} \\
\vec{B}(\vec{r})&= B_r \vec{e}_r + B_\vartheta \vec{e}_\vartheta \ , \label{435b}
\end{align}
\end{subequations}
with the components ${}^{(a)}\!A_r$, ${}^{(a)}\!A_\vartheta$, $B_r$, $B_\vartheta$ being functions exclusively of $r$ and $\vartheta$. Clearly these components must now enter the mass eigenvalue equations for the exotic spin states ($S_z=0$) which are to be deduced again from the general form (\ref{323})-(\ref{326d}) by inserting there the present special form (\ref{425a})-(\ref{425e}) of the wave amplitudes. In this way, the mass eigenvalue equations for the first particle ($a=1$) appear in the following form:
\begin{subequations}\label{436}
\begin{align}
\frac{\partial R_+}{\partial r} - \frac{i}{r e^{i\, \vartheta/2} \sqrt{\sin{\vartheta}}} \frac{\partial}{\partial \vartheta} \left[e^{i\,\vartheta/2} \sqrt{\sin{\vartheta}} \, R_+\right] + \left[\exAo + \zAo \right]\cdot R_- \nonumber \\
+ B_0S_- + \left[{}^{(2)}\!A_\vartheta + i\, {}^{(2)}\!A_r \right]\cdot R_+ + \left[B_\vartheta + i\, B_r \right]\cdot S_+ = -\frac{M_1 + M}{\hbar} c\cdot R_- \label{436a} \\[2ex] \vspace{0,5cm}
\frac{\partial R_-}{\partial r} +\frac{2}{r}R_- + \frac{i}{r e^{i\, \vartheta/2} \sqrt{\sin{\vartheta}}} \frac{\partial}{\partial \vartheta} \left[e^{i\,\vartheta/2} \sqrt{\sin{\vartheta}} \, R_-\right] - \left[ \exAo + \zAo \right]\cdot R_+ \nonumber \\
- B_0 S_+ - \left[{}^{(2)}\!A_\vartheta - i\, {}^{(2)}\!A_r \right]\cdot R_- - \left[B_\vartheta - i\, B_r \right]\cdot S_- = \frac{M_1 - M}{\hbar} c\cdot R_+ \label{436b} \ .
\end{align}
\end{subequations}
For the sake of completeness, the eigenvalue equations for the second particle may also be written down explicitly:
\begin{subequations}\label{437}
\begin{align}
\frac{\partial S_+}{\partial r} - \frac{i}{r e^{i\, \vartheta/2} \sqrt{\sin{\vartheta}}} \frac{\partial}{\partial \vartheta} \left[e^{i\,\vartheta/2} \sqrt{\sin{\vartheta}} \, S_+\right] + \left[\exAo + \eAo \right]\cdot S_- \nonumber \\
+ B_0^* R_- + \left[{}^{(1)}\!A_\vartheta + i\, {}^{(1)}\!A_r \right]\cdot S_+ + \left[B^*_\vartheta + i\, B^*_r \right]\cdot R_+ = -\frac{M_2 + M}{\hbar} c\cdot S_- \label{437a} \\[2ex]
\frac{\partial S_-}{\partial r} +\frac{2}{r}S_- + \frac{i}{r e^{i\, \vartheta/2} \sqrt{\sin{\vartheta}}} \frac{\partial}{\partial \vartheta} \left[e^{i\,\vartheta/2} \sqrt{\sin{\vartheta}} \, S_-\right] - \left[ \exAo + \eAo \right] \cdot S_+ \nonumber \\
- B^*_0 R_+ - \left[{}^{(1)}\!A_\vartheta - i\, {}^{(1)}\!A_r \right]\cdot S_- - \left[B^*_\vartheta - i\, B^*_r \right]\cdot R_- = \frac{M_2 - M}{\hbar} c\cdot S_+ \label{437b} \ .
\end{align}
\end{subequations}
\indent In order to close this system, the Poisson equations for the gauge potentials have to be supplied; but since we are satisfied for the moment with the lowest-order approximation, one can again resort to the linearized form (\ref{334a})-(\ref{334d}). However, it is important to mention here that with the present shape of the magnetic vector potentials $\vec{A}_a(\vec{r})$ (\ref{435a}) and $\vec{B}(\vec{r})$ (\ref{435b}) the magnetostatic fields $\vec{H}_a(\vec{r})$ 
\begin{subequations}\label{438}
\begin{align}
\vec{H}_1(\vec{r}) & = \vec{\nabla}\times \vec{A}_1(\vec{r}) - i\vec{B}(\vec{r}) \times \vec{B}^{\ast}(\vec{r}) \label{438a} \\
\vec{H}_2(\vec{r}) & = \vec{\nabla}\times \vec{A}_2(\vec{r}) + i\vec{B}(\vec{r}) \times \vec{B}^{\ast}(\vec{r}) \label{438b} 
\end{align}
\end{subequations}
\noindent are of purely azimuthal character, i.e.
\begin{equation}\label{439}
\vec{H}_a(\vec{r}) = {}^{(a)}H_{\varphi(r,\vartheta)} \vec{e}_{\varphi} \ ,
\end{equation}
\noindent so that these states $|S=1; S_z=0>$ of indefinite spin direction cannot carry a magnetic moment, in contrast to the states $|S=1; S_z=\pm 1>$ (see the discussion of this point in ref. \cite{u6}). Therefore only the latter states can provide a handle for an external magnetic field in order to shift the energy levels (in opposite directions) whereas the energy of the exotic spin state $|S=1; S_z=0>$ remains unchanged (Zeeman effect \cite{b7}).\\
\indent The fact that both field configurations $|S=1; S_z=0>$ and $|S=0; S_z=0>$ have vanishing $z$-component of their total spin $\vec{S}$ lets expect that their geometries will share certain common features. In order to see this more clearly, one transforms the triplet wave amplitudes $R_\pm, S_\pm$ in the following way:
\begin{subequations}\label{440}
\begin{align}
{}^{'}\!\!R_{\pm} & \doteqdot e^{i\frac{\vartheta}{2}}\sqrt{\sin \vartheta} \cdot R_{\pm} \label{440a} \\
{}^{'}\!\!S_{\pm} & \doteqdot e^{i\frac{\vartheta}{2}}\sqrt{\sin \vartheta} \cdot S_{\pm} \label{440b} \ ,
\end{align}
\end{subequations}
\noindent and additionally one defines complex "magnetic" potentials $\mathds{A}_a, \mathds{B}_{\pm}$ through
\begin{subequations}\label{441}
\begin{align}
    \mathds{A}_a & \doteqdot {}^{(a)}A_{\vartheta} + i\ {}^{(a)}A_r \label{441a} \\
\mathds{B}_{\pm} & \doteqdot B_{\vartheta} \pm i\ B_r \label{441b} \ .
\end{align}
\end{subequations}
\noindent By this arrangement, the present eigenvalue system (\ref{436a})-(\ref{437b}) is cast into the following form:
\begin{subequations}\label{442}
\begin{align}
\frac{\partial \ {}^{'}\!\!R_+}{\partial r} - \frac{i}{r} \frac{\partial}{\partial \vartheta} \ {}^{'}\!\!R_+ + \left[\exAo + \zAo \right]\cdot {}^{'}\!\!R_- + B_0 &\cdot {}^{'}\!\!S_- + \mathds{A}_2 \cdot {}^{'}\!\!R_+ + \mathds{B}_+ \cdot {}^{'}\!\!S_+ \nonumber \\
&= -\frac{M_1 + M}{\hbar} c\cdot {}^{'}\!\!R_- \label{442a} \\ 
\frac{\partial \ {}^{'}\!\!R_-}{\partial r} + \frac{2}{r} \ {}^{'}\!\!R_- + \frac{i}{r} \frac{\partial}{\partial \vartheta} \ {}^{'}\!\!R_- - \left[\exAo + \zAo \right]\cdot {}^{'}\!\!R_+ - B_0 &\cdot {}^{'}\!\!S_+ - \mathds{A}^{\ast}_2 \cdot {}^{'}\!\!R_- - \mathds{B}_- \cdot {}^{'}\!\!S_- \nonumber \\
&= \frac{M_1 - M}{\hbar} c\cdot {}^{'}\!\!R_+ \label{442b} \\
\frac{\partial \ {}^{'}\!\!S_+}{\partial r} - \frac{i}{r} \frac{\partial}{\partial \vartheta} \ {}^{'}\!\!S_+ + \left[\exAo + \eAo \right]\cdot {}^{'}\!\!S_- - B^{\ast}_0 &\cdot {}^{'}\!\!R_- + \mathds{A}_1 \cdot {}^{'}\!\!S_+ + \mathds{B}^{\ast}_- \cdot {}^{'}\!\!R_+ \nonumber \\
&= -\frac{M_2 + M}{\hbar} c\cdot {}^{'}\!\!S_- \label{442c} \\
\frac{\partial \ {}^{'}\!\!S_-}{\partial r} + \frac{2}{r} \ {}^{'}\!\!S_- + \frac{i}{r} \frac{\partial}{\partial \vartheta} \ {}^{'}\!\!S_- - \left[\exAo + \eAo \right]\cdot {}^{'}\!\!S_+ - B^{\ast}_0 &\cdot {}^{'}\!\!R_+ - \mathds{A}^{\ast}_1 \cdot {}^{'}\!\!S_- - \mathds{B}^{\ast}_+ \cdot {}^{'}\!\!R_- \nonumber \\
&= \frac{M_2 - M}{\hbar} c\cdot {}^{'}\!\!S_+ \label{442d} \ .
\end{align}
\end{subequations}
\noindent Unfortunately it is very difficult to obtain approximate solutions to this system, otherwise one could have taken the value of the energy functional upon these approximations in order to test the degeneracy hypothesis with respect to the whole triplet $|S=1; S_z=0, \pm 1>$. However, we will consider now the RST singlet states $|S=0; S_z=0>$, which obey an eigenvalue system very similar to the present one (\ref{442a})-(\ref{442d}) for which we are able to present explicit numerical solutions.

\section{Singlet Structure}

In order to see clearly both the differences and the common features of ortho- 
and para-helium, it is most instructive to inspect the latter type of helium 
along the lines of arguments being used for the preceding discussion of the 
triplet states. Especially one is interested here in seeing the reason why 
the energy of the singlet states (in lowest order) is not shifted by the 
exchange interactions in RST (and in the Hartree-Fock approach), in contrast to 
the energy of the triplet states.

\subsection{Singlet Fields}

It seems self-suggestive that for the RST analogue of the conventional states 
\\ $\ket{S=0; S_z=0}$ the single-particle spins should be anti-parallel, i.e. 
one tries for the general wave amplitudes $\aRpm, \aSpm$ (\ref{38})-
(\ref{310}) the following ansatz:
\begin{subequations}\label{51}
\begin{align}
\eRp &= - \eSp \doteqdot -\frac{i}{2} e^{i\varphi} \, R_+ \label{51a} \\
\eRm &= - \eSm \doteqdot -\frac{i}{2} e^{i\varphi} \, R_- \label{51b} \\
\zRp &= \zSp \doteqdot \frac{1}{2} \, S_+ \label{51c} \\
\zRm &= \zSm \doteqdot \frac{1}{2} \, S_- \ . \label{51d}
\end{align}
\end{subequations}
Thus the Pauli spinors $\aphipm$ (\ref{39})-(\ref{310}) become
\begin{subequations}\label{52}
\begin{align}
\ephip &= R_+ \cdot \zetappn \label{52a} \\
\ephim &= -i \, R_- \cdot \zetappe \label{52b} \\
\zphip &= S_+ \cdot \zetapmn \label{52c} \\
\zphim &= -i \, S_- \cdot \zetapme \ . \label{52d}
\end{align}
\end{subequations}
\indent Concerning the corresponding densities, one finds for the charge 
densities $\akn(\vec{r})$ (\ref{319}) formally the same expressions 
(\ref{43a})-(\ref{43b}) as for the triplet configurations $\ket{S=1; S_z=+1}$. 
Similarly the first particle's para-current $k_1(\vec{r})$ is identical to 
the ortho-current (\ref{44a}) whereas the second particle's current $k_2
(\vec{r})$ must necesseraly change sign:
\begin{equation}\label{53}
\vec{k}_2(\vec{r}) = -\frac{1}{2 \pi} \, S_+ S_- \, \sin \vartheta \, \vec{e}_
\varphi \ .
\end{equation}
However the most interesting point refers now to the exchange density $h_0
(\vec{r})$ (\ref{321}). Indeed by inserting there the present para-ansatz 
(\ref{51a})-(\ref{51d}), one finds that the exchange density vanishes 
identically
\begin{equation}\label{54}
h_0(\vec{r}) \equiv 0 \ .
\end{equation}
Consequently, one puts to zero also the time-component of the exchange 
potential $\Bmu(\vec{r})$:
\begin{equation}\label{55}
B_0(\vec{r}) \equiv 0 \ ,
\end{equation}
since this component is linked to the exchange density $h_0(\vec{r})$ via the 
Poisson equation (\ref{334b}), see the solution (\ref{416c}). But when both 
objects $h_0$ and $B_0 $ do vanish, the exchange energy $\hat{z}_a \cdot 
M_a^{(h)} \, c^2$ (\ref{343a})-(\ref{343b}) must vanish, too. However, since 
the latter agrees (in lowest order of approximation) with the exchange 
correction $\Delta E_T^{(h)}$ (\ref{354}) and therefore also with the 
conventional exchange energy $\Delta E_C$ (\ref{34}) (see the discussion 
below equation (\ref{418})), one encounters here the origin of the missing 
exchange energy in RST and in the HF approach!\\
\indent Though both approaches lead to the same effect of missing exchange 
energy for para-helium, the interpretation is however different:\\
(i) The {\it conventional interpretation} of quantum mechanics is of statistical 
nature and therefore the missing of exchange energy $\Delta E_C$ (\ref{34}) 
is traced back to the non-identity of electrons with different z-components 
$s_z$ ($=\pm\frac{1}{2}$) of their spins (thus escaping the Pauli principle). 
Such different particles are described by the Hartree approach, not the HF 
approach. As is well-known, the Hartree approximation does not take into account the 
exchange effects \cite{b7, b25}.\\
(ii) In {\it RST}, the exchange forces are considered as being due to real gauge 
interactions (similar to the electromagnetic interactions); and therefore the 
missing of the ("electric") exchange effect is interpreted as being due to 
the missing of the source for exciting the exchange potential $B_0$. \\
\indent Concerning the functional form of the gauge fields, this will be 
dictated again by the present form of the associated currents. Thus, both 
electric para-currents $\vec{k}_1(\vec{r})$ (\ref{44a}) and $\vec{k}_2
(\vec{r})$ (\ref{53}) will imply again the former shape (\ref{47a}) of the 
vector potentials $\vec{A}_a(\vec{r})$, but the ortho-form of the "magnetic" 
exchange potentials $\vec{B}(\vec{r})$ (\ref{47b}) cannot be taken over to 
the present para-case. The reason is that the exchange current $\vec{h}
(\vec{r})$ must adopt now its para-form as
\begin{equation}\label{56}
\vec{h}(\vec{r}) = i \, h(r, \vartheta) \cdot \vec{W}^*_p(\vec{r}) \ ,
\end{equation}
with the complex-valued vector field $\vec{W}_p(\vec{r})$ being defined 
through
\begin{equation}\label{57}
\vec{W}_p(\vec{r}) = -e^{i \varphi} ( \vec{e}_\vartheta + i \, \cos \vartheta 
\, \vec{e}_\varphi ) \ ,
\end{equation}
but with the strength of current $h(r, \vartheta)$ being formally the same as for 
the ortho-case (\ref{46}):
\begin{equation}\label{58}
h(r,\vartheta) = \frac{R_+\cdot S_- + R_- \cdot S_+}{4 \pi} \ .
\end{equation}
Consequently, with reference to the Poisson equation (\ref{334d}), one puts 
for the para-form of the "magnetic" exchange potential $\vec{B}$:
\begin{equation}\label{59}
\vec{B}(\vec{r}) = i \, r \, B \cdot \vec{W}_p \ ,
\end{equation}
with a real-valued exchange amplitude $B = B(r,\vartheta)$.

\subsection{Mass Eigenvalue Equations}

After the functional forms of all the para-fields are known, one inserts these 
into the general eigenvalue system (\ref{323})-(\ref{326d}) and finds then 
the mass eigenvalue system for the first particle ($a=1$) consisting again of 
a radial part
\begin{subequations}\label{510}
\begin{align}
\frac{\partial R_+}{\partial r} + \left[\exAo + \zAo \right]\cdot R_- + r 
\left[ B \cdot S_+ - A_2 \cdot R_+ \right]& \nonumber \\ 
+ \, r \cos^2 \vartheta \left[A_2\cdot R_+ + B \cdot S_+\right] &= - \frac{M_1
+M}{\hbar} c \cdot R_- \label{510a}\\[2ex]
\frac{\partial R_-}{\partial r} + \frac{2}{r}\, R_- - \left[\exAo + \zAo 
\right]\cdot R_+ + r \left[ A_2 \cdot R_- - B \cdot S_- \right]& \nonumber \\ 
- \, r \cos^2 \vartheta \left[A_2 \cdot R_- + B \cdot S_-\right] &= 
\frac{M_1-M}{\hbar} c \cdot R_+ \label{510b} \ ,
\end{align}
\end{subequations}
and of an angular part
\begin{subequations}\label{511}
\begin{align}
\frac{\partial R_+}{\partial \vartheta} &= r^2 \sin \vartheta \cos\vartheta 
\left[ A_2 \cdot R_+ + B \cdot S_+ \right] \label{511a}\\ 
\frac{\partial R_-}{\partial \vartheta} &= r^2 \sin \vartheta \cos\vartheta 
\left[ A_2 \cdot R_- + B \cdot S_- \right] \ . \label{511b}
\end{align}
\end{subequations}
\noindent Observe here that the angular part (\ref{511a})-(\ref{511b}) of the present para-system exactly agrees with the angular part (\ref{49}) and (\ref{411}) of the former ortho-system, although the exchange potential $\Bmu (\vec{r})$ looks rather different for both cases, cf. the ortho-potential $\vec{B}(\vec{r})$ (\ref{47b}) and its para-counterpart (\ref{59}). However, in contrast to the angular system, the radial equations do appear rather different, despite the fact that the first particle ($a=1$) has spin up in both cases; compare the ortho-equations (\ref{48}) and (\ref{410}) to their present para-counterparts (\ref{510a})-(\ref{510b})!\\
\indent Finally, the radial part of the second particle's eigenvalue system is found as
\begin{subequations}\label{512}
\begin{align}
\frac{\partial S_+}{\partial r} + \left[\exAo + \eAo \right]\cdot S_- + r \left[A_1 \cdot S_+ + B \cdot R_+\right]& \nonumber \\
+\, r \cos^2\vartheta \left[B \cdot R_+ - A_1 \cdot S_+ \right] &= - \frac{M_2+M}{\hbar} c \cdot S_- \label{512a} \\
\frac{\partial S_-}{\partial r} + \frac{2}{r} S_- - \left[\exAo + \eAo \right]\cdot S_+ - r \left[B \cdot R_- + A_1 \cdot S_-\right]& \nonumber \\
+\, r \cos^2\vartheta \left[A_1\cdot S_- - B \cdot R_- \right] &= \frac{M_2-M}{\hbar} c \cdot S_+ \label{512b} \ , 
\end{align}
\end{subequations} 
and the associated angular part looks as follows
\begin{subequations}\label{513}
\begin{align}
\frac{\partial S_+}{\partial \vartheta} & = r^2 \sin\vartheta \cos\vartheta \left[B\cdot R_+ - A_1 \cdot S_+ \right] \label{513a} \\
\frac{\partial S_-}{\partial \vartheta} & = r^2 \sin\vartheta \cos\vartheta \left[B\cdot R_- - A_1 \cdot S_- \right] \ . \label{513b}
\end{align}
\end{subequations}
\indent As was mentioned already in connection with the general eigenvalue equations in Sect.3, such a system must be completed by the Poisson equations for the gauge potentials in order to have a closed system. However, for the present singlet states ${}^1S_0$ we are not satisfied with the linearized form of the Poisson equations (\ref{334a})-(\ref{334d}), but will apply their exact form \cite{a41} for our subsequent numerical calculations.\\
\indent For both electrostatic potentials ${}^{(a)}A_0(r,\vartheta)$ one finds
\begin{subequations}\label{514}
\begin{align}
\Delta (r,\vartheta) {}^{(1)}A_0 & = 4\pi \alpha_s \eko - 2r^2(1+\cos^2\vartheta) \Delta_0 B^2 \label{514a} \\
\Delta (r,\vartheta) {}^{(2)}A_0 & = 4\pi \alpha_s \zko + 2r^2(1+\cos^2\vartheta) \Delta_0 B^2 \label{514b} 
\end{align}
\end{subequations}
\noindent where $\Delta (r,\vartheta)$ is the $SO(2)$ symmetric Laplacean
\begin{equation}
\Delta (r,\vartheta) = \frac{\partial^2}{\partial r^2} + \frac{2}{r}\frac{\partial}{\partial r} + \frac{1}{r}\left[\frac{\partial^2}{\partial \vartheta^2} + \cot \vartheta \frac{\partial}{\partial \vartheta}\right] \ , \label{515}
\end{equation}
\noindent and the difference of the electrostatic potentials $\Delta_0$ is defined through
\begin{equation}
\Delta_0 (\vec{r}) \doteqdot \frac{(M_1-M_2)c}{\hbar} - \left[\eAo (\vec{r}) - \zAo (\vec{r}) \right] \ . \label{516}
\end{equation}
\indent Similarly, the magnetostatic potentials $A_a(r,\vartheta)$ (\ref{47a}) must obey the following equations
\begin{equation}
\left[ \Delta (r,\vartheta) + \frac{2}{r} \frac{\partial}{\partial r} + 2\frac{\cot \vartheta}{r^2} \frac{\partial}{\partial \vartheta} \right]A_{1/2} = 4\pi\alpha_s \frac{k_{1/2}}{r} \pm \lbrace \frac{2}{\sin\vartheta} \frac{\partial}{\partial \vartheta} (\cos \vartheta B^2)+2BY_r \rbrace \label{517} \ ,
\end{equation}
\noindent with the current strengths $k_a(r,\vartheta)$ being deduced from their three-vector form (\ref{44a}) and (\ref{53}) as
\begin{subequations}\label{518}
\begin{align}
k_1 & = \frac{1}{2\pi} R_+R_- \label{518a}\\ 
k_2 & = -\frac{1}{2\pi} S_+S_- \label{518b} \ .
\end{align}
\end{subequations}
\noindent Furthermore, the radial component $Y_r$ of the "magnetic" exchange field $\vec{Y}(\vec{r})$ (\ref{346}) reads in terms of the exchange potential $B$ (\ref{59})
\begin{equation}
Y_r = \frac{1}{\sin \vartheta \cos \vartheta} \frac{\partial}{\partial \vartheta} (\cos^2 \vartheta \cdot B) + r^2 \Delta_{1/2} B\label{519}
\end{equation}
\begin{equation*}
(\Delta_{1/2} \doteqdot A_1 - A_2) \ .
\end{equation*}
\indent And finally, the Poisson equation for the "magnetic" exchange potential $B(r,\vartheta)$ is given by
\begin{align}
\left[ \Delta (r,\vartheta) + \frac{2}{r} \frac{\partial}{\partial r} - 4\frac{\cot \vartheta}{r^2} \frac{\partial}{\partial \vartheta} \right]B + \frac{1}{\cos \vartheta} \frac{\partial}{\partial \vartheta} (\sin \vartheta \Delta_{1/2} B) = 4\pi\, \alpha_s \, \frac{h}{r}& \nonumber \\- B \left( \Delta_0^2+[{}^{(1)}H_r-{}^{(2)}H_r] \right)& \ . \label{520}
\end{align}
\noindent Here, the exchange current density $h(r,\vartheta)$ is given in terms of the wave amplitudes by equation (\ref{58}) and the objects ${}^{(a)}H_r$ ($a=1,2$) are the radial components of the magnetostatic fields $\vec{H}_a(\vec{r})$.

\subsection{Ortho/Para Level Splitting}

\indent For a discussion of the energy splitting between the present para-configurations ${}^1S_0$ and the former ortho-configurations ${}^3S_1$ of Sect.4, one will resort again to the lowest-order approximation. To this end, one neglects all the interparticle interactions for the para-eigenvalue system (\ref{510a})-(\ref{513b}) and thus finds the present system being reduced to the former approximative system (\ref{414a})-(\ref{414d}) which therefore represents a common approximation to both the ortho- and para-configurations. However, both types of solutions are different with respect to their energy content $E_T$, even for the lowest-order approximation $\tilde{E}_T$ (\ref{419}). The reason is here that the present para-configurations have vanishing exchange density $h_0(\vec{r})$, cf. (\ref{54}), and therefore the exchange energy $\Delta E_T^{(h)}$ (\ref{354}) and (\ref{418}) must necessarily also vanish! Consequently, the para-levels do not receive a shift through the exchange interactions, as is the case with the ortho-levels (see fig.3 and the discussion below equation (\ref{418})). A further consequence is the fact that the RST (and HF) predictions for the energy difference between the singlet states ${}^1S_0$ and the triplet states ${}^3S_1$ are (in lowest-order approximation) {\it half} the value of the predictions of the standard theory, which, on the other hand, are in good agreement with the experiments; see the discussion of this point in ref. \cite{u6}.\\
\indent Does this mean that RST (as the relativistic generalization of the Hartree-Fock approach) can be useful only for the computation of the triplet level systems but becomes useless when one wants to consider the singlet levels? This question can be decided only by concrete calculations of the frequency of spectral lines for the transitions within the singlet level system ${}^1S_0$ and comparison to the experimental data. The point here is that the frequency of a spectral line is computed from an energy difference $E_{T_{(in)}}-E_{T_{(fin)}}$ due to the total energy $E_T$ (\ref{335}) of the initial ($in$) and final ($fin$) states; and it may well be that the influence of the (missing) exchange energy $\Delta E_T^{(h)}$, as part of the total energy $E_T$, cancels or at least is weakened when forming that energy difference. For instance, a closer inspection of the "semiclassical" (i.e. $u=0$) RST prediction of the ground state ($1s^2\ {}^{1}S_0$) energy $E_T$ has revealed that there is a deviation from the predictions of other theoretical approaches \cite{a5,a6} by not more than $1,3$ eV in the range $2\leq z_{ex}\leq 100$, see ref.\cite{u5}. Therefore one has to resort now to numerical calculations in order to decide the question of precision within the singlet level system.    

\section{Numerical Results}

For a numerical test of RST, one will first consider a sufficiently simple situation for which the computational effort remains moderate. In this respect, a considerable complication of the preceding eigenvalue system is due to the fact that the two-particle wave amplitudes $R_\pm$, $S_\pm$ are functions of both polar coordinates $\{ r, \vartheta \}$ which thus yields a two-dimensional eigenvalue problem. Therefore a possible simplification can surely be attained by considering \textit{isotropic} configurations where the wave amplitudes $\{R_\pm, S_\pm \}$ and all the gauge fields $\{ \aAo, A_a, B \}$ depend exclusively upon the radial variable $r$. However such isotropic configurations are possible as exact solutions only for the singlet states $ns^2\ {}^{1}S_0$ (Sect.V) where the spins of both particles are antiparallel and their principal quantum numbers do coincide ($n_1=n_2\doteqdot n$). But in the general case ($n_1 \neq n_2$), the angular parts (\ref{511a})-(\ref{511b}) and (\ref{513a})-(\ref{513b}) of the para-eigenvalue system demonstrate that the wave amplitudes must be expected to depend also of the polar angle $\vartheta$; unless one can find solutions for which the right-hand sides of those equations do vanish. Obviously this requirement implies the following conditions:
\begin{subequations}\label{61}
\begin{align}
&A_2 \cdot R_+ + B \cdot S_+ = 0\label{61a}\\
&A_2 \cdot R_- + B \cdot S_- = 0\label{61b}\\
&B \cdot R_+ - A_1 \cdot S_+ = 0\label{61c}\\
&B \cdot R_- - A_1 \cdot S_- = 0\label{61d}\ .
\end{align}
\end{subequations}

\subsection{Isotropic Configurations $ns^2 \ \sf {}^1S_0$}

Indeed, such isotropic solutions are possible, namely by putting
\begin{subequations}\label{62}
\begin{align}
&\eAo \equiv \zAo \doteqdot {}^{(p)}\!A_0(r) \label{62a} \\
&A_1 \equiv -A_2 \doteqdot A_p(r) \label{62b} \\
&B(r) \equiv A_p(r) \label{62c} \\
&{}^{(1)}k_0 = {}^{(2)}k_0 \doteqdot {}^{(p)}k_0(r) \label{62d} \\
&k_1= - k_2 = h \doteqdot k_p(r) \label{62e} \ .
\end{align}
\end{subequations}
Furthermore both particles are to be described by identical wave amplitudes ($R_+ \equiv S_+$; $ R_- \equiv S_-)$ because one must assume that both eigenvalues do coincide
\begin{equation}\label{63}
M_1 = M_2 \doteqdot M'' \ .
\end{equation}
This means that the principal quantum numbers $n_1, n_2$ of the conventional notation must agree ($n_1 = n_2$) so that one arrives at the RST counterparts of the conventional states  $ns^2 \ \sf {}^1S_0$. These states are the solutions of the following isotropic form of the more general system (\ref{510a})-(\ref{510b})
\begin{subequations}\label{64}
\begin{align}
\frac{dR_+}{dr} + \left[ \exAo + {}^{(p)}\!A_0 \right] \cdot R_- + 2 r B R_+ = -\frac{M''+M}{\hbar} c \, R_- \label{64a} \\
\frac{dR_-}{dr} + \frac{2}{r} R_- - \left[ \exAo + {}^{(p)}\!A_0 \right] \cdot R_+ - 2 r B R_- = \frac{M''-M}{\hbar} c \, R_+ \label{64b} \ .
\end{align}
\end{subequations}
This simple (but exact) eigenvalue system for the isotropic states must be complemented again by the isotropic Poisson equations for the gauge potentials:
\begin{subequations}\label{65}
\begin{align}
\left(\frac{d^2}{dr^2} + \frac{2}{r} \frac{d}{dr} \right) {}^{(p)}\!A_0(r) = 4\pi \, \alpha_s \, {}^{(p)}\!k_0(r) \label{65a} \\
\left(\frac{d^2}{dr^2} + \frac{4}{r} \frac{d}{dr} \right) B(r) = 4\pi \, \alpha_s \, \frac{k_p}{r} - 6B^2 (1-\frac{2}{3}r^2 B) \ , \label{65b}
\end{align}
\end{subequations}
which of course is deduced from the original system (\ref{510a})-(\ref{510b}) by means of the isotropy conditions (\ref{62a})-(\ref{62d}).\\
\indent Naturally, the assumption of isotropy will simplify also the (exact) energy functional $E_T$ (\ref{335}). The first point to observe here refers to the normalization conditions (\ref{332}). Since the "electric" exchange field strength $\vec{X}(\vec{r})$ (\ref{333a})-(\ref{333b}) is given quite generally in terms of potentials through \cite{a41}
\begin{equation}\label{66}
\vec{X}(\vec{r}) = - \vec{\nabla} B_0(\vec{r}) + i\, \Delta_0(\vec{r})\cdot \vec{B}(\vec{r}) + \, i \, B_0(\vec{r}) \left[\vec{A}_1(\vec{r}) - \vec{A}_2(\vec{r}) \right] \ ,
\end{equation}
this vector field must vanish ($\vec{X}(\vec{r}) \equiv 0$) because both the exchange potential $B_0$ and the electrostatic potential difference $\Delta_0$ (\ref{516}) is zero due to the assumption of isotropy for the singlet states $ns^2 \ \sf {}^1S_0$. Therefore both charge densities $\ako(\vec{r})$ and $\alo(\vec{r})$ (\ref{333a})-(\ref{333b}) do coincide and this puts the normalization parameter $\hat{z}_a$ (see below equation (\ref{340b})) to unity ($\hat{z}_a=1$) for both particles ($a=1,2$):
\begin{equation}\label{67}
\int d^3\vec{r} \ {}^{(p)}\!k_0 = \int dr \ r^2 \, ( R_+^{\,2} + R_-^{\,2} ) = 1 \ .
\end{equation}
\indent The next point is even more intimately connected with the property of isotropy. It refers to the relationship between the energy corrections $\Delta E_T^{(m)}$ and $\Delta E_T^{(g)}$ of the {\it magnetic} type. First recall here that on behalf of the vanishing "electric" exchange correction $\Delta E_T^{(h)}$ (\ref{342}) the exact energy functional $E_T$ (\ref{335}) reduces to
\begin{equation}
E_T \ \Rightarrow \ E_T'' = 2 \, M'' c^2 - \Delta E_T^{(e)} - \Delta E_T^{(m)} - \Delta E_T^{(g)} \ ,
\end{equation}
with the "magnetic" corrections $\Delta E_T^{(m)}$ and $\Delta E_T^{(g)}$ being given by equations (\ref{345})-(\ref{346}). However one can show \cite{a41} that {\it for the isotropic states} both "magnetic" corrections $\Delta E_T^{(m)}$ and $\Delta E_T^{(g)}$ obey the numerical relationship
\begin{equation}\label{69}
\Delta E_T^{(g)} = 2 \Delta E_T^{(m)} \ ,
\end{equation}
so that the energy functional for this type of para-states becomes further simplified to 
\begin{equation}\label{610}
E_T'' = 2 \, M'' c^2 - \Delta E_T^{(e)} - 3 \Delta E_T^{(m)} \ .
\end{equation}
\indent Here the magnetostatic correction $\Delta E_T^{(m)}$ (\ref{345}) reads in terms of the spherically symmetric wave amplitudes $R_\pm(r)$
\begin{equation}\label{611}
\Delta E_T^{(m)} = - \frac{4}{3} \, \hbar c \left\{ \int_0^\infty \!\! dr \ r^3 \, B(r) R_+(r) R_-(r) - \frac{1}{\alpha_s} \int_0^\infty \!\! dr \ r^4 \, B^3(r) [1-r^2B(r)] \right\} \ .
\end{equation}
\indent Finally it remains to specify the electrostatic energy correction $\Delta E_T^{(e)}$ for the "isotropic" functional $E_T''$ (\ref{610}). Actually, there occurs also a simplification for this contribution since the general form of electrostatic fields \cite{a41}
\begin{subequations}\label{612}
\begin{align}
\vec{E}_1(\vec{r}) &= - \vec{\nabla} \, \eAo(\vec{r}) - i \left[B_0(\vec{r}) \vec{B}^*(\vec{r}) - B_0^*(\vec{r}) \vec{B}(\vec{r}) \right]\label{612a} \\
\vec{E}_2(\vec{r}) &= - \vec{\nabla} \, \zAo(\vec{r}) + i \left[B_0(\vec{r}) \vec{B}^*(\vec{r}) - B_0^*(\vec{r}) \vec{B}(\vec{r}) \right]\label{612b} \ ,
\end{align}
\end{subequations}
is reduced for the isotropic configurations ($B_0 \equiv 0$) to the form of simple gradient fields
\begin{equation}\label{613}
\vec{E}_a(\vec{r}) \ \Rightarrow \ - \vec{\nabla} \, \aAo(\vec{r}) \ .
\end{equation}
As a consequence, the electrostatic field energy $E_R^{(e)}$ (\ref{341}) equals {\it exactly} the electrostatic mass-energies $M_a^{(e)} c^2$ (\ref{340a})-(\ref{340b}), so that the corresponding energy correction $\Delta E_T^{(e)}$ (\ref{339}) becomes
\begin{equation}\label{614}
\Delta E_T^{(e)} \Rightarrow \frac{1}{2} \sum_{a=1}^2 \hat{z}_a \cdot M_a^{(e)} c^2 = - \hbar c \int \!\! dr \ r^2 \, {}^{(p)}\!A_0(r) \left\{ R_+^{\,2}(r) + R_-^{\,2}(r) \right\} \ .
\end{equation}
This completes the "isotropic" energy functional $E_T''$ (\ref{610}) whose values ($E_T^{(n)}$, say) upon the solutions $R_\pm^{(n)}$ of the "isotropic" eigenvalue problem (\ref{64})-(\ref{65}) with mass eigenvalues $M_n''$ constitutes the spectrum for the RST states due to the conventional classification $ns^2 \ \sf {}^1S_0$.

\subsection{Relativistic Energy Difference $2s^2 \ \sf {}^1S_0$ / $1s^2 \ \sf {}^1S_0$}

Obviously the isotropic form of the RST eigenvalue problem is sufficiently simple in order to solve numerically the corresponding eigenvalue equations (\ref{64a})-(\ref{65b}) and to take the value of the {\it exact} energy functional $E_T''$ (\ref{610}) upon these solutions. This admits us to calculate, e.g., RST energy differences within the spectrum of isotropic solutions and to compare this to both the experimental data and to the analogous predictions of other theoretical approaches in the literature. As a simple demonstration, we now inspect the energy difference of the states $2s^2 \ \sf{}^1S_0$ (i.e. $n_1=n_2=2$) and $1s^2 \ \sf{}^1S_0$ (i.e. $n_1=n_2=1$), where the latter state is the ground-state of the helium-like ions and has been studied already extensively \cite{u5}. The energies to be considered refer to an idealized situation were the nucleus is thought to be infinitely heavy and the self-energy effects (and other QED corrections) are first neglected (for the inclusion of self-energy in RST, see ref.s \cite{a43,u5}). Thus denoting the semiclassical energy difference by $\Delta \!\stackrel{\;\;_\circ}{E}_{1\backslash 2}$, we have
\begin{equation}\label{615}
\Delta \!\stackrel{\;\;_\circ}{E}_{1\backslash 2} \ \doteqdot \ \stackrel{\;\;_\circ}{E} \Big\vert_{2s^2 \, \sf {}^1S_0} - \stackrel{\;\;_\circ}{E} \Big\vert_{1s^2 \, \sf {}^1S_0}
\end{equation}
where the energy functional to be used in RST is that one of the isotropic states (\ref{610}). This RST prediction may then be compared to the observational data ($\Delta_{exp}E_{1\backslash 2}$) as compiled by NIST \cite{m1}. Such a comparison is instructive because these experimental values $\Delta_{exp}E_{1\backslash 2}$ include the Lamb shift which in the present ("semiclassical") RST approach is first neglected and afterwards taken into account by switching on the RST self-interactions. But despite this neglection it is interesting to inspect also the corresponding {\it semi-classical} predictions of other theoretical approaches, especially the relativistic ($1/Z$)-expansion method \cite{a6} ($\leadsto \Delta_{1/Z} \! \stackrel{\;\;_\circ}{E}_{1\backslash 2}$) and the all-order technique in relativistic many-body perturbation theory (MBPT) \cite{a5} ($\leadsto \Delta_{all}\!\stackrel{\;\;_\circ}{E}_{1\backslash 2}$).\\
\indent The latter approaches are not directly concerned with the special energy difference $\Delta \!\stackrel{\;\;_\circ}{E}_{1\backslash 2}$ (\ref{615}) but rather with the (semiclassical) one-particle ionization energies, $\stackrel{\;\;_\circ}{J}^{(n)}_{\!1/Z}\!(2)$ and $\stackrel{\;\;_\circ}{J}^{(n)}_{\!all}\!(2)$, namely for the situation when the principle quantum number of one particle is $n$, and the other particle is in the ground state ($n=1$). However it is an easy matter to compute the desired energy difference $\Delta \!\stackrel{\;\;_\circ}{E}_{1\backslash 2}$ from those helium-like ($\stackrel{\;\;_\circ}{J}^{(n)}_{\!theo}\!(2)$) and hydrogen-like ($\stackrel{\;\;_\circ}{J}^{(n)}_{\!theo}\!(1)$) ionization energies. Indeed, when one of the two electrons is thrown out from a bound state of energy $\stackrel{\;\;_\circ}{E}^{(n)}_{theo}\!(2)$ to infinity, where its energy is just $Mc^2$, the remaining bound electron has energy $\stackrel{\;\;_\circ}{E}^{(n)}_{theo}\!(1)$; and thus the ionization energy $\stackrel{\;\;_\circ}{J}^{(n)}_{\!theo}\!(2)$ emerges as the difference of the initial and final energies:
\begin{equation}\label{616}
 \stackrel{\;\;_\circ}{J}^{(n)}_{\!theo}\!(2) = \left[Mc^2 + \stackrel{\;\;_\circ}{E}^{(n)}_{theo}\!(1) \right] - \stackrel{\;\;_\circ}{E}^{(n)}_{theo}\!(2) \ .
\end{equation}
Here the semi-classical one-particle spectrum $\stackrel{\;\;_\circ}{E}^{(n)}_{theo}\!(1)$ is due to the simplified eigenvalue problem (\ref{414a})-(\ref{414b}) which yields
\begin{equation}\label{617}
\stackrel{\;\;_\circ}{E}^{(n)}_{theo}\!(1) \equiv \tilde{M}_n c^2 = \frac{Mc^2}{\sqrt{1 + \Big(\frac{z_{ex}}{n-1+\sqrt{1-(z_{ex} \alpha_s)^2}}\Big)^2}} \ .
\end{equation}
Now one can write down the equation (\ref{616}) for both principle quantum numbers $n=1$ and $n=2$ and then finds the desired energy difference $\Delta_{theo}\!\!\stackrel{\;\;_\circ}{E}_{1\backslash 2}$ by subtracting both equations from one another:
\begin{equation}\label{618}
\Delta_{theo}\!\!\stackrel{\;\;_\circ}{E}_{1\backslash 2} = \stackrel{\;\;_\circ}{J}^{(1)}_{\!theo}\!(2) - \stackrel{\;\;_\circ}{J}^{(2)}_{\!theo}\!(2) + \stackrel{\;\;_\circ}{E}^{(2)}_{\!theo}\!(1) - \stackrel{\;\;_\circ}{E}^{(1)}_{\!theo}\!(1) \ .
\end{equation}
Here one can now insert the one-particle energies from the semiclassical spectrum (\ref{617}) and the two-particle ionization energies $\stackrel{\;\;_\circ}{J}^{(n)}_{\!theo}\!(2)$ for $n=1$ and $n=2$ from ref.s \cite{a5,a6} in order to yield ($theo \rightarrow 1/Z, all$) the energy difference $\Delta_{\!1/Z}\!\!\stackrel{\;\;_\circ}{E}_{1\backslash 2}$ of the 1/Z-expansion method and $\Delta_{\!all}\!\!\stackrel{\;\;_\circ}{E}_{1\backslash 2}$ of the all-order technique in MBPT.  Unfortunately the ionization energies $\stackrel{\;\;_\circ}{J}^{(n)}_{\!1/Z}\!(2)$ and $\stackrel{\;\;_\circ}{J}^{(n)}_{\!all}\!(2)$ are tabulated in ref.s \cite{a6,a5} exclusively for the states $1s^2 \ \sf {}^1S_0$ and $1s2s \ \sf {}^1S_0$ but not for the presently considerd $2s^2 \ \sf {}^1S_0$ states, so that we have to restrict us preliminarily to a comparison of our RST results $\Delta_{RST}\!\stackrel{\;\;_\circ}{E}_{1\backslash 2}$ to the experimental data $\Delta_{exp}\!E_{1\backslash 2}$ \cite{m1}, see table I and fig.4 \ .\\[4ex]

\begin{center}
\bf Table I
\end{center}

\subsection{Self-Interactions}

\indent One could suppose that the origin of the deviations between the present semiclassical RST results $\Delta_{RST}\!\stackrel{\;\;_\circ}{E}_{1\backslash 2}$ and the experimental values $\Delta_{exp}E_{1\backslash 2}$ (table I and fig.4) are due to the neglection of the electronic RST self-interactions. This however is not true. Indeed, we will readily demonstrate that the inclusion of the RST self-interactions is not sufficient in order to shift the semiclassical RST predictions $\Delta_{RST}\!\stackrel{\;\;_\circ}{E}_{1\backslash 2}$ (\ref{615}) towards the experimental values $\Delta_{exp}E_{1\backslash 2}$. This is in contrast to the situation with the ground state $1s^2\ {}^1S_0$, whose ionization energy deviates not more than $2$ eV from the experimental values in the whole range $2\leq z_{ex}\leq 83$ when the RST self-interactions are adequately included. The RST self-energy problem has been treated extensively in ref.s \cite{a43,u5} so that it may be sufficient to reproduce here only the main results which are relevant for the computation of the desired energy difference $\Delta_{RST}E_{1\backslash 2}$:
\begin{equation}\label{619}
\Delta_{RST}E_{1\backslash 2} = E_T \Big\vert_{2s^2 \, \sf {}^1S_0} - E_T \Big\vert_{1s^2 \, \sf {}^1S_0} \ .
\end{equation}
Since this looks quite similar to the semiclassical result (\ref{615}) without the self-interactions, it is important to clearly state now the differences.\\
\indent First let us mention that our RST self-interactions enter the theory via the fibre metric $\Kaubu$ (\ref{220}). The general form of this fibre metric is parametrized by the "self-interaction parameter" $u: \Kaubu = \Kaubu(u)$; and this parameter enters then also both the energy eigenvalue system and the energy functional $E_T$ (i.e. $E_T \Rightarrow E_T(u)$). However, for the presently considered isotropic solutions $ns^2\ {}^1S_0$, the change caused by a non-trivial value of u is minimal: First, the mass eigenvalue system (\ref{64a})-(\ref{64b}) remains invariant under a change of $u$. Next, it is easy to show that also the {\it electrostatic} Poisson equation (\ref{65a}) remains the same for a non-zero value of $u$. Thus it is only the {\it magnetostatic} Poisson equation (\ref{65b}) which changes into
\begin{equation}\label{620}
\left(\frac{d^2}{dr^2} + \frac{4}{r} \frac{d}{dr} \right) B(r) + 6 B^2(r) \left[1-\frac{2}{3}r^2B(r)\right] = 4\pi \, \alpha_s \, e^{-2u} \cdot \frac{k_p}{r} \ .
\end{equation}
And finally, the energy functional $E_T''$ (\ref{610}) remains the same but with the magnetic correction energy $\Delta E_T^{(m)}$ being modified according to
\begin{equation}
\Delta E_T^{(m)} = -\frac{4}{3} \, \hbar c \left\{ \int_0^\infty dr \ r^3 B(r) R_+(r) R_-(r) - \frac{e^{2u}}{\alpha_s} \int_0^\infty dr \ r^4 B^3(r) \left[ 1-r^2B(r) \right] \right\} \ .
\end{equation}
Thus it becomes evident, that the inclusion of the self-interactions implies for the isotropic states nothing else than replacing the coupling constant $\alpha_s$ ($=\frac{e^2}{\hbar c}$) by $\alpha_s \cdot e^{-2u}$; but only for the {\it magnetic} interactions whereas the {\it electric} interactions are unchanged! Since the \textit{non-singular} solutions $B(r)$ of the magnetic Poisson equation (\ref{620}) must vanish ($B(r)\equiv 0$) when the right-hand side becomes zero for $u\rightarrow \infty $, cf.(\ref{416d}), the magnetic effects dissapear completely and we are left with the purely electrostatic interactions. \\
\indent This effect can be easily understood by recalling the specific way, in which the self-interactions do appear in RST. First, reconsider the relationship (\ref{220a})-(\ref{220b}) between the \textit{Maxwell currents} $\jalomu $ and the \textit{RST currents} $\jalumu$ . Obviously, the link between both currents is given by the fibre metric $\Kaubu $ which for the present two-particle configurations adopts the following form \cite{a43} for non-trivial self-interaction ($u\neq 0$): 
\begin{equation}
\{\Kaubu (u)\} = \left( \begin{array}{cccc}
                                      e^u \sinh u & -e^u \cosh u & 0 & 0 \\
                                      -e^u \cosh u & e^u \sinh u & 0 & 0 \\
                                      0 & 0 & 0 & -e^{-2u} \\
                                      0 & 0 & -e^{-2u} & 0 \end{array} \right) \ .
\end{equation}
\indent Therefore the inclusion of the self-interactions (i.e. $u\neq 0$) generalizes the former relationships (\ref{318a})-(\ref{318d}) to the following form \cite{a43}
\begin{subequations}\label{623}
\begin{align}
\jeomu & = -e^{-u}\lbrace \cosh u \cdot \keumu + \sinh u \cdot \kzumu \rbrace \label{623a} \\
\jzomu & = -e^{-u}\lbrace \sinh u \cdot \keumu + \cosh u \cdot \kzumu \rbrace \label{623b} \\
\jdomu & = e^{-2u} \hsmu \label{623c} \\
\jvomu & = -e^{-2u} \hmu \label{623d} \ .
\end{align}
\end{subequations}
\noindent Thus, putting here $u \rightarrow \infty$ lets vanish the Maxwellian exchange currents $\jdomu$ and $\jvomu$ so that the exchange potential $\Bmu$ must be put to zero in order to satisfy the last two Maxwell equations (\ref{211d})-(\ref{211e}). Moreover, the first two transformation relations (\ref{623a})-(\ref{623b}) degenerate for $u \rightarrow \infty$ to the following form
\begin{subequations}\label{624}
\begin{align}
\jeomu & \Rightarrow -\frac{1}{2}(\keumu + \kzumu )\label{624a} \\
\jzomu & \Rightarrow -\frac{1}{2}(\keumu + \kzumu )\label{624b} \ ;
\end{align}
\end{subequations}
\noindent i.e. both Maxwellian charge densities become identical (cf. (\ref{62d})):
\begin{equation}
j^1\;_0 = j^2\;_0 = -\frac{1}{2}({}^{(1)}k_0 + {}^{(2)}k_0) \equiv -{}^{(p)}k_0 \ ,\label{625}
\end{equation}
\noindent but the Maxwellian three-currents $\vec{j}_a = \lbrace -j^a\;_k; k=1,2,3\rbrace $ must vanish
\begin{equation}
\vec{j}_1 = \vec{j}_2 = -\frac{1}{2}(\vec{k}_1(\vec{r}) + \vec{k}_2(\vec{r})) \Rightarrow 0 \label{626}
\end{equation}
\noindent because for the $ns^2\ {}^1S_0$ states the RST currents $\vec{k}_a(\vec{r})$ are antiparallel, cf. (\ref{62e})
\begin{equation}
\vec{k}_1(\vec{r}) = -\vec{k}_2(\vec{r}) \doteqdot \vec{k}_p(\vec{r}) \label{627}\ .
\end{equation}
\noindent Clearly, when all the Maxwellian three-currents do vanish according to (\ref{626}), then all the magnetic potentials $A_a(r)$ (\ref{62b}) and $B(r)$ (\ref{62c}) are also zero as required by the Maxwell equations (\ref{27b})-(\ref{27e}).\\
\indent Thus, the result is that for $u \rightarrow \infty$ one ends up with a configuration which has been called "electrostatic approximation" \cite{a42}, because all the magnetostatic fields have been neglected. Now it turns out that this is an exact solution, namely for inclusion of the self-interactions, but with $u \rightarrow \infty$. However, the more important point with this type of solution refers here to the magnitude of the self-interactions. It is true, for vanishing self-interactions ($u=0$) the RST predictions for the energy differences $\Delta_{RST}\!E_{1\backslash 2}$ come close to the experimental values by less than $0,3 \%$ (fig.4); but when $u$ is increased from zero to infinity, the amount of lowering of the energy $E_T$ is not sufficient in order to shift the RST prediction $\Delta_{RST}\!E_{1\backslash 2}$ sufficiently close to the experimental value $\Delta_{exp}E_{1\backslash 2}$ for small and intermediate nuclear charge numbers $z_{ex}$ ($z_{ex}\lesssim 50$, say). This is the same effect as was previously encountered in connection with the ionization energy of the helium-like ions \cite{u5} where the value of the self-interaction parameter $u$ had to be fixed for the ground-state $1s^2\ {}^1S_0$ as $u=0.03052$ in order that the RST predictions meet with the observational data for the ionization energies in the range $z_{ex}\gtrsim 50$. However, for the \textit{smaller} values of nuclear charge $z_{ex}$ ($\lesssim 50$), such an acceptable fit of the RST parameter $u$ in order to adapt the RST predictions to the experimental data is not possible, neither for the ionization energies of ref. \cite{u5} nor for the present energy differences $\Delta E_{1\backslash 2}$.\\

\begin{center}
\bf fig.4
\end{center}

\indent In order to present a numerical demonstration of this effect, consider the nuclear charge number $z_{ex}=42$, molybdenum (see table I). The two-particle ground-state energy for $u=0$ is found by numerical integration as $\stackrel{\;\;_\circ}{E}^{(1)}_{T(2)}=973\ 577.2$ eV and the excited state $2s^2\ {}^1S_0$ has energy $\stackrel{\;\;_\circ}{E}^{(2)}_{T(2)}=1\ 009\ 805.2$ eV. Thus, for vanishing self-interaction ($u=0$), one finds the RST prediction $\Delta_{RST}\!\stackrel{\;\;_\circ}{E}_{1\backslash 2} = \stackrel{\;\;_\circ}{E}^{(2)}_{T(2)}-\stackrel{\;\;_\circ}{E}^{(1)}_{T(2)}=36\ 228 $eV. Now switch on the self-interactions and let $u$ adopt its ground-state value $u_1=0.03052$ which lowers the ground-state energy $\stackrel{\;\;_\circ}{E}^{(1)}_{T(2)}$ to ${E}^{(1)}_{T(2)}\Big\vert_{u_1}=973\ 575.6$ eV, i.e. only $1.6$ eV below its value for $u=0$! Furthermore, taking the most favourable value $u_2=\infty$ for the excited state $2s^2\ {}^1S_0$ yields the corresponding energy ${E}^{(2)}_{T(2)}\Big\vert_{u_2}=1\ 009\ 804.6$ eV, which means a lowering of $\stackrel{\;\;_\circ}{E}^{(2)}_{T(2)}$ by $0.6$ eV. Consequently the desired energy difference remains practically the same: $\Delta_{RST}\!E_{1\backslash 2}\doteqdot {E}^{(2)}_{T(2)}\Big\vert_{u_2}-{E}^{(1)}_{T(2)}\Big\vert_{u_1}=36\ 229$ eV, when one switches on the self-interactions; and thus the semiclassical RST predictions cannot be sufficiently improved for the excited state $2s^2\ {}^1S_0$ by means of taking into account the self-interactions. However, for large charge numbers $z_{ex}$, the energy shift $\Delta_{RST}\!\stackrel{\;\;_\circ}{E}_{1\backslash 2} \rightarrow \Delta_{RST}\!E_{1\backslash 2}$ becomes larger (fig.5) and therefore it becomes possible for large enough $z_{ex}$ to select a parameter value $u_2$ such that the RST prediction $\Delta_{RST}\!E_{1\backslash 2}$ (\ref{619}) coincides with the (presently unknown) experimental value $\Delta_{exp}\!E_{1\backslash 2}$. A similar effect occurs with the ground-state ionization energy \cite{u5}, see figs.4 and 5.

\begin{center}
\bf fig.5
\end{center}    

Of course, a rigorous proof of the claimed coincidence of the RST predictions $\Delta_{RST}\!E_{1\backslash 2}$ with the (missing) experimental data $\Delta_{exp}\!E_{1\backslash 2}$ of ref.\cite{m1} for large $z_{ex}$ is not possible, but a rough estimate may give nevertheless a first hint upon the correctness of that extrapolating claim. Observe here that the experimental one-electron ionization energy $J^{(1)}_{\!exp\!(2)}$ of the helium ground-state for molybdenum ($z_{ex}=42$) is $J^{(1)}_{\!exp\!(2)}=23\ 791.7$ eV, see ref. \cite{m1}; and this is roughly the same order of magnitude as the present RST energy difference $\stackrel{\;\;_\circ}{\Delta}_{RST}\! E_{1\backslash 2}=36\ 175.6$ eV, see Table I. However, not only the orders of magnitude are the same, but also the relative deviations $\Delta_{RST}$ of the RST predictions from the experimental values: For the ionization energies, one defines
\begin{equation}\label{628}
\stackrel{\;\;_\circ}{\Delta}^{(1)}_{RST} = \frac{\stackrel{\;\;_\circ}{J}^{(1)}_{\!RST(2)} - J^{(1)}_{\!exp(2)}}{J^{(1)}_{\!exp(2)}}
\end{equation}
\noindent and then finds for molybdenum ($z_{ex}=42$)
\begin{equation}\label{629}
\stackrel{\;\;_\circ}{\Delta}^{(1)}_{RST}\Big\vert_{z_{ex}=42} = \frac{23\ 850.0 - 23\ 791.7}{23\ 791.7} \rightarrow 0.16 \% \ ,
\end{equation}
\noindent see ref.\cite{u5}. On the other hand, one may define the analogous deviation for the present energy differences as 
\begin{equation}\label{630}
\stackrel{\;\;_\circ}{\Delta}^{(1\backslash 2)}_{RST} = \frac{\Delta_{RST}\stackrel{\;\;_\circ}{E}_{1\backslash 2} - \Delta_{exp}\! E_{1\backslash 2}}{\Delta_{exp}\! E_{1\backslash 2}}
\end{equation}
\noindent which yields for $z_{ex}=42$ by means of table I
\begin{equation}\label{631}
\stackrel{\;\;_\circ}{\Delta}^{(1\backslash 2)}_{RST}\Big\vert_{z_{ex}=42} = \frac{36\ 228.6 - 36\ 175.6}{36\ 175.6} \rightarrow 0.15 \% \ ,
\end{equation}
\noindent and this is just the same deviation as for the ionization energies (\ref{629})! (For a comparison of both deviations $\stackrel{\;\;_\circ}{\Delta}^{(1)}_{RST}$ and $\stackrel{\;\;_\circ}{\Delta}^{(1\backslash 2)}_{RST}$ as functions of charge number $z_{ex}$, see fig.4). Consequently, it seems reasonable to assume that for large $z_{ex}$ ($z_{ex}\gtrsim 40$, say) the RST predictions $\Delta_{RST} E_{1\backslash 2}$ can be optimally fitted to the experimental values $\Delta_{exp} E_{1\backslash 2}$ by choosing for the self-interaction parameter $u_2$ again some value in the vicinity of its ground-state value $u_1=0.03052$, see ref.\cite{u5}. Table II presents an overview of the corresponding RST predictions $\Delta_{RST} E_{1\backslash 2}$ for $u_2=u_1=0.03052$ for the higher nuclear charges ($z_{ex}\gtrsim 40$). Clearly, it must be left to the outcome of the future experiments for $z_{ex}\gtrsim 40$, whether this hypothesis $u_1\approx u_2$ is actually realized in nature.

\begin{center}
\bf Table II
\end{center} 

\bibliography{bibdb_2}

\begin{figure*}
\renewcommand{\arraystretch}{0.1}
{\footnotesize
\begin{tabular}{|r||r@{.}l|r@{.}l|r@{.}l||r@{.}l|r@{.}l|r@{.}l||r@{.}l|}\hline
& \multicolumn{2}{c|}{$\stackrel{\;\;_\circ}{E}^{(1)}_{T(2)}$} & \multicolumn{2}{c|}{$\stackrel{\;\;_\circ}{E}^{(2)}_{T(2)}$}& \multicolumn{2}{c||}{$\Delta_{RST}\!\stackrel{\;\;_\circ}{E}_{1\backslash 2}$} & \multicolumn{2}{c|}{$E^{(1)}_{T(2)}$} & \multicolumn{2}{c|}{$E^{(2)}_{T(2)}$} & \multicolumn{2}{c||}{$\Delta_{RST}\! E_{1\backslash 2}$} & \multicolumn{2}{c|}{$\Delta_{exp}\! E$}\\
\raisebox{1.0ex}[-1.0ex]{$z_{\rm ex}$} & \multicolumn{2}{c|}{$u=0$} & \multicolumn{2}{c|}{$u=0$} & \multicolumn{2}{c||}{(\ref{615}),$u=0$} & \multicolumn{2}{c|}{$u=\infty $} & \multicolumn{2}{c|}{$u=\infty $} & \multicolumn{2}{c||}{$u=\infty $} & \multicolumn{2}{c|}{\cite{m1,b56}} \\\hline\hline

     2  & 1021920 & 0 & 1021978 & 2 & 58 & 3 & 1021920 & 0 & 1029782 & 8 & 58 & 32 & 57 & 87 \\
     3   & 1021800 & 9 & 1021948 & 4 & 147 & 5 & 1021800 & 9 & 1021948 & 4 & 147 & 46 & - & - \\            
     4   & 1021627 & 4 & 1021904 & 8 & 277 & 48 & 1021627 & 4 & 1021904 & 8 & 277 & 44 & - & - \\        
     5   & 1021399 & 4 & 1021847 & 6 & 448 & 24 & 1021399 & 4 & 1021847 & 6 & 448 & 28 & - & - \\        
     6   & 1021116 & 9 & 1021779 & 9 & 659 & 97 & 1021116 & 8 & 1021776 & 8 & 660 & 00 & - & - \\     
     7   & 1020779 & 8 & 1021692 & 4 & 912 & 59 & 1020779 & 7 & 1021692 & 4 & 912 & 65 & - & - \\        
     8  & 1023881 &  3 & 1021594 & 3 & 1206 & 15 & 1020388 & 0 & 1021594 & 3 & 1206 & 25 & - & - \\            
     9   & 1019941 & 8 & 1021482 & 5 & 1540 & 67 & 1019941 & 6 & 1021482 & 4 & 1540 & 84 & - & - \\               
    10   & 1019440 & 7 & 1021356 & 9 & 1916 & 19 & 1019440 & 4 & 1021359 & 9 & 1916 & 54 & - & - \\
\bf 11&\bf 1018884&\bf 8&\bf 1021217&\bf 7&\bf 2332&\bf 86&\bf 1018884&\bf 4&\bf 1021217&\bf 6&\bf 2333&\bf 23&\bf 2325 & \bf 620 \\
\bf 12&\bf 1018274&\bf 0&\bf 1021064&\bf 7&\bf 2790&\bf 66&\bf 1018273&\bf 4&\bf 1021064&\bf 6&\bf 2791&\bf 16&\bf 2782 & \bf 300 \\
\bf 13&\bf 1017608&\bf 2&\bf 1020897&\bf 8&\bf 3289&\bf 63&\bf 1017607&\bf 4&\bf 1020897&\bf 8&\bf 3290&\bf 33&\bf 3280 & \bf 460 \\   
\bf 14&\bf 1016887&\bf 2&\bf 1020717&\bf 2&\bf 3829&\bf 94&\bf 1016886&\bf 4&\bf 1020717&\bf 1&\bf 3830&\bf 75&\bf 3820 & \bf 313 \\
\bf 15&\bf 1016111&\bf 1&\bf 1020522&\bf 6&\bf 4411&\bf 55&\bf 1016120&\bf 0&\bf 1020522&\bf 6&\bf 4412&\bf 61&\bf 4401 & \bf 250 \\
\bf 16&\bf 1015279&\bf 6&\bf 1020314&\bf 2&\bf 5034&\bf 62&\bf 1015278&\bf 2&\bf 1020314&\bf 1&\bf 5035&\bf 92&\bf 5022 & \bf 820 \\
\bf 17&\bf 1014392&\bf 6&\bf 1020091&\bf 8&\bf 5699&\bf 26&\bf 1014391&\bf 0&\bf 1020091&\bf 7&\bf 5700&\bf 83&\bf 5686 & \bf 783 \\
\bf 18&\bf 1013449&\bf 9&\bf 1019855&\bf 4&\bf 6405&\bf 56&\bf 1013448&\bf 0&\bf 1019855&\bf 4&\bf 6407&\bf 40&\bf 6392 & \bf 000 \\
    20  & 1011397 & 1 & 1019340 & 5 & 7943 & 37 & 1011394 & 4 & 1019340 & 4 & 7945 & 98 & - & - \\
\bf 22&\bf 1009119&\bf 8&\bf 1018769&\bf 0&\bf 9649&\bf 11&\bf 1009116&\bf 1&\bf 1018768&\bf 8&\bf 9652&\bf 64&\bf 9631 & \bf 588 \\
\bf 23&\bf 1007896&\bf 5&\bf 1018461&\bf 7&\bf 10565&\bf 25&\bf 1007892&\bf 3&\bf 1018161&\bf 6&\bf 10569&\bf 30&\bf 10546 & \bf 591 \\
\bf 24&\bf 1006616&\bf 5&\bf 1018140&\bf 2&\bf 11523&\bf 77&\bf 1006611&\bf 7&\bf 1018140&\bf 1&\bf 11528&\bf 37&\bf 11503 & \bf 997 \\
\bf 25&\bf 1005279&\bf 5&\bf 1017804&\bf 3&\bf 12524&\bf 80&\bf 1005274&\bf 1&\bf 1017804&\bf 2&\bf 12530&\bf 03&\bf 12503 & \bf 930 \\   
\bf 26&\bf 1003885&\bf 4&\bf 1017453&\bf 9&\bf 13568&\bf 52&\bf 1003879&\bf 3&\bf 1017453&\bf 7&\bf 13574&\bf 40&\bf 13546 & \bf 265 \\
\bf 27&\bf 1002433&\bf 9&\bf 1017089&\bf 0&\bf 14655&\bf 04&\bf 1002427&\bf 1&\bf 1017088&\bf 8&\bf 14661&\bf 69&\bf 14631 & \bf 622 \\   
\bf 28&\bf 1000924&\bf 8&\bf 1016709&\bf 4&\bf 15784&\bf 59&\bf 1000917&\bf 2&\bf 1016709&\bf 2&\bf 15791&\bf 99&\bf 15759 & \bf 754 \\
\bf 29&\bf  999357&\bf 8&\bf 1016315&\bf 1&\bf 16957&\bf 29&\bf 999349&\bf 3&\bf 1016314&\bf 9&\bf 16965&\bf 59&\bf 16930 & \bf 909 \\
\bf 30&\bf  997732&\bf 6&\bf 1015905&\bf 9&\bf 18173&\bf 33&\bf 997723&\bf 2&\bf 1015905&\bf 7&\bf 18182&\bf 56&\bf 18145 & \bf 086 \\    
\bf 36&\bf  986743&\bf 2&\bf 1013135&\bf 2&\bf 26392&\bf 01&\bf 986726&\bf 7&\bf 1013134&\bf 8&\bf 26408&\bf 10&\bf 26354 & \bf 080 \\
    40  & 978212 & 8 & 1010979 & 0 & 32766 & 14 & 978189 & 9 & 1010978 & 4 & 32788 & 48 & - & - \\
\bf 42&\bf 973577&\bf 2&\bf 1009805&\bf 2&\bf 36228&\bf 6&\bf 973575&\bf 6&\bf 1009804&\bf 6&\bf 36229&\bf 0&\bf 36175&\bf 611 \\
    50  & 952478 & 7 & 1004445 & 1 & 51966 & 49 & 952433 & 0 & 1004444 & 2 & 52011 & 19 & - & - \\
    60  & 920009 & 2 & 996138 & 4 & 76129 & 19 & 919928 & 4 & 996136 & 6 & 76208 & 23 & - & - \\
    70  & 880058 & 5 & 985819 & 0 & 105760 & 47 & 879926 & 4 & 985816 & 0 & 105889 & 63 & - & - \\
    80  & 831546 & 6 & 973137 & 6 & 141591 & 00 & 831342 & 5 & 973132 & 8 & 141790 & 31 & - & - \\
    90  & 772857 & 3 & 957566 & 4 & 184709 & 14 & 772554 & 0 & 957558 & 6 & 185004 & 60 & - & - \\
   100  & 701428 & 2 & 938258 & 9 & 236830 & 70 & 700988 & 6 & 938246 & 6 & 237258 & 03 & - & - \\ \hline
               &     \multicolumn{14}{c|}{{\it all data in}\/ eV}                            \\ \hline

\end{tabular}
}\\
\flushleft{{\large {\bf Table I}:\quad{\it Comparison of Energy Differences $\Delta E_{1\backslash 2}$ }}}\\[2ex]
\qquad The energy difference $\Delta E_{1\backslash 2}$ without self-interaction (\ref{615})(4th column) and with self-interaction (\ref{619}) (7th column) is compared to the experimental values $\Delta_{exp}\!E_{1\backslash 2}$ (last column). The relative deviations $\stackrel{\;\;_\circ}{\Delta}^{(1\backslash 2)}_{RST}$ (\ref{630}) are illustrated in fig.4. \textit{In contrast to the situation for large $z_{ex}$}, the inclusion of the RST self-interactions cannot improve the predictions, because for low and intermediate charge numbers $z_{ex}$ the lowering of the energies $E^{(n)}_{T(2)}$ by the RST self-interactions is too small, see fig.5.    
\end{figure*}

\setlength{\unitlength}{1cm}
\begin{figure*}
\epsfig{file=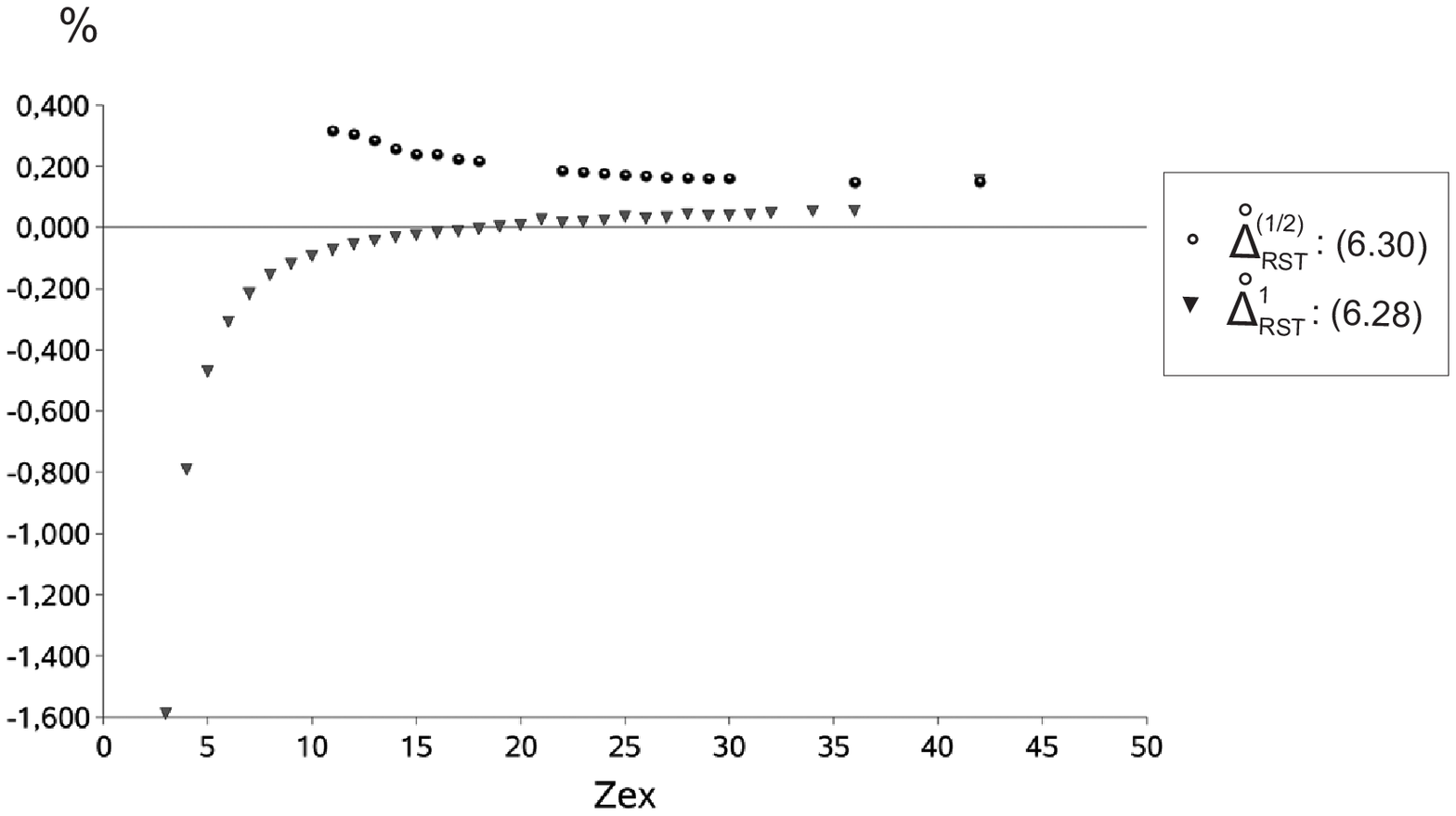, width=16cm}
\flushleft{{\large {\bf Fig.4}:\quad{\it Ionization Energies $\stackrel{\;\;_\circ}J^{(1)}_{RST(2)}$ and Energy Differences \\ \qquad \qquad $\Delta_{RST}\stackrel{\;\;_\circ}E_{1\backslash 2}$ }}}\\
\qquad The relative deviations $\stackrel{\;\;_\circ}{\Delta}^{(1)}_{RST}$ (\ref{628}) for the semiclassical ionization energy $\stackrel{\;\;_\circ}J^{(1)}_{RST(2)}$ and the deviation $\stackrel{\;\;_\circ}{\Delta}^{(1\backslash 2)}_{RST}$ (\ref{630}) for the semiclassical energy difference $\Delta_{RST}\stackrel{\;\;_\circ}E_{1\backslash 2}$ are of the same magnitude ($\lesssim 0,2\% $) for large $z_{ex}$($\gtrsim 40$). This suggests to adopt for the excited state $2s^2\ {}^1S_0$ the same value for the self-interaction parameter $u$ ($\rightarrow 0,03052$) as for the ground-state $1s^2\ {}^1S_0$, see ref. \cite{u5}.
\end{figure*}

\setlength{\unitlength}{1cm}
\begin{figure*}
\rotatebox{270}{\epsfig{file=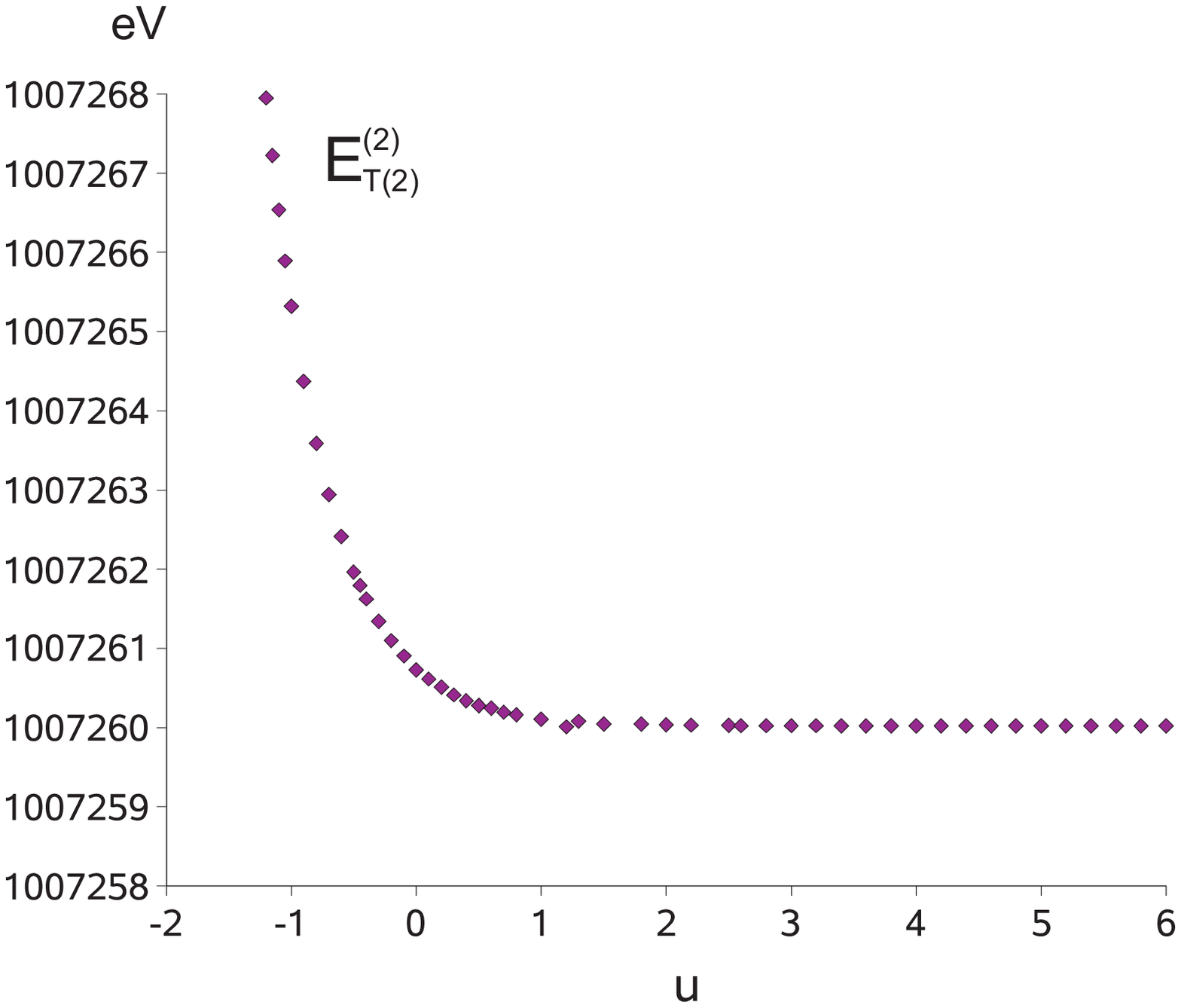,height=15cm}}
\flushleft{{\large {\bf Fig.5}:\quad{\it RST Energy $E^{(2)}_{T(2)}$ for the excited state $2s^2\ {}^{1}S_0$}}}\\
\qquad The total energy $E^{(2)}_{T(2)}$ for $z_{ex}=46$ (Pd) as a function of the self-interaction parameter $u$ decreases too weakly \textit{for small and intermediate values of $z_{ex}$} in order that the energy difference $\Delta_{RST}E_{1\backslash 2}$ (\ref{619}) can be fitted to the experimental value $\Delta_{exp}E_{1\backslash 2}$ of ref.\cite{m1} by means of a suitable choice of $u$. 
\end{figure*}

\begin{figure}
\renewcommand{\arraystretch}{0.6}
\footnotesize{
\begin{tabular}{|r||r@{.}l|r@{.}l||r@{.}l|}\hline
& \multicolumn{2}{c|}{$E^{(1)}_{T(2)}$} & \multicolumn{2}{c||}{$E^{(2)}_{T(2)}$} & \multicolumn{2}{c|}{$\Delta_{RST}E_{1\backslash 2}$, (\ref{619})} \\
\raisebox{1.0ex}[-1.0ex]{$z_{\rm ex}$} & \multicolumn{2}{c|}{$u=0.03052$} & \multicolumn{2}{c||}{$u=0.03052$} & \multicolumn{2}{c|}{$u=0.03052$} \\\hline\hline

     42 & 973575 & 64 & 1009805 & 11 & 36229 & 47 \\
     46  & 963545 & 13 & 1007260 & 69 & 43715 & 55 \\          
     50  & 952475 & 00 & 1004445 & 09 & 51969 & 09 \\       
     55  & 937122 & 63 & 1000526 & 23 & 63403 & 60 \\       
     60  & 920004 & 34 & 996138  & 26 & 76133 & 92 \\    
     65  & 901020 & 81 & 991248  & 80 & 90227 & 99 \\       
     70 & 880050 & 75 & 985818  & 84 & 105768& 09 \\           
     75  & 856948 & 10 & 979801  & 03 & 122852& 92 \\             
     80  & 831534 & 59 & 973137  & 35 & 141602& 76 \\
     85  & 803590 & 27 & 965755  & 85 & 162165& 58 \\
     90  & 772839 & 40 & 957565  & 90 & 184726& 50 \\
     95  & 738929 & 93 & 948451  & 08 & 209521& 15 \\ 
    100  & 701401 & 89 & 938258  & 17 & 236856& 28 \\ \hline
        & \multicolumn{6}{c|}{{\it all data in}\/ eV} \\\hline
\end{tabular}
}\\
\flushleft{{\large {\bf Table II}:\quad{\it RST predictions for the energy difference $\Delta E_{1\backslash 2}$ }}}\\
\qquad Adopting the same value $u=0,03052$ for both the ground state $1s^2\ {}^1S_0$ and the excited state $2s^2\ {}^1S_0$ yields the RST predictions for the energy difference $\Delta_{RST}E_{1\backslash 2}$ (\ref{619}) of these states in the limit of large $z_{ex}$($\gtrsim 40$) where observational data are not yet available.
\end{figure}
\end{document}

%% file: ellips.pstex_t
\begin{picture}(0,0)%
\includegraphics{ellips.pstex}%
\end{picture}%
\setlength{\unitlength}{4144sp}%
\begingroup\makeatletter\ifx\SetFigFont\undefined%
\gdef\SetFigFont#1#2#3#4#5{%
  \reset@font\fontsize{#1}{#2pt}%
  \fontfamily{#3}\fontseries{#4}\fontshape{#5}%
  \selectfont}%
\fi\endgroup%
\begin{picture}(7091,3351)(1801,-4930)
\put(4488,-4624){\makebox(0,0)[lb]{\smash{{\SetFigFont{12}{14.4}{\familydefault}{\mddefault}{\updefault}{\color[rgb]{0,0,0}$\Sb$}%
}}}}
\put(8206,-4592){\makebox(0,0)[lb]{\smash{{\SetFigFont{12}{14.4}{\rmdefault}{\mddefault}{\updefault}{\color[rgb]{0,0,0}$\Sa$}%
}}}}
\put(5453,-3982){\makebox(0,0)[lb]{\smash{{\SetFigFont{9}{10.8}{\rmdefault}{\mddefault}{\updefault}{\color[rgb]{0,0,0}$|S=1;S_z=0>$}%
}}}}
\put(7844,-3982){\makebox(0,0)[lb]{\smash{{\SetFigFont{9}{10.8}{\rmdefault}{\mddefault}{\updefault}{\color[rgb]{0,0,0}$|S=0;S_z=0>$}%
}}}}
\put(3421,-3976){\makebox(0,0)[lb]{\smash{{\SetFigFont{9}{10.8}{\rmdefault}{\mddefault}{\updefault}{\color[rgb]{0,0,0}$|S=1;S_z=-1>$}%
}}}}
\put(1801,-3976){\makebox(0,0)[lb]{\smash{{\SetFigFont{9}{10.8}{\rmdefault}{\mddefault}{\updefault}{\color[rgb]{0,0,0}$|S=1;S_z=+1>$}%
}}}}
\end{picture}%

%% file: splitting.pstex_t
\begin{picture}(0,0)%
\includegraphics{splitting.pstex}%
\end{picture}%
\setlength{\unitlength}{4144sp}%
\begingroup\makeatletter\ifx\SetFigFont\undefined%
\gdef\SetFigFont#1#2#3#4#5{%
  \reset@font\fontsize{#1}{#2pt}%
  \fontfamily{#3}\fontseries{#4}\fontshape{#5}%
  \selectfont}%
\fi\endgroup%
\begin{picture}(6975,4386)(2296,-5641)
\put(2701,-5461){\makebox(0,0)[lb]{\smash{\SetFigFont{9}{10.8}{\familydefault}{\mddefault}{\updefault}{\color[rgb]{0,0,0}1}%
}}}
\put(5626,-3661){\makebox(0,0)[lb]{\smash{\SetFigFont{12}{14.4}{\familydefault}{\mddefault}{\updefault}{\color[rgb]{0,0,0}ns n's}%
}}}
\put(3511,-5461){\makebox(0,0)[lb]{\smash{\SetFigFont{9}{10.8}{\familydefault}{\mddefault}{\updefault}{\color[rgb]{0,0,0}0}%
}}}
\put(8866,-5506){\makebox(0,0)[lb]{\smash{\SetFigFont{9}{10.8}{\familydefault}{\mddefault}{\updefault}{\color[rgb]{0,0,0}-1}%
}}}
\put(8101,-5506){\makebox(0,0)[lb]{\smash{\SetFigFont{9}{10.8}{\familydefault}{\mddefault}{\updefault}{\color[rgb]{0,0,0} 0}%
}}}
\put(3241,-2761){\makebox(0,0)[lb]{\smash{\SetFigFont{14}{16.8}{\familydefault}{\mddefault}{\updefault}{\color[rgb]{0,0,0}HF, RST}%
}}}
\put(7426,-1411){\makebox(0,0)[lb]{\smash{\SetFigFont{14}{16.8}{\familydefault}{\mddefault}{\updefault}{\color[rgb]{0,0,0}Standard theory}%
}}}
\put(2296,-3751){\makebox(0,0)[lb]{\smash{\SetFigFont{12}{14.4}{\familydefault}{\mddefault}{\updefault}{\color[rgb]{0,0,0}$\fSa$}%
}}}
\put(2296,-5371){\makebox(0,0)[lb]{\smash{\SetFigFont{12}{14.4}{\familydefault}{\mddefault}{\updefault}{\color[rgb]{0,0,0}$\fSb$}%
}}}
\put(4051,-5461){\makebox(0,0)[lb]{\smash{\SetFigFont{9}{10.8}{\familydefault}{\mddefault}{\updefault}{\color[rgb]{0,0,0}$\fSd$}%
}}}
\put(8011,-3751){\makebox(0,0)[lb]{\smash{\SetFigFont{12}{14.4}{\familydefault}{\mddefault}{\updefault}{\color[rgb]{0,0,0}$\fdst$}%
}}}
\put(9271,-2221){\makebox(0,0)[lb]{\smash{\SetFigFont{12}{14.4}{\familydefault}{\mddefault}{\updefault}{\color[rgb]{0,0,0}$\fSe$}%
}}}
\put(7066,-5506){\makebox(0,0)[lb]{\smash{\SetFigFont{9}{10.8}{\familydefault}{\mddefault}{\updefault}{\color[rgb]{0,0,0}$\fSc$}%
}}}
\put(9226,-5371){\makebox(0,0)[lb]{\smash{\SetFigFont{12}{14.4}{\familydefault}{\mddefault}{\updefault}{\color[rgb]{0,0,0}$\fSf$}%
}}}
\put(3151,-4516){\makebox(0,0)[lb]{\smash{\SetFigFont{12}{14.4}{\familydefault}{\mddefault}{\updefault}{\color[rgb]{0,0,0}$\fdhf$}%
}}}
\end{picture}

%% file: entwurf_newnew.bbl
\begin{thebibliography}{25}
\expandafter\ifx\csname natexlab\endcsname\relax\def\natexlab#1{#1}\fi
\expandafter\ifx\csname bibnamefont\endcsname\relax
  \def\bibnamefont#1{#1}\fi
\expandafter\ifx\csname bibfnamefont\endcsname\relax
  \def\bibfnamefont#1{#1}\fi
\expandafter\ifx\csname citenamefont\endcsname\relax
  \def\citenamefont#1{#1}\fi
\expandafter\ifx\csname url\endcsname\relax
  \def\url#1{\texttt{#1}}\fi
\expandafter\ifx\csname urlprefix\endcsname\relax\def\urlprefix{URL }\fi
\providecommand{\bibinfo}[2]{#2}
\providecommand{\eprint}[2][]{\url{#2}}

\bibitem[{\citenamefont{Cohen-Tannoudji
  et~al.}(1996)\citenamefont{Cohen-Tannoudji, Diu, and Lalo\"e}}]{b7}
\bibinfo{author}{\bibfnamefont{C.}~\bibnamefont{Cohen-Tannoudji}},
  \bibinfo{author}{\bibfnamefont{B.}~\bibnamefont{Diu}}, \bibnamefont{and}
  \bibinfo{author}{\bibfnamefont{F.}~\bibnamefont{Lalo\"e}},
  \emph{\bibinfo{title}{Quantum Mechanics}}, vol. \bibinfo{volume}{1\&2}
  (\bibinfo{publisher}{{Hermann and John Wiley \& Sons, Inc.}},
  \bibinfo{address}{Paris}, \bibinfo{year}{1996}).

\bibitem[{\citenamefont{Selleri}(1992)}]{b42}
\bibinfo{editor}{\bibfnamefont{F.}~\bibnamefont{Selleri}}, ed.,
  \emph{\bibinfo{title}{Wave\bs Particle Duality}} (\bibinfo{publisher}{Plenum
  Press}, \bibinfo{address}{New York and London}, \bibinfo{year}{1992}).

\bibitem[{\citenamefont{Mahan}(2000)}]{b30}
\bibinfo{author}{\bibfnamefont{G.~D.} \bibnamefont{Mahan}},
  \emph{\bibinfo{title}{Many Particle Physics}} (\bibinfo{publisher}{Plenum
  Press}, \bibinfo{address}{New York}, \bibinfo{year}{2000}).

\bibitem[{\citenamefont{Baym and Pethick}(1991)}]{b39}
\bibinfo{author}{\bibfnamefont{G.}~\bibnamefont{Baym}} \bibnamefont{and}
  \bibinfo{author}{\bibfnamefont{C.}~\bibnamefont{Pethick}},
  \emph{\bibinfo{title}{Landau Fermi Liquid Theory}} (\bibinfo{publisher}{Wiley
  Interscience}, \bibinfo{address}{New York}, \bibinfo{year}{1991}).

\bibitem[{\citenamefont{Savage and Das}(2000)}]{b40}
\bibinfo{editor}{\bibfnamefont{C.~M.} \bibnamefont{Savage}} \bibnamefont{and}
  \bibinfo{editor}{\bibfnamefont{M.~P.} \bibnamefont{Das}}, eds.,
  \emph{\bibinfo{title}{Bose\bs Einstein Condensation}}
  (\bibinfo{publisher}{World Scientific}, \bibinfo{address}{Singapore},
  \bibinfo{year}{2000}).

\bibitem[{\citenamefont{Gross}(1982)}]{a23}
\bibinfo{author}{\bibfnamefont{F.}~\bibnamefont{Gross}},
  \bibinfo{journal}{Phys. Rev.} \textbf{\bibinfo{volume}{C 26}},
  \bibinfo{pages}{2203} (\bibinfo{year}{1982}).

\bibitem[{\citenamefont{Landau}(1990)}]{b31}
\bibinfo{author}{\bibfnamefont{R.~H.} \bibnamefont{Landau}},
  \emph{\bibinfo{title}{Quantum Mechanics}}, vol.~\bibinfo{volume}{II}
  (\bibinfo{publisher}{John Wiley \& Sons}, \bibinfo{address}{New York},
  \bibinfo{year}{1990}).

\bibitem[{\citenamefont{Wilson}(1988)}]{b15}
\bibinfo{author}{\bibfnamefont{S.}~\bibnamefont{Wilson}},
  \emph{\bibinfo{title}{Relativistic Effects in Atoms and Molecules}},
  vol.~\bibinfo{volume}{2} of \emph{\bibinfo{series}{Methods in Computational
  Chemistry}} (\bibinfo{publisher}{Plenum Press}, \bibinfo{address}{New York},
  \bibinfo{year}{1988}).

\bibitem[{\citenamefont{Gorceix et~al.}(1987)\citenamefont{Gorceix, Indelicato,
  and Desclaux}}]{a3}
\bibinfo{author}{\bibfnamefont{O.}~\bibnamefont{Gorceix}},
  \bibinfo{author}{\bibfnamefont{P.}~\bibnamefont{Indelicato}},
  \bibnamefont{and} \bibinfo{author}{\bibfnamefont{J.~P.}
  \bibnamefont{Desclaux}}, \bibinfo{journal}{J. Phys.}
  \textbf{\bibinfo{volume}{B 20}}, \bibinfo{pages}{639/651}
  (\bibinfo{year}{1987}).

\bibitem[{\citenamefont{Plante et~al.}(1994)\citenamefont{Plante, Johnson, and
  Sapirstein}}]{a5}
\bibinfo{author}{\bibfnamefont{D.~R.} \bibnamefont{Plante}},
  \bibinfo{author}{\bibfnamefont{W.~R.} \bibnamefont{Johnson}},
  \bibnamefont{and}
  \bibinfo{author}{\bibfnamefont{J.}~\bibnamefont{Sapirstein}},
  \bibinfo{journal}{Phys. Rev.} \textbf{\bibinfo{volume}{A 49}},
  \bibinfo{pages}{3519} (\bibinfo{year}{1994}).

\bibitem[{\citenamefont{Drake}(1988)}]{a6}
\bibinfo{author}{\bibfnamefont{G.~W.} \bibnamefont{Drake}},
  \bibinfo{journal}{Can. J. Phys.} \textbf{\bibinfo{volume}{66}},
  \bibinfo{pages}{586} (\bibinfo{year}{1988}).

\bibitem[{\citenamefont{Drake}(1996)}]{b54}
\bibinfo{author}{\bibfnamefont{G.~W.~F.} \bibnamefont{Drake}},
  \emph{\bibinfo{title}{Atomic, Molecular and Optical Physics Handbook}}
  (\bibinfo{publisher}{Am. Inst. Phys.}, \bibinfo{address}{New York},
  \bibinfo{year}{1996}).

\bibitem[{\citenamefont{Pru\ss-Hunzinger and Sorg}(2003)}]{a16}
\bibinfo{author}{\bibfnamefont{S.}~\bibnamefont{Pru\ss-Hunzinger}}
  \bibnamefont{and} \bibinfo{author}{\bibfnamefont{M.}~\bibnamefont{Sorg}},
  \bibinfo{journal}{Nuov. Cim.} \textbf{\bibinfo{volume}{B 118}},
  \bibinfo{pages}{903} (\bibinfo{year}{2003}).

\bibitem[{\citenamefont{Schust et~al.}(2004)\citenamefont{Schust, Mattes, and
  Sorg}}]{a9}
\bibinfo{author}{\bibfnamefont{P.}~\bibnamefont{Schust}},
  \bibinfo{author}{\bibfnamefont{M.}~\bibnamefont{Mattes}}, \bibnamefont{and}
  \bibinfo{author}{\bibfnamefont{M.}~\bibnamefont{Sorg}},
  \bibinfo{journal}{Found. Phys} \textbf{\bibinfo{volume}{34}},
  \bibinfo{pages}{99} (\bibinfo{year}{2004}).

\bibitem[{\citenamefont{Schust et~al.}(2005)\citenamefont{Schust, Stary,
  Mattes, and Sorg}}]{a43}
\bibinfo{author}{\bibfnamefont{P.}~\bibnamefont{Schust}},
  \bibinfo{author}{\bibfnamefont{F.}~\bibnamefont{Stary}},
  \bibinfo{author}{\bibfnamefont{M.}~\bibnamefont{Mattes}}, \bibnamefont{and}
  \bibinfo{author}{\bibfnamefont{M.}~\bibnamefont{Sorg}},
  \bibinfo{journal}{Found. Phys. {\bf 35}, 1043}  (\bibinfo{year}{2005}).

\bibitem[{\citenamefont{Pru\ss-Hunzinger
  et~al.}(2005)\citenamefont{Pru\ss-Hunzinger, Stary, Mattes, and Sorg}}]{u5}
\bibinfo{author}{\bibfnamefont{S.}~\bibnamefont{Pru\ss-Hunzinger}},
  \bibinfo{author}{\bibfnamefont{F.}~\bibnamefont{Stary}},
  \bibinfo{author}{\bibfnamefont{M.}~\bibnamefont{Mattes}}, \bibnamefont{and}
  \bibinfo{author}{\bibfnamefont{M.}~\bibnamefont{Sorg}},
  \bibinfo{journal}{Nuov. Cim. {\bf B 120}, 467}  (\bibinfo{year}{2005}).

\bibitem[{m1()}]{m1}
\emph{\bibinfo{title}{\rm { National Institute of Standards and Technology
  (NIST), Atomic Spectra Database}}}, \bibinfo{howpublished}{{\it
  http://physics.nist.gov/PhysRefData/ASD/index.html}}.

\bibitem[{\citenamefont{Stary and Sorg}()}]{u6}
\bibinfo{author}{\bibfnamefont{F.}~\bibnamefont{Stary}} \bibnamefont{and}
  \bibinfo{author}{\bibfnamefont{M.}~\bibnamefont{Sorg}},
  \emph{\bibinfo{title}{{\it Ortho- and Para-Helium in Relativistic
  Schr\"odinger Theory}}}, \bibinfo{note}{preprint (2005)}.

\bibitem[{\citenamefont{Verschl et~al.}(2004)\citenamefont{Verschl, Mattes, and
  Sorg}}]{a15}
\bibinfo{author}{\bibfnamefont{M.}~\bibnamefont{Verschl}},
  \bibinfo{author}{\bibfnamefont{M.}~\bibnamefont{Mattes}}, \bibnamefont{and}
  \bibinfo{author}{\bibfnamefont{M.}~\bibnamefont{Sorg}},
  \bibinfo{journal}{Europ. Phys. J.} \textbf{\bibinfo{volume}{A 20}},
  \bibinfo{pages}{211} (\bibinfo{year}{2004}).

\bibitem[{\citenamefont{Ballentine}(1998)}]{b25}
\bibinfo{author}{\bibfnamefont{L.~E.} \bibnamefont{Ballentine}},
  \emph{\bibinfo{title}{Quantum Mechanics}} (\bibinfo{publisher}{World
  Scientific}, \bibinfo{address}{Singapore}, \bibinfo{year}{1998}).

\bibitem[{\citenamefont{Merzbacher}(1970)}]{b27}
\bibinfo{author}{\bibfnamefont{E.}~\bibnamefont{Merzbacher}},
  \emph{\bibinfo{title}{Quantum Mechanics}} (\bibinfo{publisher}{J. Wiley \&
  Sons}, \bibinfo{address}{N.Y.}, \bibinfo{year}{1970}).

\bibitem[{\citenamefont{Pru\ss-Hunzinger
  et~al.}(2004)\citenamefont{Pru\ss-Hunzinger, Mattes, and Sorg}}]{a41}
\bibinfo{author}{\bibfnamefont{S.}~\bibnamefont{Pru\ss-Hunzinger}},
  \bibinfo{author}{\bibfnamefont{M.}~\bibnamefont{Mattes}}, \bibnamefont{and}
  \bibinfo{author}{\bibfnamefont{M.}~\bibnamefont{Sorg}},
  \bibinfo{journal}{Nuov. Cim {\bf B}} \textbf{\bibinfo{volume}{119}},
  \bibinfo{pages}{277} (\bibinfo{year}{2004}).

\bibitem[{\citenamefont{Rupp and Sorg}(2003)}]{a27}
\bibinfo{author}{\bibfnamefont{S.}~\bibnamefont{Rupp}} \bibnamefont{and}
  \bibinfo{author}{\bibfnamefont{M.}~\bibnamefont{Sorg}},
  \bibinfo{journal}{Nuov. Cim. B} \textbf{\bibinfo{volume}{118}},
  \bibinfo{pages}{259} (\bibinfo{year}{2003}).

\bibitem[{\citenamefont{Schust and Sorg}()}]{a42}
\bibinfo{author}{\bibfnamefont{P.}~\bibnamefont{Schust}} \bibnamefont{and}
  \bibinfo{author}{\bibfnamefont{M.}~\bibnamefont{Sorg}},
  \emph{\bibinfo{title}{Magnetic interactions in relativistic two-particle
  systems}}, \bibinfo{howpublished}{\rm http://arxiv.org/abs/hep-th/0410023}.

\bibitem[{\citenamefont{Bashkin and Stoner}(1975)}]{b56}
\bibinfo{author}{\bibfnamefont{S.}~\bibnamefont{Bashkin}} \bibnamefont{and}
  \bibinfo{author}{\bibfnamefont{J.}~\bibnamefont{Stoner}},
  \emph{\bibinfo{title}{Atomic Energy Levels \& Grotrian Diagrams 1}}
  (\bibinfo{publisher}{North-Holland Publishing Company}, \bibinfo{address}{New
  York}, \bibinfo{year}{1975}).

\end{thebibliography}
